\documentclass[12pt]{article}

\usepackage{amssymb,amsmath}
\usepackage{epsfig}
\newcommand{\be}{\begin{equation}}
\newcommand{\ee}{\end{equation}}
\newcommand{\bea}{\begin{eqnarray}}
\newcommand{\eea}{\end{eqnarray}}

\begin{document}

\begin{center}
\bf{CP-VIOLATION AND UNITARITY TRIANGLE TEST OF THE STANDARD
MODEL}\footnote{On the basis of the lectures given to the students
of SISSA (Trieste) in 2006}
\end{center}

\begin{center}
S. M. Bilenky
\end{center}

\begin{center}
{\em  Joint Institute for Nuclear Research, Dubna, R-141980, Russia}
\end{center}
\begin{abstract}
Phenomenological issues of the $CP$ violation in the quark sector of
the Standard Model are discussed. We consider quark mixing in the
SM, standard and Wolfenstein parametrization of the $CKM$ mixing
matrix and unitarity triangle. We discuss the phenomenology of the
$CP$ violation in  $K^{0}_{L}$  and   $B_{d}^{0} ( \bar
B_{d}^{0})$-decays. The standard unitarity triangle fit of the
existing data is discussed. In appendix A we compare the
$K^{0}\leftrightarrows \bar K^{0}$, $B_{d,s}^{0}\leftrightarrows
\bar B^{0}_{d,s}$ etc oscillations with neutrino oscillations. In
Appendix B we derive the evolution equation for $M^{0}- \bar M^{0}$
system in the Weisskopf-Wigner approximation.
\end{abstract}

\section{Introduction}
Soon after the discovery of the violation of  parity $P$ and charge
conjugation $C$ in the  weak interaction \cite{Wu57} (1957)  Landau
\cite{Landau57} and Lee and Yang \cite{Lee-Yang57} suggested that
the Hamiltonian of the weak interaction is invariant under the
combined $CP$ transformation. One of the  consequence of this
suggestion was the theory of the two-component neutrino
\cite{Landau57,Lee-Yang57,Salam57} according to which the neutrino
is left-handed (right-handed) particle and antineutrino is
right-handed (left-handed) particle. The helicity of the neutrino
was measured in spectacular experiment \cite{Goldhaber58} performed
in 1958. This experiment confirmed the theory of the two-component
neutrino. It was established that neutrino is left-handed particle.

The confirmation of the theory of the two-component neutrino
strengthened belief in the hypothesis of the $CP$ invariance of the
Hamiltonian of the weak interaction. All existing data at the end of
fifties and beginning of sixties data were in agreement with this
hypothesis.

It was a big surprise for the physics community  when in the
experiment performed by Christenson, Cronin, Fitch and Turlay
\cite{Cronin}  in 1964 the decay $K^{0}_{L}\to \pi^{+}\pi^{-}$ was
observed. The observation of this decay was a proof that $CP$ is
violated.\footnote{In fact, let us consider decays of short-lived
and long-lived kaons ($K^{0}_{S}$ and $K^{0}_{L}$) into
$\pi^{+}+\pi^{-}$ in the rest frame of the kaon. Because spin of the
kaon is equal to zero, final pions have equal to zero orbital
momentum. Thus, we have $P~
|\pi^{+}~\pi^{-}\rangle=|\pi^{+}~\pi^{-}\rangle$,
$C~|\pi^{+}~\pi^{-}\rangle=|\pi^{-}~\pi^{+}\rangle=|\pi^{+}~\pi^{-}\rangle$
and $CP~|\pi^{+}~\pi^{-}\rangle=|\pi^{+}~\pi^{-}\rangle$. The decay
$K_{S}\to \pi^{+}+\pi^{-}$ is the main  decay mode of the
short-lived kaon. If $CP$  is conserved, $|K_{S}\rangle$ is the
state with $CP$-parity equal to 1. The $CP$  parity of the
orthogonal state $K_{L}$ must be equal to -1 and hence decay
$K_{L}\to \pi^{+}+\pi^{-}$ must be forbidden in the case of the $CP$
conservation.}

The  discovery of the $CP$ violation was announced  at the Rochester
conference in Dubna. In 1980 Cronin and Fitch were awarded the Nobel
Prize for this discovery.

The observed violation of $P$ and $C$ in the $\beta$ decay and other
weak decays was large. Discovered by Cronin, Fitch and others effect
of the violation of $CP$  was  very small. They found that the ratio
of the modulus of the amplitudes of the $CP$-forbidden decay
$K^{0}_{L}\to \pi^{+}+\pi^{-}$ and the $CP$-allowed decay
$K^{0}_{S}\to \pi^{+}+\pi^{-}$ was about $2\cdot10^{-3}$.

The first problem was to understand what interaction is responsible
for the $CP$ violation in  $K^{0}_{L}\to \pi+\pi$ decays. Many
hypothesis were put forward. One of the most viable idea was
proposed by Wolfenstein \cite{Wolfenstein64}. He noticed that it is
possible to explain the observed violation of the  $CP$ in decays of
$K^{0}_{L}$-meson if we assume that exist a new $|\Delta S|=2$
interaction, which  is characterized by a very small effective
interaction constant $G_{SW}\simeq 10^{-9}\,G_{F}$ ($G_{F}$ is the
Fermi constant). This interaction was called the superweak
interaction.

Measurable parameters characterizing violation of $CP$  in $K_{L}\to
\pi+\pi$ decays are $\eta_{+ -}$ and  $\eta_{0 0}$. These parameters
are, correspondingly,  ratios of the amplitudes of the decays
$K^{0}_{L}\to \pi^{+}+\pi^{-}$ and $K^{0}_{S}\to \pi^{+}+\pi^{-}$
and  $K^{0}_{L}\to \pi^{0}+\pi^{0}$ and $K^{0}_{S}\to
\pi^{0}+\pi^{0}$. If the superweak interaction is responsible for
the violation of the $CP$ in $K^{0}_{L}\to \pi+\pi$ decays in this
case
\begin{equation}\label{1}
\eta_{+ -}=\eta_{0 0}.
\end{equation}
It took many years of enormous experimental efforts \cite{NA48,KTeV}
in order to check the relation (\ref{1}). It was proved that the
relation (\ref{1}) does not valid. Thus, superweak interaction as a
possible source of the $CP$ violation in the neutral kaon decays was
excluded by these experiments.

At the time when experiments \cite{NA48,KTeV} were completed the
 Glashow \cite{Glashow61}, Weinberg
\cite{Weinberg67}, Salam \cite{Salam68} Standard Model (SM) was
established by numerous experiments. The expected in the SM
violation of the relation (\ref{1}) is very small (see
\cite{BurasI}). The data of the experiments \cite{NA48,KTeV} were in
agreement with the SM.

In 1973  Kobayashi and Maskawa \cite{Kobayashi-Maskawa73} considered
$CP$ violation in the framework of the Standard Model. In the
Standard Model violation of the $CP$ is determined by phases in the
unitary mixing matrix. In 1973 only two families of leptons and
quarks were known. It was demonstrated in \cite{Kobayashi-Maskawa73}
that it is impossible to violate $CP$ in this case. It was shown in
\cite{Kobayashi-Maskawa73} that in order to explain observed $CP$
violation we need to assume that (at least) six quarks exist.
Kobayashi and Maskawa obtained the first parametrization of the
mixing matrix in the case of three families. They showed that this
matrix is characterized by three mixing angles and one $CP$ phase.

During more than 30 years the  investigation of the $CP$  violation
was limited by the system of neutral kaons (see  book \cite{CP}).
During last 8 years with the BaBar and Belle  experiments at the
asymmetric B-factories at the SLAC and KEK a new era in the
investigation of the $CP$ violation started (see book \cite{BaBar}).
In these experiments numerous effects of the $CP$ violation in
different decays of the neutral and charged $B_{d}$-mesons were
observed. This allowed to perform the unitarity triangle test of the
SM. {\em All existing at present data are in a good agreement with
the SM and the assumption that  only three families of quarks exist
in nature}.

In this review we will  consider some phenomenological aspects of
the problem of the $CP$ violation in the quark sector. In the
section 2 we consider the SM Higgs mechanism of the mixing of
quarks. In the section 3 we consider in details quark mixing matrix
and the $CP$ violation. In the section 4 we derive the standard
parametrization of the $CKM$ mixing matrix. In section 5 we discuss
the values of the modulus of the elements of the $CKM$ matrix. In
section 6 we consider Wolfenstein parametrization of the $CKM$
matrix elements and the unitarity triangle. In the section 7 we
obtain eigenstates and eigenvalues of the effective Hamiltonian of
$K^{0}-\bar K^{0}$, $B^{0}-\bar B^{0}$, etc systems. In the section
8 we consider in details phenomenology of the $CP$ violation in
decays of $K_{L}^{0}$. In the section 9 we consider the $CP$
violation in $B^{0}-\bar B^{0}$ decays. In the section 10 we present
results of the unitarity triangle test of the Standard Model. In the
Appendix A we compare of $K^{0}\leftrightarrows\bar K^{0}$,
$B^{0}\leftrightarrows\bar B^{0}$ etc oscillations with neutrino
oscillations. In the Appendix B we derive in the Weisskopf-Wigner
approximation the evolution equation for $K^{0}-\bar K^{0}$,
$B^{0}-\bar B^{0}$ etc system.

Last years,  in connection with appearance of the B-factories,
several books \cite{Branco,Bigi}, many reviews
\cite{BaBar,Bphysics,Kleinknecht,Battaglia,Silva,Nir,Fleisher,BurasI,BurasII,SuperB}
and hundreds papers on the $CP$ violation were published. In these
books and reviews many details and many references on original
papers can be found.

I tried  to discuss here some basic questions and to derive
different relations. I hope that this review will be useful for
those who start to study this exciting field of physics.

\section{ Quark mixing in the Standard Model}
The Standard Model of the electroweak interaction is based on the
following principles (see, for example, \cite{Weinberg, Bilenky})
\begin{enumerate}
\item The local gauge $SU_{L}(2)\times U_{Y}(1)$ symmetry of the Lagrangian of
the fields of massless quarks, leptons, gauge vector bosons and
scalar Higgs bosons.
\item The spontaneous symmetry breaking. Due to  the
spontaneous  breaking of the local $SU_{L}(2)\times U_{Y}(1)$
symmetry the
 masses of $W^{\pm}$ and $Z^{0}$ bosons, mass terms of quarks and
 leptons and mass of the Higgs boson are generated.

\item Unification of the weak and electromagnetic interactions
\end{enumerate}
We will consider the quark sector of the Standard Model. The theory
is based on the assumption that the left-handed  quark fields are
transformed as $SU_{L}(2)$ doublets\footnote{The meaning of primes
will be clear later.}
\begin{eqnarray}
\psi_{1L}=\left(\begin{array}{c}u'_{L}\\d'_{L}\end{array}\right),~~~
\psi_{2L}=\left(\begin{array}{c}c'_{L}\\s'_{L}\end{array}\right),~~~
\psi_{3L}=\left(\begin{array}{c}t'_{L}\\b'_{L}\end{array}\right)
\label{3}
\end{eqnarray}
and the right-handed fields of quarks $q'_{R}$ ($q=u,d,c,s,t,b$) are
the singlets of the group.

The requirements of the local gauge $SU_{L}(2)\times U_{Y}(1)$
invariance fix the Lagrangian of the interaction of quarks and
vector bosons in the form of the sum of the charged current (CC),
neutral current (NC) and electromagnetic (EM) parts:
\begin{eqnarray}
\mathcal{L}_{I}^{CC}&=&-
\frac{g}{2\sqrt{2}}~j_{\alpha}^{CC}~W^{\alpha}+\rm{h.c.},\nonumber\\
\mathcal{L}_{I}^{NC}&=&-\frac{g}{2
\cos\theta_{W}}~j_{\alpha}^{NC}~Z^{\alpha},\nonumber\\
 \mathcal{L}_{I}^{EM}&=&-e
~j_{\alpha}^{EM}~A^{\alpha},
 \label{4}
\end{eqnarray}
where
\begin{equation}\label{5}
j_{\alpha}^{CC}=2\sum^{3}_{i=1}\bar
\psi_{iL}~\frac{1}{2}(\tau_{1}+i\tau_{2})~\gamma_{\alpha}~\psi_{iL}=
2~[\bar u'_{L}~\gamma_{\alpha}~d'_{L}+ \bar
c'_{L}~\gamma_{\alpha}~s'_{L}+\bar t'_{L}~\gamma_{\alpha}~b'_{L}]
\end{equation}
is the quark charged current,
\begin{equation}\label{7}
j_{\alpha}^{NC}=2\sum^{3}_{i=1}\bar
\psi_{iL}~\frac{1}{2}\tau_{3}~\gamma_{\alpha}~\psi_{iL}-
2\sin^{2}\theta_{W}j_{\alpha}^{EM}
\end{equation}
is the quark neutral current and
\begin{equation}\label{7}
j_{\alpha}^{EM}=\sum_{q=u,d,c,...}e_{q}~\bar q'~\gamma_{\alpha}~q'
\end{equation}
is the electromagnetic current. Here $W^{\alpha}$ is the field of
$W^{\pm}$ bosons, $Z^{\alpha}$ is the field of $Z^{0}$ bosons,
$A^{\alpha}$ is the electromagnetic field, $g$ is the electroweak
constant, $\theta_{W}$ is the weak angle, $e_{q}=2/3,-1/3$ are the
quark charges.

In the total Lagrangian of the Standard Model enter the following
$SU_{L}(2)\times U_{Y}(1)$ invariant Lagrangians of the Yukawa
interaction of  quarks and Higgs fields
\begin{equation}\label{8}
\mathcal{L}_{Y}^{\rm{down}}=-\frac{\sqrt{2}}{v}\,\sum_{i=1,2,3~q=d,s,b}
\bar\psi_{iL}~M_{iq}^{\rm{down}}~ q'_{R}~\phi +\rm{h.c.}
\end{equation}
and
\begin{equation}\label{9}
\mathcal{L}_{Y}^{\rm{up}}= -\frac{\sqrt{2}}{v}\,\sum_{i=1,2,3~
q=u,c,t}\bar\psi_{iL}~M_{iq}^{\rm{up}}~q'_{R}~\tilde{\phi}.
+\rm{h.c.}
\end{equation}
Here $M^{\rm{down}}$ and $M^{\rm{up}}$ are complex $3\times 3$
matrices,
$\phi=\left(\begin{array}{c}\phi_{+}\\\phi_{0}\end{array}\right)$ is
the Higgs doublet, $\tilde{\phi}=i\tau_{2}\phi^{*}$ and $v$ is the
constant (vacuum expectation value of the Higgs field).

If we choose
\begin{eqnarray}
\phi(x)=\left(\begin{array}{c}0\\\frac{v+\chi(x)}{\sqrt{2}}\end{array}\right),
\label{10}
\end{eqnarray}
where $\chi(x)$ is the field of neutral Higgs bosons, the symmetry
will be spontaneously broken. For the mass terms of up and down
quarks we obtain the following expressions
\begin{equation}\label{11}
\mathcal{L}_{m}^{\rm{up}}=-\overline{
U'}_{L}~M^{\rm{up}}~U'_{R}+\rm{h.c.},~~
\mathcal{L}_{m}^{\rm{down}}=-\overline{
D'}_{L}~M^{\rm{down}}~D'_{R}+\rm{h.c.},
\end{equation}
where
\begin{eqnarray}
U'_{L,R}=\left(\begin{array}{c}u'_{L,R}\\c'_{L,R}\\t'_{L,R}\end{array}\right),~~~
D'_{L,R}=\left(\begin{array}{c}d'_{L,R}\\s'_{L,R}\\b'_{L,R}\end{array}\right).
\label{12}
\end{eqnarray}

 The complex matrices $M^{\rm{up}}$ and $M^{\rm{down}}$ can be diagonalized by
the biunitary transformations
\begin{equation}\label{13}
M^{\rm{up}}=V^{\rm{up}}_{L}~m^{\rm{up}}~V^{\rm{up}\dagger}_{R},~~~
M^{\rm{down}}=V^{\rm{down}}_{L}~m^{\rm{down}}~V^{\rm{down}\dagger}_{R}.
\end{equation}
Here $V_{L,R}^{\rm{up}}$ and $V_{L,R}^{\rm{down}}$ are unitary
matrices and $m^{\rm{up}}$ and $ m^{\rm{down}}$ are diagonal
matrices with positive diagonal elements.

From (\ref{11}) and (\ref{13}) we find
\begin{equation}\label{14}
\mathcal{L}_{m}^{\rm{up}}=-\bar
U~m^{\rm{up}}~U,~~~\mathcal{L}_{m}^{\rm{down}}=-\bar
D~m^{\rm{down}}~D.
\end{equation}
Here
\begin{eqnarray}
U=U_{L}+ U_{R}=\left(\begin{array}{c}u\\c\\t\end{array}\right),~~~
D=D_{L}+ D_{R}=\left(\begin{array}{c}d\\s\\b\end{array}\right),
\label{15}
\end{eqnarray}
\begin{eqnarray}
m^{\rm{up}}=\left(\begin{array}{ccc}m_{u}& 0&0\\0 &m_{c}& 0\\0& 0&
m_{t}\end{array}\right),~~~
m^{\rm{down}}=\left(\begin{array}{ccc}m_{d}& 0&0\\0 &m_{s}& 0\\0& 0&
m_{b}\end{array}\right)\label{15a}
\end{eqnarray}
 and
\begin{equation}\label{16}
U_{L,R}=V_{L,R}^{\rm{up}\dag}~U'_{L,R},~~~D_{L,R}=V_{L,R}^{\rm{down}\dag}~D'_{L,R}.
\end{equation}

From (\ref{14}), (\ref{15}) and (\ref{16}) we obtain the standard
mass terms for up and down quarks
\begin{equation}\label{17}
\mathcal{L}_{m}^{\rm{up}}(x)=-\sum_{q=u,c,t}m_{q}~\bar q
(x)\,q(x),~~~
\mathcal{L}_{m}^{\rm{down}}(x)=-\sum_{q=d,s,b}m_{q}~\bar q(x) \,q
(x)\
\end{equation}
Thus, $q(x)$ is the field of the $q$-quarks with the mass $m_{q}$
($q=u,d,c,s,t,b$). The left-handed and right-handed fields of quarks
with definite masses and primed quark fields, which have definite
transformation properties, are connected by the unitary
transformations (\ref{16}).

Let us consider now the charged current  of the quarks. From
(\ref{5}) and (\ref{16}) we find
\begin{equation}\label{18}
j_{\alpha}^{CC}=2~\bar U'_{L}~\gamma_{\alpha}~D'_{L}= 2~\bar
U_{L}~\gamma_{\alpha}~V~D_{L}= 2~[\bar
u_{L}~\gamma_{\alpha}~d^{\rm{mix}}_{L}+ \bar
c_{L}~\gamma_{\alpha}~s^{\rm{mix}}_{L}+\bar
t_{L}~\gamma_{\alpha}~b^{\rm{mix}}_{L}].
\end{equation}
Here
\begin{equation}\label{19}
V=(V_{L}^{\rm{up}})^{\dag}~V_{L}^{\rm{down}}
\end{equation}
and
\begin{equation}\label{20}
d^{\rm{mix}}_{L}=\sum_{d_1=d,s,b} V_{ud_1}~d_{1L},~~
s^{\rm{mix}}_{L}=\sum_{d_1=d,s,b}
V_{cd_1}~d_{1L},~~b^{\rm{mix}}_{L}=\sum_{d_1=d,s,b}V_{td_1}~d_{1L}.
\end{equation}
From (\ref{19}) it follows that $V$ is unitary matrix\footnote{We
assume  that there are no additional heavy families of quarks.}
\begin{equation}\label{21}
V^{\dagger}V=1.
\end{equation}
From (\ref{18}) and (\ref{20}) we conclude that \emph{fields of down
quarks enter into CC of the SM  in the form of the "mixed "
combinations $d^{\rm{mix}}_{L}$, $s^{\rm{mix}}_{L}$,
$b^{\rm{mix}}_{L}$}. The unitary 3$\times$3 mixing matrix $V$ is
called Cabibbo \cite{Cabibbo}-Kobayashi-Maskawa
\cite{Kobayashi-Maskawa73} ($CKM$) mixing matrix. We will see later
that \emph{ the violation of the $CP$ invariance is determined in
the SM by the matrix $V$}.

Let us consider now the electromagnetic current. From (\ref{7}) we
have
\begin{equation}\label{22}
j_{\alpha}^{EM}=\frac{2}{3}~(\bar U'_{L}~\gamma_{\alpha}~U'_{L}
+\bar U'_{R}~\gamma_{\alpha}~U'_{R})- \frac{1}{3}~(\bar
D'_{L}~\gamma_{\alpha}~D'_{L} +\bar D'_{R}~\gamma_{\alpha}~D'_{R}).
\end{equation}
Taking into account the unitarity of the matrices
$V_{L,R}^{\rm{up}}$ and $V_{L,R}^{\rm{down}}$, we find
\begin{equation}\label{23}
j_{\alpha}^{EM}=\frac{2}{3}~(\bar U_{L}~\gamma_{\alpha}~U_{L} +\bar
U_{R}~\gamma_{\alpha}~U_{R})- \frac{1}{3}~(\bar
D_{L}~\gamma_{\alpha}~D_{L} +\bar D_{R}~\gamma_{\alpha}~D_{R})=
\sum_{q=u,d,c,...}e_{q}~\bar q~\gamma_{\alpha}~q,
\end{equation}
where $e_{u,c,t}=\frac{2}{3}$ and $e_{d,s,b}=-\frac{1}{3}$. Thus, we
come to the standard expression for the electromagnetic current
which is diagonal in the quark flavors.

Let us consider  the neutral current.  We have
\begin{eqnarray}j_{\alpha}^{NC}&=&2\sum^{3}_{i}\bar
\psi_{iL}~\frac{1}{2}\tau_{3}~\gamma_{\alpha}~\psi_{iL}
-2\sin^{2}\theta_{W}j_{\alpha}^{EM} \nonumber\\&=&\bar
U'_{L}~\gamma_{\alpha}~U'_{L}-\bar
D'_{L}~\gamma_{\alpha}~D'_{L}-2\sin^{2}\theta_{W}j_{\alpha}^{EM}
\nonumber\\& =& \sum_{u_1=u,c,t}\bar u_{1L}~\gamma_{\alpha}~u_{1L}-
\sum_{d_1=d,s,b}\bar
d_{1L}~\gamma_{\alpha}~d_{1L}-2\sin^{2}\theta_{W}j_{\alpha}^{EM}.\label{24}
\end{eqnarray}
Thus,  the neutral current of the SM is also diagonal in the quark
flavors. Only the charged current changes flavor of the quarks
($s\to u +W^{-}$ etc). We will show later that the electromagnetic
and NC interactions of the SM automatically conserve $CP$. The $CP$
invariance can be violated only by the flavor-changing $CC$
interaction.

\section{Mixing matrix}
We will consider here general properties of the  unitary mixing
matrix $V$. Let us calculate first the number of the angles and
phases which characterize the unitary mixing matrix $V$ in the
general $n\times n $ case.

The unitary matrix $V$ can be presented in the form $V=e^{iH}$,
where $H$ is the hermitian matrix. Such matrix is characterized by
$n $(diagonal~ elements) +2~($\frac{n^{2}-n}{2}$) (nondiagonal~
elements)=$n^{2}$ real parameters.

The number of the angles which characterize $n\times n $ unitary
matrix coincides with the number of parameters which characterize
$n\times n $ orthogonal matrix $O$ ($O^{T}O=1$). Such matrix can be
presented in the form $O=e^{A}$, where $A^{T}=-A$ . The
antisymmetric matrix $A$ is characterized by
$\frac{n(n-1)}{2}(\rm{nondiagonal~ elements})$ real parameters.
Thus, the number of the angles which characterize the unitary matrix
is equal to
\begin{equation}\label{25}
n_{\rm{angles}}=\frac{n(n-1)}{2}.
\end{equation}
Other parameters of the matrix $V$ are phases. The number of the
phases is equal to
\begin{equation}\label{26}
n_{\rm{phases}}=n^{2}-\frac{n(n-1)}{2}=\frac{n(n+1)}{2}.
\end{equation}
The number of {\em physical phases}, which characterize mixing
matrix, is significantly smaller than $n_{\rm{phases}}$.

The mixing matrix enter into CC together with the quark fields:
\begin{equation}\label{27}
j_{\alpha}^{CC}=2\sum_{u_{1}=u,c,t~d_{1}=d,s,b}\bar
u_{1L}~\gamma_{\alpha}~V_{u_{1}d_{1}}~d_{1L}.
\end{equation}
The free Lagrangian of quark fields is invariant under the
transformation
\begin{equation}\label{28}
    q(x)\to e^{i\alpha_{q}}~q(x),~~q=u,d,...
\end{equation}
where $\alpha_{q}$ is an arbitrary constant phase. quark fields are
We will take this fact into account in the calculation of the number
of physical phases in the mixing matrix $V$.

The unitary matrix can be presented in the form
\begin{equation}\label{29}
V=S^{\dag}(\alpha)~\tilde{V}~S(\beta),
\end{equation}
where $S(\alpha)$ and  $S(\beta)$ are diagonal phase matrices
($S_{u_{1}u_{2}}(\alpha)=\delta_{u_{1}u_{2}}~e^{i\alpha_{u_{1}}};~~
S_{d_{1}d_{2}}(\beta)=\delta_{d_{1}d_{2}}~e^{i\beta_{d_{1}}}$) and
$\tilde{V}$ is an unitary matrix. There are $2(n-1)+1$ independent
phases $\alpha_{u_{1}}$ and $\beta_{d_{1}}$\footnote{We must take
into account that only difference of common phases of $S(\beta)$ and
$S(\alpha)$ enters into (\ref{29})}.

The  phase factors $e^{i\alpha_{u_{1}}}$ and $e^{i\beta_{d_{1}}}$
can be included into quark fields. Thus, the number of measurable,
physical phases  which characterize unitary mixing matrix
$\tilde{V}$ is equal to
\begin{equation}\label{30}
n_{\rm{phases}}^{\rm{phys}}=\frac{n(n+1)}{2}-(2n-1)=\frac{(n-1)(n-2)}{2}.
\end{equation}
Let us obtain now  the constraints on the mixing matrix which follow
from the requirements of the $CP$ invariance of the  $CC$
interaction. For the $CC$ Lagrangian we have
\begin{eqnarray}
\mathcal{L}_{I}^{CC}(x)&=&-\frac{g}{\sqrt{2}}\sum_{u_{1}=u,c,t~
d_1=d,s,b}
\bar u_{1L}(x)~\gamma^{\alpha}~V_{u_{1}\,d_1}~d_{1L}(x)~W_{\alpha}(x)\nonumber\\
&-&\frac{g}{\sqrt{2}}\sum_{u_{1}=u,c,t~ d_1=d,s,b} \bar
d_{1L}(x)~\gamma^{\alpha}~V^{*}_{u_{1}\,d_1}~u_{1L}(x)~W^{\dag}_{\alpha}(x),
\label{31}\end{eqnarray} where $V$ is the $3\times3$ unitary $CKM$
 mixing matrix (we suppressed tilde).

The $CP$ is conserved if  Lagrangian satisfies the following
condition
\begin{equation}\label{32}
V_{CP}~\mathcal{L}_{I}^{CC}(x)~V^{-1}_{CP}=\mathcal{L}_{I}^{CC}(x'),
\end{equation}
where $V_{CP}$ is the operator of the $CP$  conjugation and
$x'=(x^{0},-\vec{x})$.

For the left-handed quark field $q_{L}(x)$ we have
\begin{equation}\label{33}
V_{CP}~q_{L}(x)~V^{-1}_{CP}=e^{-2i\alpha_{q}}~\gamma^{0}~C~\bar
q^{T}_{L}(x').
\end{equation}
Here $\alpha_{q}$ is an arbitrary phase and $C$ is the matrix of the
charge conjugation, which satisfies the relations
\begin{equation}\label{34}
C~\gamma^{T}_{\alpha}~C^{-1}=-\gamma_{\alpha},~~C^{T}=-C.
\end{equation}
Taking into account that  phases of quark fields are arbitrary, we
can   include  phase factor $e^{i\alpha_{q}}$ into the field $q(x)$.
We  obtain in this case
\begin{equation}\label{35}
V_{CP}~q_{L}(x)~V^{-1}_{CP}=\gamma^{0}~C~\bar q^{T}_{L}(x').
\end{equation}
From  (\ref{34}) from (\ref{35}) we also have
\begin{equation}\label{36}
V_{CP}~\bar q_{L}(x)~V^{-1}_{CP}=-q^{T}_{L}(x')~C^{-1}\gamma^{0}.
\end{equation}
Let us consider now the current $\bar
u_{1L}(x)~\gamma_{\alpha}~d_{1L}(x)$. From (\ref{34}), (\ref{35})
and (\ref{36}) we find
\begin{eqnarray}
V_{CP}~\bar u_{1L}(x)~\gamma_{\alpha}~d_{1L}(x)~V^{-1}_{CP}&=&
-u^{T}_{1L}(x')~C^{-1}\gamma^{0}~\gamma_{\alpha}~\gamma^{0}~C~\bar
d_{1L}(x')\nonumber\\&=&-\delta_{\alpha}~\bar
d_{1L}(x')~\gamma_{\alpha}~u_{1L}(x'). \label{37}
\end{eqnarray}
Here $\delta=(1,-1,-1,-1)$ is the sign factor. Notice that in  the
relation (\ref{37}) we took into account anticommutator properties
of the fermion fields.

Under the $CP$ transformation the field of the vector $W^{\pm}$
bosons is transformed as follows
\begin{equation}\label{38}
V_{CP}~W_{\alpha}(x)~V^{-1}_{CP}=-e^{-2i\beta_{W}}~\delta_{\alpha}~W^{\dag}_{\alpha}(x'),
\end{equation}
where $\beta_{W}$ is an arbitrary phase. Taking into account that
phase of the nonhermitian $W_{\alpha}(x)$ field is  arbitrary, we
can include phase factor $e^{i\beta_{W}}$ into the $W$ field. In
this case we have
\begin{equation}\label{39}
V_{CP}~W_{\alpha}(x)~V^{-1}_{CP}=-\delta_{\alpha}~W^{\dag}_{\alpha}(x').
\end{equation}
With the help of (\ref{31}), (\ref{37}) and  (\ref{39}) we find
\begin{eqnarray}
V_{CP}~\mathcal{L}_{I}^{CC}(x)~V^{-1}_{CP}&=&
-\frac{g}{\sqrt{2}}\sum_{u_{1}, d_1}
\bar d_{1L}(x')~\gamma^{\alpha}~V_{u_{1}\,d_1}~u_{1L}(x')~W^{\dag}_{\alpha}(x')\nonumber\\
&-&\frac{g}{\sqrt{2}}\sum_{u_{1}, d_1}  \bar
u_{1L}(x')~\gamma^{\alpha}~V^{*}_{u_{1}\,d_1}~d_{1L}(x')~W_{\alpha}(x').
\label{40}
\end{eqnarray}
From (\ref{31}), (\ref{32}) and (\ref{40}) we conclude that in the
case of the $CP$ invariance the $CKM$ mixing matrix $V$ is real:
\begin{equation}\label{41}
V_{u_{1}\,d_1}=V_{u_{1}\,d_1}^{*}
\end{equation}
We will comment now this condition. The first term of the CC
Lagrangian (\ref{31}) is responsible for the flavor-changing
transition
\begin{equation}\label{42}
    d_1 \to u_1 +W^{-}, ~~d_1=d,s,b,~~u_1=u,c,t.
\end{equation}
Amplitude of this transition is equal to $V_{u_{1}\,d_1}$. The
second term of the Lagrangian (\ref{31}) is responsible for the
$CP$-conjugated transition
\begin{equation}\label{43}
    \bar d_1 \to \bar u_1 +W^{+}, ~~\bar d_1=\bar d,\bar s,\bar b,~~\bar u_1=
    \bar u,\bar c,\bar t.
\end{equation}
 Because  the Lagrangian is hermitian the
amplitude of the transition (\ref{43}) is equal to
$V^{*}_{u_{1}\,d_1}$. If the $CP$ invariance  holds the amplitude of
transition (\ref{42}) is  equal to the amplitudes of $CP$-conjugated
transition (\ref{43}).

As we have shown the number of the physical phases in the $CKM$
mixing matrix is given by (\ref{30}). For $n=2$ the mixing matrix is
real. Thus, for two families of quarks the unitarity of the mixing
matrix assures invariance  of the Lagrangian of interaction of the
quarks and $W$-bosons under $CP$ transformation.\footnote{In order
to explain in the framework of the SM observed violation of the $CP$
invariance we need to assume that (at least) three families of
quarks exist in nature. This was original argument of  Kobayashi and
Maskawa \cite{Kobayashi-Maskawa73} in favor of the existence of the
third family of quarks. When this argument was presented  only two
families of quarks were known.}

For $n=3$ number of  measurable phases in the mixing matrix is equal
to one.\footnote{The minimal number of families at which the CC
Lagrangian of the SM can violate $CP$ is equal to three. This
minimal number is equal to the number of SM families of quarks and
leptons which exist in nature. In fact, it was established by the
experiments on the measurement of the width of the decay $Z\to
\nu+\bar\nu$ that the number of flavor neutrinos is equal to three.
(see \cite{PDG}). This means that the number of the lepton families
is equal to three. For the SM to be renormalizable the number of the
quark families must be also equal to three.} It follows from
(\ref{41}) that in the case of the $CP$ invariance this phase must
be equal to zero.

We have considered the CC part of the SM interaction Lagrangian.
 Let us discuss now the neutral current and
electromagnetic interactions. From (\ref{34}), (\ref{35}) and
(\ref{36}) for the left-handed current we have
\begin{equation}\label{44}
V_{CP}~\bar q_{L}(x)~\gamma_{\alpha}~q_{L}(x)~V^{-1}_{CP}
=-\delta_{\alpha}~\bar q_{L}(x')~\gamma_{\alpha}~q_{L}(x').
\end{equation}
Analogously, for right-handed current we obtain
\begin{equation}\label{45}
V_{CP}\bar q_{R}(x)~\gamma_{\alpha}~q_{R}(x)~V^{-1}_{CP}
=-\delta_{\alpha}~\bar q_{R}(x')~\gamma_{\alpha}~q_{R}(x').
\end{equation}
Taking into account that
\begin{equation}\label{46}
V_{CP}~Z^{\alpha}(x)~V^{-1}_{CP}=-\delta_{\alpha}~Z^{\alpha}(x'),~~
V_{CP}~A^{\alpha}(x)~V^{-1}_{CP}=-\delta_{\alpha}~A^{\alpha}(x')
\end{equation}
from  (\ref{4}), (\ref{23}) and (\ref{24}) we find
\begin{equation}\label{47}
V_{CP}~\mathcal{L}_{I}^{NC}(x)~V^{-1}_{CP}=\mathcal{L}_{I}^{NC}(x'),~~~
V_{CP}~\mathcal{L}_{I}^{EM}(x)~V^{-1}_{CP}=\mathcal{L}_{I}^{EM}(x').
\end{equation}
Thus, the SM Lagrangians of the NC and electromagnetic interactions
are automatically invariant under $CP$ transformation. This is
connected with the fact that the electromagnetic and neutral current
interactions of the SM are diagonal in the quark flavors.

We have chosen $CP$ phase factors of quark and $W$ fields equal to
one and determined $CP$ transformations by the relations (\ref{35})
and (\ref{39}). In this case $CKM$ matrix  is characterized by three
angles and one phase responsible for the violation of the $CP$
invariance. It is of interest to characterize $CP$ violation in a
rephrasing-invariant way \cite{Jarlskog}.

Let us consider quantities
\begin{equation}\label{48}
Q^{d_{1}d_{2}}_{u_{1}u_{2}}=V_{u_{1}d_{1}}~V_{u_{2}d_{2}}~V^{*}_{u_{1}d_{2}}~
V^{*}_{u_{2}d_{1}}
\end{equation}
invariant under  phase transformation
\begin{equation}\label{49}
V_{u_{i}d_{k}}\to
e^{-i\alpha_{u_{i}}}~V_{u_{i}d_{k}}~e^{i\beta_{d_{k}}},
\end{equation}
where $\alpha_{u_{i}}$ and $\beta_{d_{k}}$ are arbitrary phases. It
is evident that
\begin{equation}\label{50}
(Q^{d_{1}d_{2}}_{u_{1}u_{2}})^{*}=Q^{d_{2}d_{1}}_{u_{1}u_{2}}=
Q^{d_{1}d_{2}}_{u_{2}u_{1}}.
\end{equation}
If we  determine the $CP$ conjugation by the relations (\ref{33})
and (\ref{38}) with arbitrary $CP $ phases of  the quark and $W$
fields from  the $CP$ invariance of the CC Lagrangian we find
\begin{equation}\label{51}
e^{2i\alpha_{u_1}}~V_{u_{1}~d_{1}}~e^{-2i\alpha_{d_1}}e^{-2i\beta_{W}}=
V^{*}_{u_{1}~d_{1}}.
\end{equation}
It follows from (\ref{48}) and (\ref{51})  that in the case of the
$CP$ invariance the quantities $Q^{d_{1}d_{2}}_{u_{1}u_{2}}$ are
real:
\begin{equation}\label{52}
Q^{d_{1}d_{2}}_{u_{1}u_{2}}= V^{*}_{u_{1}
d_{1}}~~V^{*}_{u_{2}d_{2}}~
V_{u_{1}d_{2}}~V_{u_{2}d_{1}}=(Q^{d_{1}d_{2}}_{u_{1}u_{2}})^{*}.
\end{equation}
Let us introduce the quantities
\begin{equation}\label{53}
J^{d_{1}d_{2}}_{u_{1}u_{2}}= \rm{Im}~Q^{d_{1}d_{2}}_{u_{1}u_{2}}.
\end{equation}
In the case of the  $CP$ invariance we have
\begin{equation}\label{54}
J^{d_{1}d_{2}}_{u_{1}u_{2}}=0.
\end{equation}
In the general case of the $CP$ violation from (\ref{50}) we obtain
the following relations
\begin{equation}\label{55}
    J^{d_{1}d_{2}}_{u_{1}u_{2}}=-J^{d_{2}d_{1}}_{u_{1}u_{2}},~~
    J^{d_{1}d_{2}}_{u_{1}u_{2}}=-J^{d_{1}d_{2}}_{u_{2}u_{1}}.
\end{equation}
Thus, $J^{d_{1}d_{2}}_{u_{1}u_{2}}\neq 0$ only if $d_{1}\neq d_{2}$
and $u_{1}\neq u_{2}$.

Further, from the unitarity of the mixing matrix we find
\begin{equation}\label{56}
\sum_{d_1}Q^{d_{1}d_{2}}_{u_{1}u_{2}}=\delta_{u_{1}u_{2}}~V_{u_{2}d_{2}}~
V^{*}_{u_{1}d_{2}},~~\sum_{u_1}Q^{d_{1}d_{2}}_{u_{1}u_{2}}=\delta_{d_{1}d_{2}}~
V_{u_{2}d_{2}}~ V^{*}_{u_{2}d_{1}}.
\end{equation}
From these relations we have
\begin{equation}\label{57}
\sum_{d_{1}}J^{d_{1}d_{2}}_{u_{1}u_{2}}=0,~~~\sum_{u_{1}}J^{d_{1}d_{2}}_{u_{1}u_{2}}=0.
\end{equation}
Let us consider first the simplest case of two families. We have in
this case
\begin{equation}\label{58}
J^{ds}_{uc}=0.
\end{equation}
This result corresponds to the absence of the physical phases in the
mixing matrix for n=2.

We will consider now the case of three families. From the first
relation ({57}) we have
\begin{equation}\label{59}
J^{sd}_{u_{1}u_{2}}+J^{bd}_{u_{1}u_{2}}=0,~
J^{ds}_{u_{1}u_{2}}+J^{bs}_{u_{1}u_{2}}=0,~
J^{db}_{u_{1}u_{2}}+J^{sb}_{u_{1}u_{2}}=0.
\end{equation}
It follows from (\ref{59}) and (\ref{55}) that the following cycling
relations hold
\begin{equation}\label{60}
J^{ds}_{u_{1}u_{2}}=J^{sb}_{u_{1}u_{2}}=J^{bd}_{u_{1}u_{2}}.
\end{equation}
From the second relation (\ref{57}) we obtain following equations
\begin{equation}\label{61}
J^{d_{1}d_{2}}_{cu}+J^{d_{1}d_{2}}_{tu}=0,~
J^{d_{1}d_{2}}_{uc}+J^{d_{1}d_{2}}_{tc}=0,~
J^{d_{1}d_{2}}_{ut}+J^{sd}_{ct}=0.
\end{equation}
From these relations and (\ref{55}) we find
\begin{equation}\label{62}
J^{d_{1}d_{2}}_{uc}=J^{d_{1}d_{2}}_{ct}=J^{d_{1}d_{2}}_{tu}.
\end{equation}
From (\ref{60}) and  (\ref{62}) we obtain the following relations
\begin{equation}\label{63}
J^{d s }_{u c}=J^{d s }_{ ct}=J^{d s }_{t u }=J^{s b }_{u c}=J^{b d
 }_{u c}=...=J.
\end{equation}
Other nonzero  $J^{d_{1}d_{2}}_{u_{1}u_{2}}$  differ from $J$ by
sign ($J^{b s }_{u c}= -J$ etc). Thus, in the case of three families
exist only one independent rephrasing invariant quantity. This
result is determined by  the fact that for n=3 there is only one
physical phase in the mixing matrix. The quantity $J$ is called
Jarskog invariant.

\section{Standard parametrization of the $CKM$ mixing matrix}
Several parameterizations of the unitary $CKM$ mixing matrix $V$
were proposed in literature. We will obtain here the so called
standard parametrization \cite{PDG} which is based on the three
Euler rotations.

Let us consider three orthogonal and normalized vectors
\begin{equation}\label{64}
|d\rangle,~~ |s\rangle ~~\rm{and}~~|b\rangle .
\end{equation}
In order to obtain three general "mixed" vectors we will perform the
three Euler rotations. The first rotation will be performed at the
angle $\theta_{12}$ around the vector $|b\rangle $. New orthogonal
and normalized vectors are
\begin{eqnarray} |d\rangle'=& c_{12}~|d\rangle
+s_{12}~|s\rangle\nonumber\\
|s\rangle'=& -s_{12}~|d\rangle +c_{12}~|s\rangle\nonumber\\
|b\rangle'=&|b\rangle, \label{65}
\end{eqnarray}
where $c_{12}=\cos\theta_{12}$ and $s_{12}=\sin\theta_{12}$. In the
matrix form (\ref{65}) can be written as follows
\begin{equation}\label{66}
|D\rangle'=V'~|D\rangle.
\end{equation}
Here
\begin{eqnarray}\label{67}
|D\rangle'=\left(%
\begin{array}{c}
 |d\rangle'\\
 |s\rangle'\\
  |b\rangle'\\
\end{array}%
\right),~~~~
|D\rangle=\left(%
\begin{array}{c}
 |d\rangle\\
 |s\rangle\\
  |b\rangle\\
\end{array}%
\right)
\end{eqnarray}
and
\begin{eqnarray}\label{68}
V'=\left(%
\begin{array}{ccc}
  c_{12} & s_{12}& 0\\
  -s_{12} & c_{12} & 0\\
  0 & 0 & 1\\
\end{array}%
\right) \end{eqnarray} Let us perform now the  second rotation at
the angle $\theta_{13}$ around the vector $|s \rangle'$. At this
step we will introduce the $CP$ phase $\delta$, connected with the
rotation of the vector of the third family $|b\rangle$.  We will
obtain the following three orthogonal vectors:
\begin{eqnarray}\label{69}|d\rangle''=&c_{13}~|d\rangle'
+s_{13}e^{-i\delta}~|b\rangle'\nonumber\\
|s\rangle''=&|s\rangle'\nonumber\\
|b\rangle''=& -s_{13}e^{i\delta}~|d\rangle' +c_{13}~|b\rangle'.
\end{eqnarray}
In the matrix form we have
\begin{equation}\label{70}
|D\rangle''=V''~|D\rangle'.
\end{equation}
Here
\begin{eqnarray}\label{71}
V''=\left(%
\begin{array}{ccc}
  c_{13} & 0& s_{13}e^{-i\delta}\\
  0 & 1 & 0\\
  -s_{13}e^{i\delta} & 0 &c_{13}  \\
\end{array}%
\right). \end{eqnarray} Finally, let us perform rotation around the
vector $|d\rangle''$ at the angle $\theta_{23}$. New orthogonal
vectors are
\begin{eqnarray} |d\rangle'''=&|d\rangle''\nonumber\\
 |s\rangle'''=&c_{23}|~s\rangle''
+s_{13}~|b\rangle''\nonumber\\
|b\rangle'''=& -s_{23}~|s\rangle'' +c_{23}~|b\rangle''\label{72}
\end{eqnarray}
We have
\begin{equation}\label{73}
|D'''\rangle=V'''~|D''\rangle.
\end{equation}
Here
\begin{eqnarray}\label{74}
V'''=\left(%
\begin{array}{ccc}
  1 & 0& 0\\
  0 & c_{23}  & s_{23} \\
  0 & -s_{23}  &c_{23}  \\
\end{array}%
\right). \end{eqnarray} From (\ref{66}), (\ref{70}) and(\ref{73}) we
find
\begin{equation}\label{75}
    |D'''\rangle=V~|D\rangle,
\end{equation}
where
\begin{equation}\label{76}
    V=V'''~V''~V'.
\end{equation}
It is obvious that  $V$ is the unitary matrix.

 Thus, the general
3$\times$3 unitary mixing matrix has the form
\begin{eqnarray}\label{77}
V=
\left(%
\begin{array}{ccc}
  1 & 0& 0\\
  0 & c_{23}  & s_{23} \\
  0 & -s_{23}  &c_{23}  \\
\end{array}%
\right)
\left(%
\begin{array}{ccc}
  c_{13} & 0& s_{13}e^{-i\delta}\\
  0 & 1 & 0\\
  -s_{13}e^{i\delta} & 0 &c_{13}  \\
\end{array}%
\right)
\left(%
\begin{array}{ccc}
  c_{12} & s_{12}& 0\\
  -s_{12} & c_{12} & 0\\
  0 & 0 & 1\\
\end{array}%
\right).
\end{eqnarray}
From (\ref{77}) we find
\begin{eqnarray}
V=\left(\begin{array}{ccc}c_{13}c_{12}&c_{13}s_{12}&s_{13}e^{-i\delta}\\
-c_{23}s_{12}-s_{23}c_{12}s_{13}e^{i\delta}&
c_{23}c_{12}-s_{23}s_{12}s_{13}e^{i\delta}&c_{13}s_{23}\\
s_{23}s_{12}-c_{23}c_{12}s_{13}e^{i\delta}&
-s_{23}c_{12}-c_{23}s_{12}s_{13}e^{i\delta}&c_{13}c_{23}
\end{array}\right).
\label{78}
\end{eqnarray}
In the standard parametrization  the 3$\times$3  mixing matrix is
characterized by three Euler angles $\theta_{12}$, $\theta_{23}$ and
$\theta_{13}$ and one phase $\delta$. We have seen before that in
the case of the $CP$ conservation $V^{*}=V$. Thus, in this case
$\delta=0$.

Let us calculate  in the standard parametrization of the $CKM$
mixing matrix the invariant $J$ given by (\ref{63}). From (\ref{78})
we have
\begin{equation}\label{79}
J=c_{12}c_{23}c^{2}_{13}s_{12}s_{23}s_{13}\sin \delta.
\end{equation}
As we have seen in the previous section in the case of the $CP$
conservation the Jarlskog invariant $J$ is equal to zero. It follows
from experimental data that all mixing angles are different from
zero. (see below). The rephrase invariant condition of the $CP$
conservation has the form: $\sin \delta=0$.

\section{Modulus of the elements of $CKM$ matrix}

The values of the modulus of the $CKM$ matrix elements  were
determined from the data of different experiments (see
\cite{CeccucciPDG}).

The highest accuracy was reached in the measurement of {\bf the
element $|V_{u d}|$.} There are three sources of information about
this element: i) The superallowed $0^{+} \to 0^{+}$ $\beta$-decay of
nuclei. ii) The neutron decay. iii) The $\beta$-decay of pion
$\pi^{+}\to \pi^{0}e^{+}\nu_{e}$.

Only  vector current gives contribution to the matrix element of the
$0^{+} \to 0^{+}$ $\beta$-transition. From the isotopic invariance
and the $CVC$ follows that matrix element of  $0^{+} \to 0^{+}$
transition between components of  isotopic triplet is given by
\begin{equation}\label{80}
|\langle p'|V_{\alpha}|p \rangle|=N~|V_{u
d}|\sqrt{2}~(p+p')_{\alpha},
\end{equation}
where $p$ and $p'$ are momenta of initial and final nuclei and $N$
is the normalization factor. The nuclear Coulomb effects and
radiative corrections, which violate this relation, must be taken
into account. From the most precise measurements of the $ft$ values
of nine nuclei the following average value  was obtained
\cite{Hardy,Savard}
\begin{equation}\label{81}
|V_{u d}|=0.97377 \pm 0.00027
\end{equation}
It is necessary to notice, however, that  the $Q$-value of
$\rm{^{46}V}$ was recently remeasured \cite{Savard}. The new value
leads to an increase of  the $f$-factor which gives  2.7 $\sigma$
decrease  of the value of $|V_{u d}|$ with respect to the average
value (\ref{81}).

The element  $|V_{u d}|$ can be  determined also from data of the
experiments on the measurements of the lifetime of the neutron
$\tau_{n}$ and from the ratio of axial and vector constants $g_{A}$.
The constant $g_{A}$ can be obtained from the data of the
experiments on the measurement of the asymmetry of electrons in
decay of  polarized neutrons. From the world averages values of
$\tau_{n}$ and $g_{A}$ \cite{PDG}
\begin{equation}\label{82}
\tau_{n}=885.7\pm 0.8 ~\rm{sec},~~g_{A}=-1.2695\pm 0.0029
\end{equation}
for the element  $|V_{u d}|$ it was found the value \cite{Marciano}
\begin{equation}\label{83}
|V_{u d}|=0.9746\pm 0.0004\pm 0.0018\pm 0.0002.
\end{equation}
Here the first (second) error is due to the error  of $\tau_{n}$
($g_{A}$) and the third error  is due to the uncertainty in the
calculations of radiative corrections. As it is seen from (\ref{83})
the dominant uncertainty is due to the error of the constant
$g_{A}$.

Finally, the value of the element $|V_{u d}|$ was obtained  from the
measurement of the branching ratio of the decay $\pi^{+}\to
\pi^{0}e^{+}\nu_{e}$. Only vector $CC$  current gives contribution
to the hadronic matrix element of this process. From the $CVC$ and
isotopic invariance it follows that matrix element of the hadronic
vector current is given by the relation (\ref{80}). The problem of
the calculation of the radiative corrections is much more simpler in
the pion case than in the nuclear case. However, the branching ratio
of the pion $\beta$-decay is very small ($B(\pi^{+}\to
\pi^{0}e^{+}\nu_{e})\simeq 10^{-8}$). As a result, the accuracy of
the determination of the element $|V_{u d}|$ from the measurement of
this branching ratio is much worse than from the measurement of the
$ft$ values of the nuclear $0^{+}\to 0^+$ $\beta$-decays. In
\cite{Pocanic} it was found the value
\begin{equation}\label{84}
|V_{u d}|=0.9728\pm 0.0030.
\end{equation}

The value of {\bf the  element $|V_{us}|$ } was obtained from the
measurement of the widths of the decays $K_{L}\to
\pi^{\pm}l^{\mp}\nu_{l}$ ($l=e,\mu$) and $K^{+}\to
\pi^{0}e^{+}\nu_{e}$. Only CC vector current gives contribution to
the hadronic part of the matrix elements of these decays. The matrix
element is characterized by the two form factors and has the form
\begin{equation}\label{85}
\langle p'|V_{\alpha}|p \rangle=N ~V_{u d} ~\left
(f_{+}(Q^{2})~(p+p')_{\alpha}+ f_{-}(Q^{2})~(p-p')_{\alpha}\right ).
\end{equation}
Here $p$ and $p'$ are momenta of kaon and pion, $Q^{2}=-(p'-p)^{2}$
and $N$ is the standard normalization factor. Taking into account
the results of the measurements of the form factors $f_{\pm}(Q^{2})$
and recent measurements of the branching ratios of the decays
$K_{L}\to \pi e \nu$ and $K_{L}\to \pi \mu \nu$
\cite{Alexopoulos,Ambrosino,Lai} for the element $|V_{u s}|$ the
following value was found  \cite{Marciano}
\begin{equation}\label{86}
|V_{u s}|=0.2257 \pm 0.0021
\end{equation}
This  result was obtained  with the chiral perturbation value
\cite{Leutwyler} $f_{+}(0)=0.961\pm 0.008$ was used.

The value of the parameter $|V_{u s}|$ can be also obtained  from
the measurement of the widths of the decays $K^{+}\to
\mu^{+}\nu_{\mu}$ and $\pi^{+}\to \mu^{+}\nu_{\mu}$. Using  for the
ratio of the decay constants the value
\begin{equation}\label{87}
    \frac{f_{K}}{f_{\pi}}=1.198 ^{+0.016}_{-0.005}\pm 0.003,
\end{equation}
which was obtained in the  lattice calculations \cite{Bernard}, for
the matrix element $|V_{u s}|$ it was found \cite{Marciano}
\begin{equation}\label{88}
|V_{u s}|=0.2245^{+0.0012}_{-0.0031}.
\end{equation}

 The value of the element $|V_{u s}|$ can be also inferred  from the
analysis of data on the investigation  of the hyperon decays. From
these data it was found \cite{CabibboI}
\begin{equation}\label{89}
|V_{u s}|=0.2250 \pm 0.0027
\end{equation}
Finally,  an information about the value of the parameter $|V_{u
s}|$ can be obtained from the data of the experiments on the
investigation of the decays $\tau^{\pm}\to
\nu_{\tau}+\rm{hadrons}(S=\pm 1)$. From these data the following
value of the matrix element $ |V_{u s}|$ was found \cite{Gamiz}
\begin{equation}\label{90}
|V_{u s}|=0.2208 \pm 0.0034
\end{equation}
Thus, the values of the element $|V_{u s}|$, determined from the
different experimental data and with different theoretical inputs,
are compatible.

From the unitarity of the $CKM$ matrix $V$ we have
\begin{equation}\label{91}
|V_{u d}|^{2}+|V_{u s}|^{2}+|V_{u b}|^{2}=1
\end{equation}
The last term gives negligible contribution to this relation (see
later). From (\ref{83}) and (\ref{86}) it was found \cite{Marciano}
\begin{equation}\label{92}
|V_{u d}|^{2}+|V_{u s}|^{2}+|V_{u b}|^{2}=0.9992\pm0.0005\pm0.0009,
\end{equation}
where the first error is due to the error of $|V_{u d}|$ and the
second one is due to the error of $|V_{u s}|$. Thus, the values
(\ref{83}) and (\ref{86}) of the parameters $|V_{u d}|$ and $|V_{u
s}|$ saturate the unitarity relation (\ref{91}).

{\bf The element $|V_{c d}|$} can be determined from the data on the
production of the muon pairs in the processes of interaction of
$\nu_{\mu}$ and $\bar\nu_{\mu}$ with nucleons.\footnote{One muon is
produced in a process of interaction of neutrino (antineutrino) with
nucleon and another in decay of produced charmed particle.} From
these data was found \cite{CeccucciPDG}
\begin{equation}\label{93}
|V_{cd}|=0.230 \pm 0.011.
\end{equation}
The element $|V_{c d}|$ can be also obtained from the data on the
study of the decays $D\to \pi l\nu_l$  if the corresponding form
factors are known. Using lattice calculations of the form factors
\cite{Aubin} it was found \cite{Artuso}
\begin{equation}\label{94}
|V_{cd}|=0.213 \pm 0.008 \pm 0.021,
\end{equation}
where the dominant error is the theoretical one.

{\bf The value of the element $|V_{c s}|$} was determined from the
data on the investigation of the decays $D\to K l\nu_l$. Using the
lattice calculations of the  form factors \cite{Aubin} it was found
the value \cite{Artuso}
\begin{equation}\label{95}
|V_{cs}|=0.957 \pm 0.017 \pm 0.093,
\end{equation}
where the second (theoretical) error is the largest one.

An model independent information about the element $|V_{c s}|$ can
be obtained from the data on the study of the decay $W^+\to c+ \bar
s$. From the LEP data it was found the value \cite{Abren}
\begin{equation}\label{96}
|V_{cs}|=0.94^{+0.32}_{-0.26}\pm 0.13.
\end{equation}

{\bf The value of the element $|V_{c b}|$} was determined from the
data on the investigation of the semileptonic inclusive decays $\bar
B\to X_{c}l\bar\nu_l$ and exclusive $\bar B\to D(D^{*}) l\bar\nu_l$
decays. Analysis of the inclusive data is based on the operator
product expansion theory \cite{Bigi,Manohar}. From the LEP and
$B$-factories data it was found the following average value
\cite{KowalewskiPDG}
\begin{equation}\label{97}
|V_{cb}|=(41.7\pm 0.7)\cdot 10^{-3}.
\end{equation}
Analysis of the exclusive data is based on the heavy quark effective
theory \cite{Isgur,Shifman}. The  average value
\begin{equation}\label{98}
|V_{cb}|=(40.9\pm 1.8)\cdot 10^{-3}
\end{equation}
which was found from analysis of the exclusive data
\cite{KowalewskiPDG} is compatible with (\ref{97}).

{\bf The value of the element $|V_{u b}|$} can be obtained from the
study of semileptonic inclusive decay
\begin{equation}\label{99}
\bar B\to X_{u}l\bar\nu_l
\end{equation}
and exclusive decay
\begin{equation}\label{100}
\bar B\to \pi l\bar\nu_l
\end{equation}
The suppression of the background from  the $CKM$ enhanced inclusive
decay $\bar B\to X_{c}l\bar\nu_l$ is the main problem in the
investigation of the decay (\ref{99}). The following average value
of $|V_{u b}|$ was obtained from different inclusive measurements
\cite{KowalewskiPDG}
\begin{equation}\label{101}
 |V_{u b}| = (4.40 \pm 0.20 \pm 0.27)\cdot 10^{-3} (\rm{inclusive}).
\end{equation}
In  the exclusive decay (\ref{100}) both final charged particles are
detected. This leads to the better suppression of the background
than in the inclusive case. However, the branching ratio of the
exclusive decay (which is known at present with the accuracy $\sim$
8\%) is much smaller than the branching ratio of the inclusive
decay. The hadronic matrix element of the process (\ref{100}) is
given by
\begin{equation}\label{102}
\langle p'|V_{\alpha}|p \rangle=N ~V_{u b}
~\left(f_{+}(q^{2})~(p+p'-\frac{m^{2}_{B}-m^{2}_{\pi}}{q^{2}})~q_{\alpha}+
f_{0}(q^{2})~\frac{m^{2}_{B}-m^{2}_{\pi}}{q^{2}}~q_{\alpha}\right),
\end{equation}
where $q=p-p'$ and $f_{+}(q^{2})$, $f_{0}(q^{2})$ are the form
factors.

The calculation of the form factors $f_{+}(q^{2})$ and
$f_{0}(q^{2})$ is the main problem in the determination of $|V_{u
b}|$ from the exclusive data. Using lattice calculations
\cite{Okamoto,Shigemitsu} the following value was found
\cite{KowalewskiPDG}
\begin{equation}\label{103}
 |V_{u b}| = (3.84^{+0.67}_{-0.49})\cdot 10^{-3} (\rm{exclusive}).
\end{equation}
This value is compatible with (\ref{101}). From (\ref{101}) and
(\ref{103}) the following weighted average of the matrix element
$|V_{u b}|$ was obtained \cite{KowalewskiPDG}
\begin{equation}\label{104}
 |V_{u b}| = (4.31 \pm 0.39)\cdot 10^{-3}(\rm{exclusive}).
\end{equation}
{\bf The element $|V_{t d}|$} can be determined from the measurement
of the mass difference of $B_{d}^{0}$ mesons.  The major
contribution to the box diagram which determine mass differences
$\Delta m_{q}$ ($q=d,s$) gives the virtual $t$-quark. We have (see,
for example, \cite{BurasI})
\begin{equation}\label{105}
\Delta
m_{q}=\frac{G^{2}_{F}}{6\pi^{2}}~m_{B_{q}}m^{2}_{W}~(f^{2}_{B_{q}}\hat{B}_{B_{q}})~
\eta_{B} S_{0}(x_{t})~|V_{tb}V^{*}_{tq}|^{2}.
\end{equation}
Here  $f_{B_{q}}$ is the  decay constant and $\hat{B}_{B_{q}}$ is so
called B-factor. The factor $\eta_{B}$ is due to short distance
$QCD$ corrections ($\eta_{B}=0.55\pm0.01$) and $S_{0}(x_{t})$ is
known function of $x_{t}=\frac{m^{2}_{t}}{m^{2}_{t}}$.

For the mass difference $ \Delta m_{d}$ the following value was
obtained \cite{Barbero}
\begin{equation}\label{106}
\Delta m_{d}=(0.507 \pm 0.004)~\rm{ps}^{-1}.
\end{equation}
Assuming $|V_{tb}|=1$ and taking into account the lattice result
\cite{Gray,Aoki}
\begin{equation}\label{107}
f_{B_{d}}\sqrt{\hat{B}_{B_{d}}}=(244 \pm 11 \pm 24) \rm {MeV}
\end{equation}
for the element $|V_{t d}|$ it was found the value \cite{Okamoto}
\begin{equation}\label{108}
|V_{t d}|=(7.4 \pm 0.8)~10^{-3}.
\end{equation}
Recently the mass difference of $B_{s}^{0}$ mesons was measured.
Using the CDF value \cite{CDFdeltas}
\begin{equation}\label{109}
\Delta m_{s}=(17.31^{+0.33}_{-0.18} \pm 0.07)~\rm{ps}^{-1}
\end{equation}
and the lattice result
\begin{equation}\label{110}
\frac{f_{B_{s}}\sqrt{\hat{B}_{B_{s}}}}{f_{B_{d}}\sqrt{\hat{B}_{B_{d}}}}= 1.21
\pm 0.04 ^{+0.04}_{-0.01}
\end{equation}
it was obtained \cite{CDFdeltas}
\begin{equation}\label{111}
\frac{|V_{t d}|}{|V_{t s}|}= 0.208 ^{+0.008}_{-0.006}
\end{equation}
The value of the {\bf element $|V_{t s}|$} can be  found from the
unitarity relation $V_{c b}V^{*}_{c s}+V_{t b }V^{*}_{t s}+V_{u
b}V^{*}_{u s}=0$. It was obtained \cite{CeccucciPDG}
\begin{equation}\label{112}
|V_{t s}|=(40.6 \pm 2.7)~10^{-3}.
\end{equation}
Finally, an information about {\bf the element $|V_{t b}|$} can be
inferred from the measurement of the ratio $\frac{B(t\to
Wb)}{\sum_{q=d,s,b}B(t\to Wq)}=|V_{t b}|^{2} $. From the Fermilab
data \cite{Acosta,AbazovD0} it was found the following 95 \% CL
lower bound \cite{CeccucciPDG}
\begin{equation}\label{113}
|V_{t b}|>0.78.
\end{equation}

\section{Wolfenstein parameters.
Unitarity triangle}

From the values of the modulus of elements of the $CKM$ matrix,
which we discussed in the previous section, it follows that quark
mixing angles are small and exist a hierarchy  of  mixing between
different families. In fact, in the standard parametrization of the
$CKM$ matrix we have
\begin{equation}\label{114}
V_{ud}= c_{13}c_{12},~~ V_{us}=
c_{13}s_{12},~~V_{cb}=c_{13}s_{23},~~~V_{ub}= s_{13}e^{-i\delta}.
\end{equation}
From these relations we find
\begin{equation}
s_{12} = \frac{| V_{us}|}{\sqrt{|V_{ud}|^{2}+|V_{us}|^{2}}},~~~
s_{23} = \frac{| V_{cb}|}{\sqrt{|V_{ud}|^{2}+|V_{us}|^{2}}},~~~
s_{13}=|V_{ub}|. \label{115}
\end{equation}
From (\ref{81}), (\ref{86}), (\ref{97}), (\ref{101}) and (\ref{115})
for the parameters $ s_{ik}$ we find
\begin{equation}\label{115a}
s_{12}\sim 2\cdot 10^{-1},~~s_{23}\sim 4\cdot 10^{-2},~~s_{13}\sim
4\cdot 10^{-3}.
\end{equation}
Let us introduce the parameter
\begin{equation}\label{116}
\lambda=s_{12}.
\end{equation}
We have
\begin{equation}\label{117}
s_{23} \simeq \lambda^{2},~~~s_{13} \simeq \frac{1}{2}~\lambda^{3}.
\end{equation}
Thus, exist  a \emph{hierarchy of angles of the mixing} between
different quark families. The strength of the coupling between the
families is determined by the degree of the  parameter $\lambda $.

 Wolfenstein \cite{Wolfensteinparam} proposed
a parametrization of the mixing matrix which take into account this
hierarchy. Instead of $s_{12}$, $s_{23}$ and $s_{13}e^{-i\delta}$ he
introduced four real parameters $\lambda,~A,~\rho ~\rm{and}~\eta$ by
the following relations
\begin{equation}
s_{12}=\lambda,~  s_{23}= A \lambda^{2},~s_{13}e^{-i\delta}= A
\lambda^{3} (\rho- i\eta). \label{118}
\end{equation}
Let us   develop  elements of the $CKM$ matrix over the small
parameter $\lambda$. Keeping terms of the order of $\lambda^{5}$ for
the $CKM$ mixing matrix $V$ we have
\begin{eqnarray}
V=\left(\begin{array}{ccc}1-\frac{1}{2}\lambda^{2}-\frac{1}{8}\lambda^{4}&\lambda&
A\lambda^{3}(\rho-i\eta)\\
-\lambda+\frac{1}{2}A^{2}\lambda^{5}(1-2(\rho+i\eta))&
1-\frac{1}{2}\lambda^{2}-\frac{1}{8}\lambda^{4}(1+4A^{2})&A\lambda^{2}\\
A\lambda^{3}(1-(1-\frac{1}{2}\lambda^{2})(\rho+i\eta))&
-A\lambda^{2}+\frac{1}{2}A\lambda^{4}(1-2(\rho+i\eta))&1-\frac{1}{2}A^{2}\lambda^{4}
\end{array}\right)
\label{119}\end{eqnarray}
 We will obtain now the so called unitarity
triangle relation. This relation follows from the condition of the
unitarity of the mixing matrix
\begin{equation}\label{120}
V^{\dag}~V =1.
\end{equation}
For the three families of the quarks from (\ref{120}) we have
\begin{equation}
\sum_{u_{1}=u,c,t} V^{*}_{u_{1}d_{1}}V_{u_{1}
d_{2}}=\delta_{d_{1}d_{2}}.
 \label{121}
\end{equation}
From (\ref{121}) we obtain the following relations
\begin{equation}\label{122}
\sum_{u_{1}=u,c,t} |V_{u_{1}d}|^{2}=1,~~\sum_{u_{1}=u,c,t}
|V_{u_{1}s}|^{2}=1,~~\sum_{u_{1}=u,c,t} |V_{u_{1}b}|^{2}=1.
\end{equation}
and
\begin{equation}
\sum_{u_{1}=u,c,t} V_{u_{1} d}V^{*}_{u_{1} s}=0,~~\sum_{u_{1}=u,c,t}
V_{u_{1} s}V^{*}_{q b}=0, ~~\sum_{u_{1}=u,c,t} V_{u_{1}
d}V^{*}_{u_{1} b}=0.\label{123}
\end{equation}
Let us consider the relations (\ref{123}). In the first relation the
first and the second terms are of the order $\lambda$ and the third
one is of the order $\lambda^{5}$. Thus, in this relation the main
contribution give terms which connect only two families (the first
and the second). In the second relation (\ref{123}) the first term
is of the order $\lambda^{4}$ and the second and the third terms are
of the order $\lambda^{2}$. In this relation the main contribution
also give terms which connect only two families (the second and
third). The only relation in which all terms are of the same
($\lambda^{3}$) order is the third relation (\ref{123}). It has the
form
\begin{equation}
V_{u d}V^{*}_{u b}+ V_{c d}V^{*}_{c b}+ V_{t d}V^{*}_{tb}=0.
\label{124}
\end{equation}
Let us  now expand different terms of  (\ref{124}) over the powers
of the small  parameter $\lambda$.  We have \cite{Burasparam}
\begin{equation}
V_{u d}V^{*}_{u b} =c_{13}c_{12}s_{13}e^{i\delta}= A \lambda^{3}
(\bar \rho+ i\bar \eta) + O(\lambda^{7})
 \label{125}
\end{equation}
where
\begin{equation}
\bar \rho = (1 -\frac{1}{2}\lambda^{2})~\rho,~~~ \bar \eta= (1
-\frac{1}{2}\lambda^{2})~\eta. \label{126}
\end{equation}
For the second term of the relation (\ref{124}) we find
\begin{equation}
V_{c d}V^{*}_{c b}=(-s_{12}c_{23} -c_{12} s_{23}s_{13}e^{i\delta}) ~
c_{13}s_{23}= - A \lambda^{3}+ O(\lambda^{7}).\label{127}
\end{equation}
Finally, for the third term of (\ref{124}) we obtain
\begin{equation}
V_{t d}V^{*}_{tb}=(s_{23}s_{12} -c_{23} c_{12}s_{13}e^{i\delta})~
 c_{13}c_{23}=
\simeq A \lambda^{3}~ (1- (\bar \rho+ i\bar \eta))+ O(\lambda^{7}).
\label{128}
\end{equation}
We see from the  relations (\ref{125}), (\ref{127}) and (\ref{128})
that up to small terms of the order of $\lambda^{7}$ all terms in
(\ref{124}) are proportional to $ A\lambda^{3}$.

Let us rewrite the relation (\ref{124}) in the form
\begin{equation}
\frac{V_{u d}V^{*}_{u b}}{(-V_{c d}V^{*}_{c b})} + \frac{V_{t
d}V^{*}_{t b}}{(-V_{c d}V^{*}_{c b})}=1. \label{129}
\end{equation}
We have\footnote{It is obvious that the ratios of the products of
the $CKM$ matrix elements in (\ref{129}) are invariant under phase
transformation (\ref{49}).}
\begin{equation}\label{130}
\frac{V_{u d}V^{*}_{u b}}{(-V_{c d}V^{*}_{c b})}= \bar \rho+ i\bar
\eta =\sqrt{\bar \rho^{2}+\bar \eta^{2}} ~e^{i\gamma}
\end{equation}
and
\begin{equation}\label{131}
\frac{V_{t d}V^{*}_{t b}}{(-V_{c d}V^{*}_{c b})} = 1- (\bar \rho+
i\bar \eta)= \sqrt{(1-\bar \rho)^{2}+\bar \eta^{2}}~e^{-i\beta}.
\end{equation}
Thus, the unitarity relation (\ref{129}) takes the form
\begin{equation}
(\bar \rho+ i\bar \eta) + (1- (\bar \rho+ i\bar \eta)) =1.
\label{132}
\end{equation}
This relation can be presented as a triangle in the complex ($\bar
\rho,~~ \bar \eta$) plane (see Fig.1). It is called the unitarity
triangle.

\begin{figure}[htbp]
  \begin{center}
    \leavevmode
\epsfxsize=4in
    \epsfbox{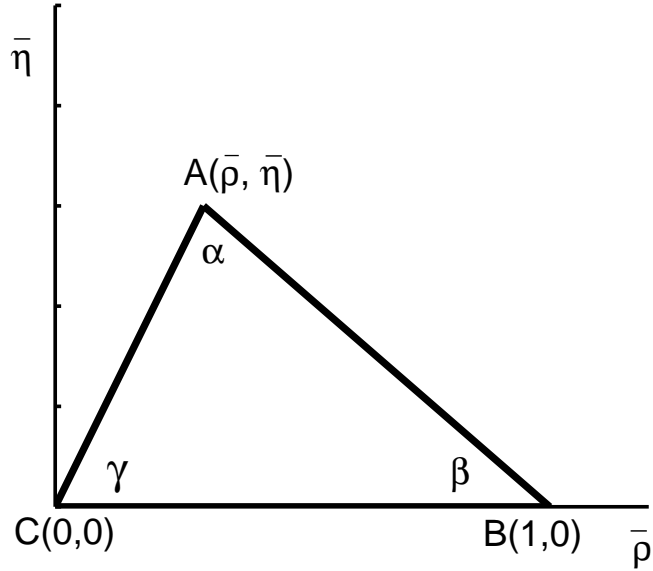}
  \end{center}
  \caption{The unitarity triangle. The angles $\alpha \equiv\phi_{2},~
  \beta\equiv\phi_{1},~\gamma\equiv\phi_{3}$ are shown }
  \label{fig:1}
\end{figure}

From (\ref{131}) and (\ref{132}) for the angles $\gamma$ and $\beta$
we have \footnote {Other notations for the angles of the unitarity
triangle, which are often used in literature,
 are: $\phi_{1}\equiv\beta;~\phi_{2}\equiv\alpha;~ \phi_{3}\equiv\gamma$.}
\begin{equation}\label{133}
    \gamma=\arg {(-\frac{V_{u d}V^{*}_{u b}}
    {V_{c d}V^{*}_{c b}})},~~\beta=\arg{(-\frac{V_{c d}V^{*}_{c b}}{V_{t d}V^{*}_{t
    b}})}.
\end{equation}
From (\ref{78})  and  (\ref{130}) it follows that the angle $\gamma$
coincides with the $CKM$ angle $\delta$. For the angle $\alpha$ we
find
\begin{equation}\label{134}
\alpha=\pi-\beta-\gamma=\arg{(-\frac{V_{t d}V^{*}_{t b}}{V_{u
d}V^{*}_{u
    b}})}.
\end{equation}

The square of the unitarity triangle is equal to
   $$ \tilde{S}=\frac{1}{2}~1 \cdot \bar \eta$$
From  (\ref{130}) we find
\begin{equation}\label{135}
-\frac{V_{u d}V_{cb}V^{*}_{u b}V^{*}_{c d}} {|V_{c d}|^{2}
|V_{cb}|^{2}}=\bar \rho+ i\bar \eta.
\end{equation}
From this relation we have
\begin{equation}\label{136}
\bar \eta=\frac{J}{|V_{c d}|^{2} |V_{cb}|^{2}},
\end{equation}
where $J$ is the Jarskog  invariant (\ref{63}). Thus, the square of
the unitarity triangle is given by
\begin{equation}\label{137}
\tilde{S}=\frac{1}{2}~\frac{J}{|V_{c d}|^{2} |V_{cb}|^{2}}.
\end{equation}
For the square of the unitarity triangle, given by the relation
(\ref{129}), we have \cite{Jarlskog}
\begin{equation}\label{138}
S=\frac{1}{2}~J
\end{equation}
The relations (\ref{123})  are conditions of orthogonality of the
columns of the matrix $V$. Additional three relations can be
obtained from the conditions of orthogonality of the lines of the
matrix $V$. We can  see easily that the only relation in which all
terms are of the same $\lambda^{3}$ order is the condition of the
orthogonality of the first and the third lines:
\begin{equation}\label{139}
\sum_{d_{1}=d,s,b} V_{u d_{1}}V^{*}_{t d_{1}}=0
\end{equation}
This relation after expansion of different terms over the powers of
the parameter $\lambda$ also takes the form of the relation
(\ref{132}).

\section{Eigenstates and eigenvalues of the effective Hamiltonian
of the $M^{0}-\bar M^{0}$ system}

We will obtain here  eigenstates and eigenvalues of the effective
$2\times 2$ nonhermitian Hamiltonian $\mathcal{H}$ of the
$M^{0}-\bar M^{0}$ system ($M^{0}= K^{0}, B^{0}_{d,s},...)$ which we
derived in the Appendix B. We have
\begin{equation}\label{144}
\mathcal{H}~a_{H,L}=\mu_{H,L}~a_{H,L}.
\end{equation}
Here
\begin{equation}\label{145}
\mu_{H,L}=m_{H,L}-\frac{i}{2}~\Gamma_{H,L},
\end{equation}
and
\begin{eqnarray}\label{145a}
a_{H,L}=
\left(%
\begin{array}{c}
 a_{H,L}(1)  \\
  a_{H,L}(2)\\
\end{array}%
\right).
\end{eqnarray}
If  the wave function at the initial time $t=0$  is equal to
$a_{H,L}$, at $t\geq0$ we have
\begin{equation}\label{146}
a_{H,L}(t)=e^{-im_{H,L}t-\frac{1}{2}\Gamma_{H,L}t}~a_{S,L}.
\end{equation}
Thus, $m_{H,L}$ and $\Gamma_{S,L}$ are  masses and  total decay
widths of $M^{0}_{H,L}$-bosons, particles which are described by the
functions  $a_{H,L}$.  We will use the index $H$ for the heavier
particle and the index $L$ for the lighter particle. Thus, we have
$m_{H}>m_{L}$. For the vectors of the states of $M^{0}_{H,L}$ we
have
\begin{equation}\label{147}
|M^{0}_{H,L}\rangle =\sum_{\alpha=1,2
}a_{H,L}(\alpha)|\alpha\rangle,
\end{equation}
where $|1\rangle \equiv |M^{0}\rangle$ and $|2\rangle \equiv |\bar
M^{0}\rangle$ are states of $M^{0}$ and $\bar M^{0}$ particles (in
the rest frame system).

Assuming  the $CPT$ invariance of the  Hamiltonian, we can present
the effective Hamiltonian of the $M^{0}-\bar M^{0}$ system in the
form
\begin{equation}\label{148}
\mathcal{H}=\mathcal{H}_{11}+ \mathcal{H}^{\rm{nd}},
\end{equation}
where
\begin{eqnarray}\label{149}
\mathcal{H}^{\rm{nd}}=
\left(%
\begin{array}{cc}
  0 & \mathcal{H}_{12} \\
  \mathcal{H}_{21} & 0\\
\end{array}%
\right).
\end{eqnarray}
From (\ref{148}) and (\ref{149}) we have
\begin{equation}\label{150}
\mathcal{H}^{\rm{nd}}~a_{H,L}=\kappa_{H,L}~a_{H,L}.
\end{equation}

Here
\begin{equation}\label{151}
\kappa_{H,L}=\mu_{H,L}-\mathcal{H}_{11}
\end{equation}
For the eigenvalues $\kappa_{S,L}$ of the Hamiltonian
$\mathcal{H}^{\rm{nd}}$ we obviously have
\begin{equation}\label{152}
\kappa_{H,L}=\mp \sqrt{\mathcal{H}_{12}~\mathcal{H}_{21}}
\end{equation}
Further from (\ref{150}) and (\ref{152}) we find that $a_{H,L}(2)$
and $a_{H,L}(1)$ are connected by the relation
\begin{equation}\label{153}
a_{H,L}(2)=\mp
\sqrt{\frac{\mathcal{H}_{21}}{\mathcal{H}_{12}}}~a_{H,L}(1),
\end{equation}
where $a_{H,L}(1)$ is an arbitrary constant.

The equation (\ref{144}) have the following solutions
\begin{eqnarray}\label{154}
a_{H,L}=
\left(%
\begin{array}{c}
  1 \\
  \mp \sqrt{\frac{\mathcal{H}_{21}}{\mathcal{H}_{12}}}\\
\end{array}%
\right)~a_{H,L}(1)
\end{eqnarray}
with
\begin{equation}\label{155}
\mu_{H,L}=\mathcal{H}_{11}\mp
\sqrt{\mathcal{H}_{12}~\mathcal{H}_{21}}.
\end{equation}
Three physical complex parameters $\mu_{H,L}$ (masses and total
decay width of $M^{0}_{H,L}$) and  parameter
$\sqrt{\frac{\mathcal{H}_{21}}{\mathcal{H}_{12}}}$, which
characterize mixing  of $M^{0}$ and $\bar M^{0}$, correspond to
three complex matrix elements of the matrix $\mathcal{H}$ (
$\mathcal{H}_{11}$, $\mathcal{H}_{12}$ and $\mathcal{H}_{21}$).

Let us choose
\begin{equation}\label{156}
a_{H,L}(1)=\frac{\sqrt{\mathcal{H}_{12}}}{\sqrt{|\mathcal{H}_{12}|+
    |\mathcal{H}_{21}|}}=p
\end{equation}
We have
\begin{eqnarray}\label{157}
a_{H,L}=
\left(%
\begin{array}{c}
  p \\
  \mp q\\
\end{array}%
\right),
\end{eqnarray}
where
\begin{equation}\label{158}
    q=\frac{\sqrt{\mathcal{H}_{21}}}
{\sqrt{|\mathcal{H}_{12}|+
    |\mathcal{H}_{21}|}}
\end{equation}
With this choice we have
\begin{equation}\label{159}
|p|^{2}+|q|^{2}=1~~ \rm{and} ~~(a^{\dag}_{H,L}a_{H,L})=1.
\end{equation}
The states of $M^{0}_{H,L}$  are given by the following relations
\begin{equation}\label{160}
|M^{0}_{H}\rangle =p~|M^{0}\rangle -q~|\bar M^{0}\rangle,~~
|M^{0}_{L}\rangle =p~|M^{0}\rangle +q~|\bar M^{0}\rangle.
\end{equation}
If $CP$  is conserved in this case
$\mathcal{H}_{21}=\mathcal{H}_{12}$ and $q= p$. For the eigenstates
of the Hamiltonian we have
\begin{equation}\label{161}
|M^{0}_{2,1}\rangle =\frac{1}{\sqrt{2}}(|M^{0}\rangle \mp |\bar
M^{0}\rangle).
\end{equation}
Let us make the following remark. We  have chosen  phases  of the
states $|M^{0}\rangle$ and $|\bar M^{0}\rangle$  in such a way that
(see Appendix B)
\begin{equation}\label{162}
CP~|M^{0}\rangle = |\bar M^{0}\rangle
\end{equation}
The states  $|M^{0}\rangle$ and $|\bar M^{0}\rangle$ are eigenstates
of the Hamiltonians of the strong and electromagnetic interactions.
These interactions conserve quark flavors. This means that it is
impossible to distinguish states $|M^{0}\rangle$ and $|\bar
M^{0}\rangle$ from the states
\begin{equation}\label{163}
|M^{0}\rangle'=e^{i\alpha}|M^{0}\rangle,~~|\bar
M^{0}\rangle'=e^{-i\alpha}|\bar M^{0}\rangle,
\end{equation}
where $\alpha$ is an arbitrary phase.

If for the states of  $M^{0}$ and $\bar M^{0}$  we will use
$|M^{0}\rangle'$ and $|\bar M^{0}\rangle'$ in this case we have
\begin{equation}\label{164}
    p'=e^{-i\alpha}p,~~q'=e^{i\alpha}q.
\end{equation}
The states of $M^{0}_{H,L}$ are invariant under the change of the
basis. In fact, we have
\begin{equation}\label{165}
    |M^{0}_{H,L}\rangle'=p'~|M^{0}\rangle' \mp q'~|\bar
    M^{0}\rangle'=p~|M^{0}\rangle \mp q~|\bar
    M^{0}\rangle=|M^{0}_{H,L}\rangle.
\end{equation}

\section{$CP$ violation in the decays of $K^{0}_{L}$-meson}

The observation of the decay $K^{0}_{L}\to \pi^{+}\pi^{-}$ marked
the discovery of the $CP$ violation \cite{Cronin}. During more than
30 years the study of  decays of  neutral kaons   was the only
source of the information about the $CP$ violation. In this section
we will consider in some details effects of the $CP$ violation in
decays of $K^{0}_{L}$-mesons (see \cite{WolfensteinPDG}).

The branching ratios of   main decay modes of $K^{0}_{S,L}$-mesons
are presented in the Table I \cite{PDG}

\begin{center}
 Table I
\end{center}
\begin{center}
Branching ratios for main decay channels of $K^{0}_{S,L}$-mesons.
\end{center}
\begin{center}
\begin{tabular}{|cccc|}
\hline $K^{0}_{S}$-decay channels & Branching ratio &
$K^{0}_{L}$-decay channels & Branching ratio
\\
\hline $K^{0}_{S}\to \pi^{+}  \pi^{-}$ & $68.95 \pm 0.14 \%$ &
$K^{0}_{L}\to \pi^{+}  e^{-} \nu_{e}$ & $38.8 \pm 0.27 \%$
\\
$K^{0}_{S}\to \pi^{0}  \pi^{0}$ & $31.05 \pm 0.14 \%$ & $K_{L}\to
\pi^{+}  \mu^{-} \nu_{\mu}$ & $27.19 \pm 0.25 \%$
\\
$K^{0}_{S}\to \pi^{+}  \pi^{-} \pi^{0} $ & $(3.2 \pm 1.2) 10^{-7}$ &
$K^{0}_{L}\to 3\pi^{0} $ & $21.05 \pm 0.23 \%$
\\
------
&
------
& $K^{0}_{L}\to\pi^{+}  \pi^{-} \pi^{0}  $ & $12.59 \pm 0.19 \%$
\\
\hline
\end{tabular}
\end{center}

As it is seen from the Table I  $K^{0}_{S}$-meson decays mainly into
two pions and $K^{0}_{L}$-meson decays mainly into three particles:
three pions and pion, lepton, neutrino. Because the phase-space
factor in the case of  the decay into two particles is much larger
than in the case of the decay into three particles, the time of life
of $K_{L}$ is much larger than the time of life of $K_{S}$
\cite{PDG}:
\begin{equation}\label{166}
\tau_{L}= \frac{1}{\Gamma_{L}}= (5.18 \pm 0.04)\cdot 10^{-8}
~\rm{sec},~ \tau_{S}= \frac{1}{\Gamma_{S}}=( 0.8953\pm 0.006) \cdot
10^{-10} ~\rm{sec}.
\end{equation}
For the ratio of the widths of $K^{0}_{S}$ and $K^{0}_{L}$ we have
\begin{equation}\label{167}
\frac{\Gamma_{S}}{\Gamma_{L}}\simeq 580.
\end{equation}

For the masses of $K^{0}_{S}$ and $K^{0}_{L}$ it was found the value
\cite{PDG}
\begin{equation}\label{168}
m_{S,L}= (497.648\pm 0.022)~ \rm{MeV}
\end{equation}
For the difference of the masses of $K^{0}_{L}$ and $K^{0}_{S}$
mesons the following value was found \cite{PDG}
\begin{equation}\label{169}
\Delta m= m_{L}- m_{S}= (0.5992\pm 0.0010)\cdot 10^{10}\hbar~ s^{-1}= (3.483\pm 0.006)\cdot 10^{-12}~ \rm{MeV}
\end{equation}
Let us notice the following approximate empirical  relation
\begin{equation}\label{170}
\frac{1}{2}~\Gamma_{S} \simeq \Delta m.
\end{equation}

We will consider  the $CP$ forbidden decays
\begin{equation}\label{171}
K^{0}_{L}\to \pi^{+} + \pi^{-} ~~\rm{and} ~~~K^{0}_{L}\to \pi^{0} +
\pi^{0}.
\end{equation}
For the states of  $K_{L,S}$-mesons we have
\begin{equation}\label{172}
|K^{0}_{L}\rangle =p~|K^{0}\rangle -q~|\bar K^{0}\rangle,~~
|K^{0}_{S}\rangle =p~|K^{0}\rangle +q~|\bar K^{0}\rangle,
\end{equation}
where the parameters $p$ and $q$ are given by (\ref{156}) and
(\ref{158}), respectively.

The $|K^{0}_{L,S}\rangle$ states  can be presented in another form.
Let us introduce the complex parameter $\bar\epsilon$\footnote{ The
parameter $\bar\epsilon$ characterizes $CP$ violation in the
 $K^{0}_{L,S}$ states. Let us  stress, however, that $\bar\epsilon$
depends on arbitrary phases of the $|K^{0}\rangle$ and $|\bar
K^{0}\rangle$ states.}
\begin{equation}\label{174}
\bar\epsilon=\frac{1-\frac{q}{p}}{1+\frac{q}{p}}= \frac{p-q}{p+q}
\end{equation}
From (\ref{172}) and (\ref{174}) for the normalized
$|K^{0}_{L,S}\rangle $ states we find
\begin{equation}\label{175}
|K^{0}_{L,S}\rangle
=\frac{1}{\sqrt{2~(1+|\bar\epsilon|^{2})}}~[(1+\bar\epsilon)|K^{0}\rangle
\mp(1-\bar\epsilon)|\bar K^{0}\rangle].
\end{equation}
 If $CP$ is conserved, in this case  $\bar\epsilon=0$ and for
the states of the long-lived and the short-lived kaons we have
\begin{equation}\label{176}
|K^{0}_{2,1}\rangle =\frac{1}{\sqrt{2}}(|K^{0}\rangle \mp |\bar
K^{0}\rangle).
\end{equation}
The states $|K^{0}_{2,1}\rangle $ are eigenstates of the operator of
the $CP$ conjugation:
\begin{equation}\label{177}
  CP~|K^{0}_{2,1}\rangle=\mp |K^{0}_{2,1}\rangle.
\end{equation}
The states $|K^{0}_{L,S}\rangle$ can be presented in the form
\begin{equation}\label{178}
|K^{0}_{L}\rangle
=\frac{1}{\sqrt{(1+|\bar\epsilon|^{2})}}(|K_{2}^{0}\rangle
+\bar\epsilon|~\bar K_{1}^{0}\rangle),~~ |K^{0}_{S}\rangle
=\frac{1}{\sqrt{(1+|\bar\epsilon|^{2})}}(|K_{1}^{0}\rangle
+\bar\epsilon|~\bar K_{2}^{0}\rangle)
\end{equation}
Let us introduce the measurable parameters
\begin{equation}\label{179}
\eta_{+-}=\frac{\langle \pi^{+}    \pi^{-}|T|   K^{0}_{L}\rangle
}{\langle \pi^{+}    \pi^{-}|T|   K^{0}_{S}\rangle},~~~
\eta_{00}=\frac{\langle \pi^{0}    \pi^{0}|T|   K^{0}_{L}\rangle
}{\langle \pi^{}    \pi^{}|T|   K^{0}_{S}\rangle},
\end{equation}
 which characterize the $CP$ violation in the decays
(\ref{171}).\footnote{ In fact, the states $|\pi^{+} \pi^{-}\rangle$
and $|\pi^{0} \pi^{0}\rangle$ are eigenstates of the  operator of
the $CP$ conjugation with eigenvalues equal to 1. If $CP$ is
conserved, the state of the long-lived kaon is $|K^{0}_{2}\rangle $,
which is eigenstate of $CP$ with eigenvalue equal to -1. Thus,  in
the case of the $CP$ conservation $\eta_{+-}=\eta_{00}=0$.} In
(\ref{179}) $T$-matrix is connected with the $S$-matrix by the
relation $S=1+iT$.

The complex parameters  $\eta_{+-}$ and $\eta_{00}$  can be
presented in the form
\begin{equation}\label{180}
\eta_{+-}=|\eta_{+-}|~e^{i\phi_{+-}},~~~\eta_{00}=|\eta_{00}|~e^{i\phi_{00}}.
\end{equation}
From analysis of the experimental data it was found \cite{PDG}
\begin{eqnarray}\label{181}
|\eta_{+-}|&=&(2.288 \pm 0.014)\cdot10^{-3},~|\eta_{+-}|=(2.276\pm 0.014)\cdot10^{-3}\nonumber\\
\phi_{+-}&=&(43.52 \pm 0.06)^\circ,~\phi_{+-}=(43.50 \pm
0.06)^\circ.
\end{eqnarray}
The spin of the kaon is equal to zero, Thus,  in the $K^{0}_{L,S}$-
decays two pions are produced in the $ S$-state. Total isotopic spin
$I$ of two pions  takes the values 0,1,2. However, from the
Bose-Einstein statistics it follows that  state with $I=1$ is
forbidden. Hence, the states of two pions, produced in the decays of
$K_{L,S}$  are superpositions of states with the total isotopic spin
equal to 0 and 2. We have
\begin{eqnarray}\label{182}
| \pi^{+}  \pi^{-} \rangle& =&\sqrt{\frac{2}{3}}~| 0 \rangle+\sqrt{\frac{1}{3}}~| 2 \rangle\nonumber\\
| \pi^{0}  \pi^{0} \rangle& =&\sqrt{\frac{1}{3}}~| 0 \rangle-\sqrt{\frac{2}{3}}~| 2 \rangle,
\end{eqnarray}
where $| I \rangle$ is the state of the two pions with the angular
momentum equal to zero  and the total isotopic spin equal to $I$.

The presentation of the states of two pions as a superposition of
states with definite total isotopic spin  will allow us to take into
account the approximate $|\Delta I| =\frac{1}{2} $ rule, which is
valid for the nonleptonic decays of the strange particles. For
example, according to this  rule the ratio of the total widths of
the decays $K_{S}\to \pi^{+} \pi^{-}$ and $K_{S}\to \pi^{0} \pi^{0}$
must be equal to 2. We see from the Table I that this prediction is
satisfied with the accuracy of about 10\%.

Let us consider now the parameters $\eta_{+-}$ and $\eta_{00}$. From
(\ref{179}) and  (\ref{182}) we have
\begin{equation}\label{184}
\eta_{+-}=\frac{ \langle 0|T|   K^{0}_{L}\rangle
+\frac{1}{\sqrt{2}}~ \langle 2|T| K^{0}_{L}\rangle}{ \langle 0|T|
K^{0}_{S}\rangle +\frac{1}{\sqrt{2}}~ \langle 2|T|
K^{0}_{S}\rangle}.
\end{equation}
The amplitude $ \langle 0|T|   K^{0}_{S}\rangle $ is  $CP$-allowed
and is allowed by the $|\Delta I| =\frac{1}{2} $) rule. If we divide
numerator and denominator of (\ref{184}) by this "large" amplitude
in the linear over small parameters approximation we find
\begin{equation}\label{185}
\eta_{+-}\simeq \epsilon +\epsilon',
 \end{equation}
where
\begin{equation}\label{186}
\epsilon= \frac{\langle 0|T|   K^{0}_{L}\rangle}{ \langle 0|T|
K^{0}_{S}\rangle}
 \end{equation}
and
\begin{equation}\label{187}
\epsilon'= \frac{1}{\sqrt{2}}~\left( \frac{\langle 2|T|
K^{0}_{L}\rangle}{ \langle 0|T|   K^{0}_{S}\rangle}
-\epsilon~\frac{\langle 2|T|   K^{0}_{S}\rangle}{ \langle 0|T|
K_{S}\rangle}\right).
 \end{equation}
Analogously, for  $\eta_{00}$ we obtain
\begin{equation}\label{188}
\eta_{00}\simeq \epsilon -2~\epsilon',
 \end{equation}
Thus, we can characterize the $CP$ violation in the decays
(\ref{171}) by the parameters $\epsilon$ and $\epsilon'$.
\footnote{Let us notice that parameters $\epsilon$ and $\epsilon'$
do not depend on arbitrary phases of $| K^{0}\rangle$ and $|\bar
K^{0}\rangle.$} From (\ref{187}),  we can expect that $|\epsilon'|
\ll |\epsilon|$. As we will see later, experimental data confirm
this expectation.

All existing data on the investigation of effects of the $CP$
violation in decays of $K^{0}_{L}$ are described by the Standard
Model with three families of quarks. It is interesting, however,  to
mention other alternatives. Historically it was important the
hypothesis of a superweak interaction \cite{Wolfenstein64}. It was
suggested in \cite{Wolfenstein64} that effects of the $CP$ violation
in the decays (\ref{171}) can be explained by existence of a new
interaction which violates $CP$ and changes the strangeness by two
units.

In order to explain the idea of the superweak model, let us consider
 the relation (\ref{174}). Taking into account (\ref{156}) and
(\ref{158}) we find
\begin{equation}\label{189}
\bar \epsilon= \frac{
\mathcal{H}_{12}-\mathcal{H}_{21}}{(\sqrt{\mathcal{H}_{12}}+
\sqrt{\mathcal{H}_{21}})^{2}}
\end{equation}
Obviously we have
\begin{equation}\label{190}
(\sqrt{\mathcal{H}_{12}}+ \sqrt{\mathcal{H}_{21}})^{2}
=\bar\epsilon^{2} (\sqrt{\mathcal{H}_{12}}+
\sqrt{\mathcal{H}_{21}})^{2}+4
\sqrt{\mathcal{H}_{12}\mathcal{H}_{21}}.
\end{equation}
From (\ref{155}) and (\ref{190}) we find
\begin{equation}\label{191}
(\sqrt{\mathcal{H}_{12}}+ \sqrt{\mathcal{H}_{21}})^{2} =-\frac{2
(\lambda_{L}-   \lambda_{S})}{1-\bar\epsilon^{2}}
\end{equation}
Thus,  we have
\begin{equation}\label{192}
\frac{\bar\epsilon}{ 1-  \bar\epsilon^{2} }= - \frac{
\mathcal{H}_{12}-\mathcal{H}_{21}} { 2 (\lambda_{L}-   \lambda_{S})
}
\end{equation}
Taking into account that  $|\bar\epsilon|\sim 2\cdot 10^{-3}$ and
$\frac{\Gamma_{L}}{\Gamma_{S}}\ll 1 $, from (\ref{192}) we find the
following relation
\begin{equation}\label{193}
\bar\epsilon\simeq - \frac{ \mathcal{H}_{12}-\mathcal{H}_{21}} { 2(
\Delta m +\frac{i}{2} \Gamma_{S}) },
\end{equation}
where $\Delta m=m_{L}- m_{S}$.

It is shown in the Appendix B that in the  effective Hamiltonian
$\mathcal{H}$ enters linear in interaction term (see (\ref{B32})).
Taking into account that the usual weak interaction Hamiltonian
changes the strangeness by one unit, we have
\begin{equation}\label{194}
\langle K^{0}|~H_{W}~|\bar K^{0}\rangle= \langle K^{0}|~H_{SW}~|\bar
K^{0}\rangle\not=\langle \bar K^{0}|~H_{SW}~|K^{0}\rangle,
\end{equation}
where $H_{SW}$ is a  Hamiltonian which violates $CP$ and changes the
strangeness by two units.
 It is evident
from (\ref{193}) that if  such  interaction exists the parameter
$\bar\epsilon$ is different from zero.

In order to estimate the effective constant $G_{SW}$, which
characterize the interaction $H_{SW}$,  we will use the relation
(\ref{193}). Taking into account that $\Delta m \simeq \frac{1}{2}~
\Gamma_{S}$ we have
\begin{equation}\label{195}
|\bar\epsilon|\simeq\frac{G_{SW}}{ G^{2}_{F}m^{2}_{K}}\approx
10^{-3},
\end{equation}
where $G_{F}\simeq 10^{-5}\frac{1}{m^{2}_{p}}$ is the Fermi constant
and $m_{K}$ is the mass of the kaon. Thus, effects of the $CP$
violation in the decays of $K_{L}$ can be explained if the constant
of $|\Delta S| =2$  interaction which violate $CP$ is given by $
G_{SW}\approx 10^{-9}~ G_{F}$, i.e. is much smaller than the Fermi
constant. This is the reason why this interaction is called
superweak.

Let us consider the parameters $\eta_{+-}$ and $\eta_{00}$ in the
case of the superweak interaction. From (\ref{178}) and (\ref{179})
we have
\begin{equation}\label{196}
\eta_{+-}=\frac{ \langle \pi^{+}\pi^{-}|T|   K_{2}\rangle +
\bar\epsilon \langle \pi^{+}\pi^{-}|T|   K_{1}\rangle } {\langle
\pi^{+}\pi^{-}|T|   K_{1}\rangle +\bar\epsilon \langle
\pi^{+}\pi^{-}|T|   K_{2}\rangle }.
\end{equation}
In the superweak model   $ \langle \pi^{+}\pi^{-}|T|
K_{2}\rangle\simeq 0$. We  have
\begin{equation}\label{197}
\eta_{+-}=\bar \epsilon.
\end{equation}
Analogously, for the decay $K_{L} \to \pi^{0}\pi^{0}$ we find
\begin{equation}\label{198}
\eta_{00}=\bar \epsilon.
\end{equation}
Thus, if the superweak interaction is the origin of the effects of
the $CP$ violation, observed in the decays $K_{L} \to \pi\pi $, we
would have
\begin{equation}\label{199}
\eta_{+-}=\eta_{00}.
\end{equation}
From  (\ref{185}), (\ref{188}) and (\ref{199}) we conclude that in
case of the superweak interaction $\epsilon'=0$.

 Taking into account linear in
$\frac{\epsilon'}{\epsilon}$ terms, from (\ref{185}) and (\ref{188})
 we have in the general case
\begin{equation}\label{200}
|\eta_{+-}|^{2}\simeq
|\epsilon|^{2}(1+2~\rm{Re}\frac{\epsilon'}{\epsilon}),~
|\eta_{00}|^{2}\simeq
|\epsilon|^{2}(1-4~\rm{Re}~\frac{\epsilon'}{\epsilon}).
\end{equation}
From this relations we find
\begin{equation}\label{201}
\rm{Re}~\frac{\epsilon'}{\epsilon}=\frac{1}{6}(1-\frac{|\eta_{00}|^{2}}{|\eta_{+-}|^{2}}).
\end{equation}
The ratio $\frac{|\eta_{00}|^{2}}{|\eta_{+-}|^{2}}$
 was measured in spectacular NA48 \cite{NA48} and KTeV \cite{KTeV}
 experiments.
It was found from the data of these experiments that
\begin{equation}\label{202}
\rm{Re}~\frac{\epsilon'}{\epsilon}=(14.7 \pm
2.2)~10^{-4}~(\rm{NA48}),~~ \rm{Re}~\frac{\epsilon'}{\epsilon}=
(20.7 \pm 2.8)~10^{-4}~( \rm{KTeV}).
\end{equation}
Thus, it was proved that the parameter $\epsilon'$ is different from
zero and is much smaller than the parameter $\epsilon$. Therefore,
it was proved that effects of the $CP$ violation, observed in the
decays (\ref{171}), can not be explained by the superweak
interaction. It was shown that the measured value of the parameter
$\rm{Re}~\frac{\epsilon'}{\epsilon}$ can be explained by the SM
(see\cite{BurasII}).

 We will consider  now  the expressions (\ref{186})
and (\ref{187}) for $\epsilon$ and $\epsilon'$. Neglecting quadratic
in small parameters terms,  for $\epsilon$ we obtain the following
expression
\begin{equation}\label{203}
\epsilon \simeq \bar \epsilon +\frac{\langle 0|T| K_{2}\rangle}{
\langle 0|T|   K_{1}\rangle}.
\end{equation}
For the parameter $\epsilon'$ we find
\begin{eqnarray}\label{204}
\epsilon'&=&\frac{1}{\sqrt{2}}~\left[\frac{\langle 2|T|   K_{2}\rangle}{ \langle 0|T|   K_{1}\rangle}
+(\bar \epsilon-\epsilon)\frac{\langle 2|T|   K_{1}\rangle}{ \langle 0|T|   K_{1}\rangle}\right]\nonumber\\
&=&\frac{1}{\sqrt{2}}~\frac{\langle 2|T|   K_{1}\rangle}{ \langle
0|T|   K_{1}\rangle} \left[\frac{\langle 2|T|   K_{2}\rangle}{
\langle 2|T|   K_{1}\rangle} -\frac{\langle 0|T|   K_{2}\rangle}{
\langle 0|T|   K_{1}\rangle}\right],
\end{eqnarray}
where
\begin{equation}\label{205}
\langle I|T|   K_{1,2}\rangle =\frac{1}{\sqrt{2}}~(\langle I|T|
K^{0}\rangle\pm \langle I|T|  \bar K^{0}\rangle),~~ I=0,2.
\end{equation}
Let us consider the matrix elements $\langle I|T| K^{0}\rangle$ and
$\langle I|T| \bar  K^{0}\rangle$. From  the unitarity of the
S-matrix we have
\begin{equation}\label{206}
S^{\dagger}~T=T^{\dagger}.
\end{equation}
From this relation  we find
\begin{equation}\label{207}
\langle I|S^{\dagger}~T|   K^{0}\rangle =
\sum_{n}\langle I|S^{\dagger}|  n\rangle
\langle n|T| K^{0} \rangle= \langle K^{0} |T| I \rangle^{*}.
\end{equation}
In the sum over intermediate states $|n\rangle$ enter $| I \rangle$,
$| \pi\pi\pi \rangle$, $| \pi\pi\gamma \rangle$ and other states.
The main contribution  gives  the two-pion state $| I \rangle$: the
state of three pions is forbidden by the conservation of $G$-parity
in the strong interaction, contributions of other states are
suppressed by phase space factor and by $\alpha$. We have
\begin{equation}\label{208}
\langle I|S^{\dagger}| I\rangle =e^{-2i\delta_{I}},
\end{equation}
where $\delta_{I}$ is the phase of the $\pi-\pi$-scattering in the
state with the total isotopic spin equal to $I$, the angular
momentum  equal to zero and  the energy in the  center of mass
system equal to $m_{K}$.

Further, from the  $CPT$ invariance it follows that
\begin{equation}\label{209}
\langle K^{0} |T| I \rangle=\langle I|T| \bar  K^{0}\rangle.
\end{equation}
Thus, from the unitarity of the $S$-matrix and the $CPT$ invariance
we find that the matrix elements $\langle I|T|   K^{0}\rangle$ and
$\langle I|T| \bar  K^{0}\rangle$ are connected by the following
relation
\begin{equation}\label{210}
e^{-2i\delta_{I}}\,\langle I   |T|  K^{0}\rangle=\langle I|T| \bar
K^{0}\rangle^{*}.
\end{equation}
Let us introduce the complex amplitudes $A_{I}$ and $\bar A_{I}$ in
the following way
\begin{equation}\label{211}
\langle I   |T|  K^{0}\rangle= e^{i\delta_{I}}A_{I},~~ \langle I |T|
\bar K^{0}\rangle= e^{i\delta_{I}}\bar A_{I}
\end{equation}
From the relation  (\ref{210}) we find that
\begin{equation}\label{212}
\bar A_{I}= A^{*}_{I}.
\end{equation}
Thus, we have
\begin{equation}\label{213}
\langle I   |T|  K^{0}\rangle= e^{i\delta_{I}}A_{I},~~ \langle I |T|
\bar K^{0}\rangle= e^{i\delta_{I}} A^{*}_{I}.
\end{equation}
In the case of the  $CP$ conservation we have that
\begin{equation}\label{214}
\langle I   |T|  K^{0}\rangle=\langle I   |T| \bar K^{0}\rangle
\end{equation}
and
\begin{equation}\label{215}
 A_{I}= A^{*}_{I}.
\end{equation}
Let us return now back to relations (\ref{203}) and (\ref{204}).
Taking into account (\ref{213}) we find
\begin{equation}\label{216}
\epsilon =\bar\epsilon +i\frac{\rm{Im} A_{0}}{\rm{Re} A_{0} }.
\end{equation}
and
\begin{equation}\label{217}
\epsilon'=\frac{1}{\sqrt{2}} e^{i(\delta_{2}-\delta_{0} +\frac{\pi}{2})}\frac{\rm{Re} A_{2}}{\rm{Re} A_{0}}
\left[\frac{\rm{Im} A_{2}}{\rm{Re} A_{2}}-\frac{\rm{Im} A_{0}}{\rm{Re} A_{0}}\right]
\end{equation}
From these relations we can conclude the following:
\begin{enumerate}\item \begin{equation}\label{218}
 \phi_{\epsilon'}=\delta_{2}-\delta_{0} +\frac{\pi}{2},
\end{equation}
where $ \phi_{\epsilon'}=\rm{arg}~\epsilon'$ is the phase of the
parameter $\epsilon'$. From analysis of the  $\pi-\pi$ scattering
data it was obtained \cite{Calangelo}
\begin{equation}\label{219}
\delta_{2}-\delta_{0} +\frac{\pi}{2}=(42.3\pm 1.5) ^\circ
\end{equation}
\item \begin{equation}\label{220}
\rm{Re}~\epsilon=\rm{Re}~\bar\epsilon
\end{equation} The parameter $\bar\epsilon$ depends on the choice of arbitrary
phase of the $|K^{0}\rangle$ and $|\bar K^{0}\rangle$ states. We see
from (\ref{220}) that $\rm{Re}~\bar\epsilon$ is rephrase invariant
quantity.
\end{enumerate}
 For the phase of the parameter $\epsilon$ the following relation
 holds
\begin{equation}\label{221}
\phi_{\epsilon} \simeq \arctan\frac{2\Delta m}{\Gamma_{S}},
\end{equation}
 where $ \phi_{\epsilon}=\rm{arg}~\epsilon$.
This relation is based on the Bell-Steinberger unitarity relation
which we derive now. We have
\begin{equation}\label{222}
\mathcal{H}~a_{L}=\lambda_{L}~a_{L},~~a^{\dag}_{S}~\mathcal{H}^{\dag}=
\lambda^{*}_{S}~a^{\dag}_{S}.
\end{equation}
If we multiply the first equation by $a^{\dag}_{S}$ from the left
and the second one by $a_{L}$ from the right and subtract from the
first relation the second one we find
\begin{equation}\label{223}
(a^{\dag}_{S}\Gamma
a_{L})=i(\lambda_{L}-\lambda^{*}_{S})~(a^{\dag}_{S}a_{L}),
\end{equation}
This relation can be rewritten in the form
\begin{equation}\label{224}
  \langle K_{S}|\Gamma|K_{L}\rangle=
  i(\lambda_{L}-\lambda^{*}_{S})~\langle K_{S}|K_{L}\rangle
\end{equation}
The relation (\ref{224}) is  the Bell-Steinberger unitarity
relation. For the left-hand side of this equation we have
\begin{equation}\label{225}
\langle K_{S}|\Gamma|K_{L}\rangle=2\pi \sum_{i}\langle
K_{S}|H_{W}|i\rangle~\langle i|H_{W}|K_{L}\rangle ~\delta(E_{i}-m).
\end{equation}
In the sum over intermediate states $|i\rangle$ main contribution
give  two-pion states. Taking into account these states we have
\begin{equation}\label{226}
\langle K_{S}|\Gamma|K_{L}\rangle \simeq \eta_{+-}~\Gamma(K_{S}\to
\pi^{+}\pi^{-}) + \eta_{00}~\Gamma(K_{S}\to \pi^{0}\pi^{0}).
\end{equation}
Now, according to $|\Delta I|=1/2$ rule  we have
\begin{equation}\label{226a}
\Gamma(K_{S}\to \pi^{+}\pi^{-}) \simeq 2~\Gamma(K_{S}\to
\pi^{0}\pi^{0}).
\end{equation}
From (\ref{185}), (\ref{188}), (\ref{226}) and (\ref{226a}) we find
\begin{equation}\label{227}
\langle K_{S}|\Gamma|K_{L}\rangle \simeq (\frac{2}{3} ~\eta_{+-}+
\frac{1}{3}~\eta_{00})~\Gamma_{S}=\epsilon~\Gamma_{S},
\end{equation}
where $\Gamma_{S}$ is the total width of $K^{0}_{S}$-meson. Further,
we find
\begin{equation}\label{228}
\langle
K_{S}|K_{L}\rangle=\frac{2\rm{Re}~\bar\epsilon}{1+|\bar\epsilon|^{2}}
\simeq 2\rm{Re}~\bar\epsilon=2\rm{Re}~\epsilon
\end{equation}
From  (\ref{224}), (\ref{227}) and (\ref{228}) we obtain the
following relation
\begin{equation}\label{229}
 \epsilon~\Gamma_{S}=2(i\Delta
m+\frac{1}{2}\Gamma_{S})~\rm{Re}~\epsilon
\end{equation}
If we take the real part of (\ref{229}) we obtain identity. From the
imaginary part of (\ref{229}) we find the following relation
\begin{equation}\label{230}
\frac{\rm{Im}~\epsilon}{\rm{Re}~\epsilon} = \tan
\phi_{\epsilon}=\frac{2\Delta m}{\Gamma_{S}}.
\end{equation}
Thus, the phase of the parameter $\epsilon$ is given by  the
relation (\ref{221}).

The mass difference $\Delta m$ and the width $\Gamma_{S}$ are
connected by the empirical relation $\Delta m \simeq
\frac{1}{2}\Gamma_{S}$. Thus, $\phi_{\epsilon} \simeq \pi/4$. The
experimental data are in an agreement with this prediction of the
theory. We have \cite{PDG}
\begin{equation}\label{231}
\phi_{\epsilon} =(43.5\pm 0.7)^\circ.
\end{equation}
From (\ref{219}) and (\ref{231}) it follows that phases of the
parameters $\epsilon$ and $\epsilon'$ are approximately equal
\begin{equation}\label{231a}
\phi_{\epsilon} \simeq \phi_{\epsilon'}.
\end{equation}

Up to now we considered effects of the $CP$ violation in the
two-pion decays of $K^{0}_{L}$-meson. Effects of the $CP$ violation
were observed also in the semi-leptonic decays
\begin{equation}\label{232}
K^{0}_{L}\to \pi^{-}l^{+}\nu_{l},~~K^{0}_{L}\to
\pi^{+}l^{-}\bar\nu_{l}.
\end{equation}
Let us determine the $CP$ asymmetry
\begin{equation}\label{233}
A_{L} = \frac{\Gamma(K^{0}_{L}\to \pi^{-}l^{+}\nu_{l})-
\Gamma(K^{0}_{L}\to \pi^{+}l^{-}\bar\nu_{l}) }{  \Gamma(K^{0}_{L}\to
\pi^{-}l^{+}\nu_{l})+ \Gamma(K^{0}_{L}\to \pi^{+}l^{-}\bar\nu_{l})
},
\end{equation}
where $\Gamma(K_{L}\to \pi^{-}l^{+}\nu_{l})$ and
$\Gamma(K^{0}_{L}\to \pi^{+}l^{-}\bar\nu_{l})$ are the total widths
of the decays $K^{0}_{L}\to \pi^{-}l^{+}\nu_{l}$  and $K_{L}\to
\pi^{+}l^{-}\bar\nu_{l}$. If  the $CP$ is conserved the  asymmetry
$A_{L}$ is equal to zero. In fact, in this case initial state  is
the eigenstate of the operator of the $CP$ conjugation and final
states are the $CP$ conjugated states. The probabilities of the
transitions to such states must be equal in the case of the $CP$
conservation.

Let us consider the asymmetry $A_{L}$. The semi-leptonic decay of
the $K^{0}$-meson, which is
 the bound state of $\bar s$ and $d$ quarks, is due to the transition $\bar
s\to \bar u +l^{+}+\nu_{l}$. Analogously the decay  of the $\bar
K^{0}$-meson, which is the bound state of the $ s$ and $\bar d$
quarks, is due the transition $s\to  u +l^{-} +\bar\nu_{l}$. Thus,
the decay $K^{0}\to \pi^{-}l^{+}\nu_{l}$ is allowed and decay $\bar
K^{0}\to \pi^{-}l^{+}\nu_{l}$ is forbidden and the decay $\bar
K^{0}\to \pi^{+}l^{-} \bar\nu_{l}$ is allowed and $K^{0}\to
\pi^{+}l^{-} \bar\nu_{l}$  is forbidden. This corresponds to the
$\Delta Q= \Delta S $ rule.  Further, from the $CPT$ invariance it
follows that
\begin{equation}\label{234}
\langle\pi^{+}l^{-} \bar\nu_{l}|T|\bar K^{0}\rangle = \langle
K^{0}|T|\pi^{-}l^{+}\nu_{l}\rangle \simeq\langle\pi^{-}l^{+}
\nu_{l}|T| K^{0}\rangle ^{*},
\end{equation}
where we took into account that in the Born approximation
$T=T^{\dag}$.

From (\ref{175}) and (\ref{234}) we find
\begin{equation}\label{235}
A_{L} = \frac{|1+\bar\epsilon|^{2}-|1-\bar\epsilon|^{2}} {
|1+\bar\epsilon|^{2}+|1-\bar\epsilon|^{2}} =\frac{2~\rm{Re}
\bar\epsilon}{1+|\bar\epsilon|^{2}}\simeq 2\rm{Re}~ \bar\epsilon.
\end{equation}
Now, taking into account (\ref{220}), we finally have
\begin{equation}\label{236}
A_{L} \simeq2\rm{Re}~\epsilon=2|\epsilon|\cos\phi_{\epsilon}
\end{equation}
From experimental data for the asymmetry $A_{L}$ it was found the
value \cite{PDG}
\begin{equation}\label{237}
A_{L}= (3.32 \pm 0.06)\cdot 10^{-3}.
\end{equation}
From  (\ref{219}) and (\ref{237}) for the parameter $|\epsilon|$ was
found the value
\begin{equation}\label{238}
|\epsilon|=(2.232 \pm 0.007)\cdot 10^{-3}
\end{equation}
In order to connect $|\epsilon|$  with parameters, characterizing
$CKM$ mixing matrix, it is necessary to calculate quark box diagrams
which determine the amplitude of $K^{0}\to \bar K^{0} $ transition.
Taking into account the $QCD $ corrections for the parameter
$|\epsilon|$ it was found the following expression
\begin{equation}\label{239}
|\epsilon|=A^{2}~a~\bar\eta~[A^{2}b~(1-\bar \rho)+c],
\end{equation}
where  $a, b$ and $c$ are given in \cite{BurasII}. The equation
(\ref{239}) gives hyperbola in $\bar \rho,\bar\eta$ plane. It is
used in the standard unitarity triangle fit which we will discuss
later.

\section{$CP$ violation and mixing in  $B^{0}- \bar B^{0}$  system}

We will consider in this section effects of the $CP$ violation in
decays of the mixed $B^{0}-\bar B^{0}$ system. These effects were
investigated in details in the  BaBar and the Belle experiments at
the asymmetric B-factories and in the D0 and the CDF experiments at
the Fermilab. At the B-factories $B^{0}_{d}$ and $\bar B^{0}_{d}$
mesons are resonansly produced in decays of $\Upsilon(4S)$.

The states of $B^{0}_{H}$ and $B^{0}_{L}$ mesons, particles with
definite masses and widths, are given by the following relations
\begin{equation}\label{240}
|B^{0}_{H,L}\rangle =p~|B^{0}\rangle \mp q~|\bar B^{0}\rangle,
\end{equation}
where $p$ and $q$ are connected with non diagonal elements of the
effective Hamiltonian by the relations (\ref{156}) and (\ref{157}).
Let us stress that we have choosen  arbitrary phases of the states
$|B^{0}\rangle$ and $|\bar B^{0}\rangle$ in such a way that $|\bar
B^{0}\rangle= CP~|B^{0}\rangle$.

The states   $|B^{0}_{H,L}\rangle $ are eigenstates of the effective
Hamiltonian $\mathcal{H}$ with eigenvalues
\begin{equation}\label{241}
\mu_{H,L}=m_{H,L}-i\frac{1}{2}~\Gamma_{H,L}.
\end{equation}
Here $m_{H,L}$ and $\Gamma_{H,L}$ are masses and total decay widths
of $B^{0}_{H,L}$ mesons. Because of the large difference in the
lifetimes of the short-lived and long-lived kaons it is possible to
produce beams of $K^{0}_{L}$-mesons. In the case of the $B^{0}$
mesons the situation is different. The lifetimes of $B^{0}_{H}$ and
$B^{0}_{L}$ are quite close. Only mixtures of $B^{0}_{H}$ and
$B^{0}_{L}$ can be studied in experiments.

Let us obtain first the mixed states which are the result of the
evolution of the initial (at $t=0$)  $|B^{0}\rangle$ and $|\bar
B^{0}\rangle $ states. From (\ref{240}) we have
\begin{equation}\label{242}
|B^{0}\rangle =\frac{1}{2p}~(|B_{H}^{0}\rangle +|
B_{L}^{0}\rangle),~~~|\bar B^{0}\rangle
=\frac{1}{2q}~(-|B_{H}^{0}\rangle +| B_{L}^{0}\rangle).
\end{equation}
From (\ref{242}) we find
\begin{equation}\label{243}
|B^{0}(t)\rangle=\frac{1}{2p}~(e^{-i\mu_{H}t}|B_{H}^{0}\rangle +
e^{-i\mu_{L}t}|B_{L}^{0}\rangle)=g_{+}(t)~|B^{0}\rangle
-\frac{q}{p}~g_{-}(t)~| \bar B^{0}\rangle
\end{equation}
and
\begin{equation}\label{244}
|\bar B^{0}(t)\rangle=\frac{1}{2q}~(-e^{-i\mu_{H}t}|B_{H}^{0}\rangle
+
e^{-i\mu_{L}t}|B_{L}^{0}\rangle)=-\frac{p}{q}~g_{-}(t)~|B^{0}\rangle
+g_{+}(t)~|\bar B^{0}\rangle.
\end{equation}
Here
\begin{equation}\label{245}
g_{\pm}(t)=\frac{1}{2}~(e^{-i\mu_{H}t}\pm e^{-i\mu_{L}t}).
\end{equation}
Let us present $\mu_{H,L}$ in the form
\begin{equation}\label{246}
\mu_{H}=\mu +\frac{1}{2}~\Delta\mu,~~~\mu_{L}=\mu
-\frac{1}{2}~\Delta\mu.
\end{equation}
Here
\begin{equation}\label{247}
\mu=\frac{\mu_{H}+\mu_{L}}{2}=m-i\frac{1}{2}~\Gamma,~~\Delta\mu=\mu_{H}-\mu_{L}
=\Delta m -i\frac{1}{2}~\Delta \Gamma,
\end{equation}
where
\begin{equation}\label{248}
 m=\frac{m_{H}+m_{L}}{2}
 ,~\Gamma=\frac{\Gamma_{H}+\Gamma_{L}}{2},~
\Delta m=m_{H}-m_{L},~\Delta\Gamma=\Gamma_{H}-\Gamma_{L}.
\end{equation}
Notice that by the definition $\Delta m>0$. From (\ref{245})  and
(\ref{246}) for the functions $g_{\pm}(t)$ we obtain the following
expressions
\begin{equation}\label{249}
g_{\pm}(t)=\frac{1}{2}~e^{-i\mu t}(e^{-i\frac{1}{2}\Delta\mu t}\pm
e^{i\frac{1}{2}\Delta\mu t}).
\end{equation}
Let us consider the decays of $B^{0}$ and $\bar B^{0}$ into a state
$|f\rangle$ which is the eigenstate of the operator of the $CP$
conjugation
\begin{equation}\label{250}
CP~|f\rangle=\pm |f\rangle.
\end{equation}
From (\ref{243}) for the transition amplitude we find
\begin{equation}\label{251}
\langle f|T| B^{0}(t)\rangle=\langle f|T| B^{0}\rangle~
(g_{+}(t)-\lambda_{f}~g_{-}(t)),
\end{equation}
where
\begin{equation}\label{252}
\lambda_{f}=\frac{q}{p}~\frac{\langle f|T| \bar
B^{0}\rangle}{\langle f|T| B^{0}\rangle}.
\end{equation}
For  the transition amplitude $\langle f|T| \bar B^{0}(t)\rangle$ we
have
\begin{equation}\label{253}
\langle f|T| \bar B^{0}(t)\rangle=\frac{p}{q}~  \langle f|T|
B^{0}\rangle~(-g_{-}(t)+\lambda_{f}~g_{+}(t)).
\end{equation}
From (\ref{251}) we find that  the transition probability is given
by the  expression
\begin{equation}\label{254}
\Gamma(B^{0}(t)\to f)=\Gamma(B^{0}\to f)~(|g_{+}(t)|^{2}+
|\lambda_{f}|^{2}~|g_{-}(t)|^{2}-2~\rm{Re}\lambda_{f}~g_{-}(t)g^{*}_{+}(t)).
\end{equation}
Further,  from (\ref{249}) we have
\begin{equation}\label{255}
|g_{\pm}(t)|^{2}=\frac{1}{2}~e^{-\Gamma
t}(\cosh\frac{1}{2}\Delta\Gamma t \pm \cos\Delta m t)
\end{equation}
and
\begin{equation}\label{256}
g_{-}(t)g^{*}_{+}(t)=-\frac{1}{2}~e^{-\Gamma
t}~(\sinh\frac{1}{2}\Delta\Gamma t +i\sin\Delta m t).
\end{equation}
From (\ref{254}), (\ref{255}) and (\ref{256}) for the transition
probability $\Gamma(B^{0}(t)\to f)$ we obtain the following
expression
\begin{eqnarray}\label{257}
\Gamma(B^{0}(t)\to f)&=&\frac{1}{2}~e^{-\Gamma t}~\Gamma(B^{0}\to
f)~(1+|\lambda_{f}|^{2})~(\cosh\frac{1}{2}\Delta\Gamma t
\nonumber\\
&+&C_{f}\cos\Delta m t +D_{f} \sinh\frac{1}{2}\Delta\Gamma t
-S_{f}\sin\Delta m t),
\end{eqnarray}
where
\begin{equation}\label{258}
C_{f}= \frac{1-|\lambda_{f}|^{2}}{(1+|\lambda_{f}|^{2})},~
D_{f}=\frac{2~\rm{Re}~\lambda_{f}}{(1+|\lambda_{f}|^{2})},~
S_{f}=\frac{2~\rm{Im}~\lambda_{f}}{(1+|\lambda_{f}|^{2})}.
\end{equation}
It is obvious from (\ref{258}) that parameters $C_{f}$, $D_{f}$ and
$S_{f}$ satisfy the  relation
\begin{equation}\label{259}
C^{2}_{f}+D^{2}_{f}+ S^{2}_{f}=1
\end{equation}
For the transition probability $\Gamma(\bar B^{0}(t)\to f)$ from
(\ref{253}) we find
\begin{equation}\label{260}
\Gamma(\bar B^{0}(t)\to f)=|\frac{p}{q}|^{2}~\Gamma(B^{0}\to
f)~(|g_{-}(t)|^{2}+
|\lambda_{f}|^{2}~|g_{+}(t)|^{2}-2\rm{Re}\lambda_{f}~g_{+}(t)g^{*}_{-}(t)).
\end{equation}
From (\ref{255}), (\ref{256}) and (\ref{260}) for the probability
$\Gamma(\bar B^{0}(t)\to f)$ we obtain the following expression
\begin{eqnarray}\label{261}
\Gamma(\bar B^{0}(t)\to f)&=&\frac{1}{2}~e^{-\Gamma
t}|\frac{p}{q}|^{2}~\Gamma(B^{0}\to
f)~(1+|\lambda_{f}|^{2})~(\cosh\frac{1}{2}\Delta\Gamma t
\nonumber\\
&-&C_{f}\cos\Delta m t +D_{f} \sinh\frac{1}{2}\Delta\Gamma t
+S_{f}\sin\Delta m t).
\end{eqnarray}
If $CP$ is conserved  $\langle f|T| B^{0}\rangle=\pm\langle f|T|
\bar B^{0}\rangle$, $p=q$ and $\lambda_{f}=\pm 1$. In this case we
have: $\Gamma(B^{0}(t)\to f)=\Gamma(\bar B^{0}(t)\to f)=
e^{-\Gamma_{H} t}(e^{-\Gamma_{L} t})~\Gamma(B^{0}\to f). $

The quantity $\lambda_{f}$, which determine the time dependence of
the probabilities $\Gamma( B^{0}(t)\to f)$ and $\Gamma(\bar
B^{0}(t)\to f)$, does not depend on arbitrary  phases of the states
of $B^{0}$, $\bar B^{0}$  and  $f$. In fact, let us consider the
states
\begin{equation}\label{262}
|B^{0}\rangle'=e^{i\alpha}|B^{0}\rangle,~|\bar
B^{0}\rangle'=e^{-i\alpha}|\bar
B^{0}\rangle,~|f\rangle'=e^{i\beta}|f\rangle,
\end{equation}
where $\alpha$ and $\beta$ are arbitrary constants. From (\ref{156})
and (\ref{158}) we have
\begin{equation}\label{263}
    q'=e^{i\alpha}q,~~p'=e^{-i\alpha}p.
\end{equation}
From (\ref{252}), (\ref{262}) and (\ref{263}) we find
\begin{equation}\label{264}
\lambda_{f}'=\frac{q'}{p'}~\frac{'\langle f|T| \bar
B^{0}\rangle'}{'\langle f|T| B^{0}\rangle'}=\lambda_{f}.
\end{equation}

Let us consider now the matrix element $\Gamma_{12}$. From
(\ref{B35}) we have
\begin{equation}\label{265}
\Gamma_{12}=2\pi~\sum_{i}\langle B^{0}|H_{W}| i\rangle \langle
i|H_{W}|\bar B^{0} \rangle \delta (E-m_{B})
\end{equation}
From this expression  follows that  the contribution to
$\Gamma_{12}$ give intermediate states $| i\rangle $ in which both
$B^{0}$ and $\bar B^{0}$ mesons can decay. In the Standard Model
 transitions to such states are strongly
suppressed (see, for example, \cite{BurasII}). Thus, in the SM we
have
\begin{equation}\label{266}
|\Gamma_{12}| \ll |M_{12}|.
\end{equation}
From (\ref{155}) we find
\begin{equation}\label{267}
    \Delta \mu = \Delta m -i\frac{1}{2}\Delta \Gamma =2 |M_{12}|
\sqrt{(1-\frac{i}{2}\frac{\Gamma_{12}}{M_{12}})
(1-\frac{i}{2}\frac{\Gamma^{*}_{12}}{M^{*}_{12}}})
\end{equation}
Taking into account (\ref{266}), from this expression we obtain
\begin{equation}\label{268}
\Delta m -i\frac{1}{2}\Delta \Gamma = 2~|M_{12}|(1-\frac{i}{2}
\rm{Re}~\frac{\Gamma_{12}}{M_{12}})+O(|\frac{\Gamma_{12}}{M_{12}}|^{2})
\end{equation}
Thus, we have
\begin{equation}\label{269}
\Delta m\simeq 2~ |M_{12}|,~~~\Delta \Gamma\simeq
2~\rm{Re}~\frac{\Gamma_{12}}{M_{12}}~|M_{12}|.
\end{equation}
Let us consider now the mixing parameter $\frac{q}{p}$.  We have
\begin{equation}\label{270}
 \frac{q}{p}=\sqrt{\frac{H_{21}}{H_{12}}}=-\frac{2~H_{21}}{\Delta\mu}.
\end{equation}
Neglecting terms of the order $O(|\frac{\Gamma_{12}}{M_{12}}|^{2})$,
from (\ref{268}) and (\ref{270}) we find
\begin{equation}\label{271}
\frac{q}{p}\simeq
-\frac{M^{*}_{12}~(1-\frac{i}{2}\frac{\Gamma^{*}_{12}}{M^{*}_{12}})}
{|M_{12}|(1-\frac{i}{2}\rm{Re}~\frac{\Gamma_{12}}{M_{12}})}= \simeq
-\frac{M^{*}_{12}}{|M_{12}|}(1-\frac{1}{2}~\rm{Im}~\frac{\Gamma_{12}}{M_{12}}).
\end{equation}
Let us determine the $CP$ asymmetry in the case of the decays of
$B^{0}$ and $\bar B^{0}$      into the  state $f$ which is the
eigenstate of the operator of the $CP$ conjugation
\begin{equation}\label{272}
A_{f}^{CP}(t)= \frac{\Gamma(\bar B^{0}(t)\to f)-\Gamma( B^{0}(t)\to
f)}{\Gamma(\bar B^{0}(t)\to f)+\Gamma(B^{0}(t)\to f)}.
\end{equation}
In the SM in the case of the  $B_{d}$-mesons
\begin{equation}\label{273}
    \frac{\Delta\Gamma_{d}}{\Gamma_{d}}\ll 1.
\end{equation}
For example, in \cite{Nierste} it was found
\begin{equation}\label{274}
\frac{\Delta\Gamma_{d}}{\Gamma_{d}}=(40.9^{+8.9}_{-9.9})\cdot10^{-4}.~~
\frac{\Delta\Gamma_{s}}{\Gamma_{s}}=0.127 \pm 0.024.
\end{equation}
We will consider $B^{0}_{d}- \bar B^{0}_{d}$-system. We can neglect
in (\ref{257}) and (\ref{261}) $\Delta\Gamma_{d}$. We can also
neglect $\rm{Im}~\frac{\Gamma_{12}}{M_{12}}$  in (\ref{271}). Thus,
we have $|\frac{q}{p}|\simeq 1$ and  from (\ref{257}) and
(\ref{261}) for the asymmetry we find the expression
\begin{equation}\label{276}
A_{f}^{CP}(t)=-C_{f}\cos\Delta m t  +S_{f}\sin\Delta m t.
\end{equation}
In conclusion  we will consider the following decays
\begin{equation}\label{277}
B^{0}_{d}(\bar B^{0}_{d})\to J/\Psi +K^{0}_{S,L}.
\end{equation}
These decay modes are called golden by the reasons which will be
clear later.

Final $J/\Psi$  and $K^{0}_{S,L}$ particles are in  the state with
$l=1$. Neglecting in the matrix elements of the decay small terms of
the order of $\sim 10^{-3}$ we can put $ |K_{S}\rangle\simeq
|K_{1}\rangle$ and $|K_{L}\rangle\simeq |K_{2}\rangle$. Thus, we
find
\begin{equation}\label{278}
    CP~|J/\Psi~ K^{0}_{S,L}\rangle=\eta_{S,L}~|J/\Psi~
    K^{0}_{S,L}\rangle,
\end{equation}
where  $\eta_{S,L}=\mp 1$.

Matrix elements of the processes $\bar B^{0}_{d}(B^{0}_{d})\to
J/\Psi +K^{0}_{S.L}$ are determined by decays of the $b$-quark,
which are governed by the tree  and penguin  electroweak diagrams.
If we take into account $QCD$ corrections, the matrix elements of
the process   $\bar B^{0}_{d}\to J/\Psi +K^{0}_{S,L}$ is given by
the relation (see reviews \cite{BurasII, Fleisher})
\begin{eqnarray}\label{279}
\langle J/\Psi~ K^{0}_{S,L}|T|\bar B^{0}_{d}\rangle
&=&\frac{G_{F}}{\sqrt{2}}\sum_{q=u,c}V_{qb}V^{*}_{qs}  (\sum_{k=1,2}
C_{k}(\mu)~\langle J/\Psi~
K^{0}_{S,L}|O^{qs}_{k}|\bar B^{0}_{d}\rangle \nonumber\\
& +&\sum_{k=3}^{10} C_{k}~(\mu)\langle J/\Psi~
K^{0}_{S,L}|O^{s}_{k}|\bar B^{0}_{d}\rangle  ).
\end{eqnarray}
Here $C_{k}(\mu)$ are real Wilson coefficients, $O^{qs}_{k}$ are
4-quark  current-current operators and $O^{s}_{k}$ are 4-quark
penguin operators (for the definitions see, for example,\cite{
Fleisher}).

For the matrix element of the process $ B^{0}_{d}\to J/\Psi
+K^{0}_{S,L}$ we have
\begin{eqnarray}\label{280}
\langle J/\Psi~ K^{0}_{S,L}|T|B^{0}_{d}\rangle
&=&\frac{G_{F}}{\sqrt{2}}\sum_{q=u,c}V^{*}_{qb}V_{qs} (\sum_{k=1,2}
C_{k}(\mu)~\langle J/\Psi~
K^{0}_{S,L}|(O^{qs}_{k})^{\dag}|B^{0}_{d}\rangle \nonumber\\
& +&\sum_{k=3}^{10} C_{k}~(\mu)\langle J/\Psi~
K^{0}_{S,L}|(O^{s}_{k})^{\dag}|B^{0}_{d}\rangle  ).
\end{eqnarray}
Further, we have
\begin{equation}\label{281}
(O^{qs}_{k})^{\dag}=(CP)^{-1}~O^{qs}_{k}~CP,~~~(O^{s}_{k})^{\dag}=(CP)^{-1}~
O^{s}_{k}~CP.
\end{equation}
From (\ref{278}), (\ref{280}) and (\ref{281}) we find
\begin{eqnarray}\label{282}
\langle J/\Psi~ K^{0}_{S,L}|T|B^{0}_{d}\rangle
&=&\eta_{S,L}\frac{G_{F}}{\sqrt{2}}\sum_{q=u,c}V^{*}_{qb}V_{qs}
(\sum_{k=1,2} C_{k}(\mu)~\langle J/\Psi~
K^{0}_{S,L}|O^{qs}_{k}|\bar B^{0}_{d}\rangle \nonumber\\
& +&\sum_{k=3}^{10} C_{k}~(\mu)\langle J/\Psi~
K^{0}_{S,L}|O^{s}_{k}|\bar B^{0}_{d}\rangle  ).
\end{eqnarray}
Let us compare now the matrix elements $\langle J/\Psi~
K^{0}_{S,L}|T|\bar B^{0}_{d}\rangle$ and $\langle J/\Psi~
K^{0}_{S,L}|T|B^{0}_{d}\rangle$. The ratio of these matrix elements
which enter into the expression for the parameter $\lambda_{J/\Psi
K^{0}_{S,L}}$ (see (\ref{252})) depends  on the $CKM$ matrix
elements and  on the hadronic matrix elements. However,
$|V_{ub}V^{*}_{us}| \simeq 10^{-2}~|V_{cb}V^{*}_{cs}|$. If we
neglect in (\ref{279}) and (\ref{282}) the contribution of the terms
proportional to $|V_{ub}V^{*}_{us}| $ and take into account that the
product $V^{*}_{cb}V_{cs}$ is real, we come to the following result
\begin{equation}\label{283}
\frac{\langle J/\Psi~ K^{0}_{S,L}|T|\bar B^{0}_{d}\rangle}{\langle
J/\Psi~ K^{0}_{S,L}|T| B^{0}_{d}\rangle}\simeq \eta_{S,L}.
\end{equation}

For the parameter $\lambda_{J/\Psi K^{0}_{S,L}}$ we find
\begin{equation}\label{287}
    \lambda_{J/\Psi K^{0}_{S,L}}\simeq \eta_{S,L}~\frac{q}{p}
\end{equation}
Thus, in the case of the decays (\ref{277}) the parameter
$\lambda_{J/\Psi K^{0}_{S,L}}$ (practically) does not depend on
hadronic uncertainties of the decay matrix elements.

 The mixing parameter $\frac{q}{p}$ is
given by the relation $\frac{q}{p}\simeq
-\frac{M^{*}_{12}}{|M_{12}}|$ (see (\ref{191})).
 Main contribution to the box diagrams which determine
matrix element $ M_{12}$  gives the virtual $t$-quark. We have
\begin{equation}\label{288}
\frac{q}{p}\simeq -e^{\arg({V^{*}_{tb}V_{td})^{2}}}
\end{equation}
From (\ref{132}), (\ref{287}) and (\ref{288}) we find
\begin{equation}\label{289}
\lambda_{J/\Psi K^{0}_{S,L}}\simeq -\eta_{S,L} e^{2i\beta}.
\end{equation}
Thus,   we have
\begin{equation}\label{290}
    C_{J/\Psi K^{0}_{S,L}}\simeq 0,~~S_{J/\Psi K^{0}_{S,L}}\simeq -\eta_{S,L} \sin 2
    \beta.
\end{equation}
From (\ref{276}) and (\ref{290}) for the asymmetry  $A_{J/\Psi
K^{0}_{S}}^{CP}(t)$ we find the following expression
\begin{equation}\label{291}
    A_{J/\Psi K^{0}_{S}}^{CP}(t)=\sin 2 \beta~\sin\Delta m t.
\end{equation}
Asymmetry $A_{J/\Psi K^{0}_{L}}^{CP}(t)$ differs by  sign from
$A_{J/\Psi K^{0}_{S}}^{CP}(t)$. We have
\begin{equation}\label{292}
    A_{J/\Psi K^{0}_{L}}^{CP}(t)=-\sin 2 \beta~\sin\Delta m t.
\end{equation}
We came to an important conclusion: the measurement of $t$
dependence of the $CP$ asymmetries in the decays  $\bar B_{d}^{0} (
B_{d}^{0}) \to J/\Psi K^{0}_{S,L}$ allow to determine the angle
$\beta$ \emph{in a model independent way} \cite{Sandra}.

The asymmetries  $A^{J/\Psi K^{0}_{L,S}}_{CP}(t)$ were measured by
the BaBar collaboration in experiments at the asymmetric B-factory
at SLAC and by the Belle collaboration in  experiments at the
asymmetric B-factory at KEK. In these experiments the first evidence
of the $CP$ violation in $B_{d}^{0}(\bar B_{d}^{0})$ decays was
found and the value of the parameter $\sin 2 \beta$ was determined
\cite{BaBarI,BelleI}. Recently the results of the  measurement of
the parameter $\sin 2 \beta$ in the experiments which were performed
during
 1999-2006 were published \cite{BaBarII,BelleII}.

At the asymmetric B-factories $B_{d}^{0}$-mesons are produced  in
the decay $\Upsilon(4S)\to B_{d}^{0}+\bar B_{d}^{0}$. Flavor of a
particle is determined by the tagging another particle. The proper
time $t$ in (\ref{276}) and other equations is given by difference
between the proper time of reconstructed and  tagged $B^{0}$ mesons:
$t=t_{\rm{rec}}-t_{\rm{tag}}$. Because the $B_{d}^{0}$ and $\bar
B_{d}^{0}$ mesons are practically at rest in $\Upsilon(4S)$ rest
frame we have $t_{\rm{rec}}-t_{\rm{tag}}=\frac{
z_{\rm{rec}}-z_{\rm{tag}}}{\beta\gamma c}$, where  $z_{\rm{rec}}$
and $z_{\rm{tag}}$ are positions of corresponding decay vertices and
$\beta\gamma $ is the Lorenz boost  of $\Upsilon(4S)$.

In the BaBar experiment $(347.5\pm 3.8)\cdot10^{6} \Upsilon(4S)\to
B_{d}^{0}+\bar B_{d}^{0}$ decays were detected. For analysis were
used decays determined by the transition $b\to c \bar c s$. The
decays into the following egenstates of the $CP$ operator  were
analyzed: $J/\Psi K^{0}_{S}$, $J/\Psi K^{0}_{L}$, $\Psi(2S)
K^{0}_{S}$, $\chi_{c1} K^{0}_{S}$, $\eta_{c} K^{0}_{S}$ and $J/\Psi
K^{*0}_{S}$.

From analysis of the data the following result was obtained
\cite{BaBarII}\footnote{The first error is statistical and second is
systematical.}
\begin{equation}\label{293}
   \sin 2 \beta= 0.710 \pm 0.034 \pm 0.019,~~~C=0.070\pm 0.026\pm
    0.018.
\end{equation}
In the Belle experiment $535 \cdot10^{6} ~\Upsilon(4S)\to
B_{d}^{0}+\bar B_{d}^{0}$ decays were detected. From analysis of the
decays into $J/\Psi K^{0}_{S}$ and  $J/\Psi K^{0}_{L}$ states it was
found \cite{BelleII}.
\begin{equation}\label{294}
   \sin 2 \beta= 0.642 \pm 0.031 \pm 0.017,~~~C=-0.018\pm 0.021\pm
    0.014.
\end{equation}
Thus, the parameter $\sin 2 \beta$ is known today with accuracy
about 5\%. This model independent result is very important for the
unitarity triangle fit of the experimental data which we will
discuss in the next section.

\section{Unitarity triangle test of the Standard Model}

Several groups \cite{UTfitI,UTfitII,CKMfitter,HFAG}  analyze
experimental data with the aim to perform  the unitarity triangle
test of the Standard Model and to search for effects of beyond the
SM physics. Different groups use different statistical methods of
the analysis of experimental data. We will present here some results
of the UTfit collaboration \cite{UTfitI,UTfitII} which use the
Bayesian method. Other groups obtain similar results.

In the standard unitarity triangle fit  the results of the
measurement of the following  quantities are used
\begin{equation}\label{295}
|\frac {V_{ub}} {V_{cb}}|,~~\Delta m_{d},~~\frac{\Delta
m_{d}}{\Delta m_{s}},~~\epsilon ~~\rm{and}~~\sin 2 \beta.
\end{equation}
From (\ref{119}) we have
\begin{equation}\label{296}
|\frac {V_{ub}} {V_{cb}}|=\lambda~\sqrt{\rho^{2}+\eta^{2}}=
\frac{\lambda}{1-\frac{1}{2}\lambda^{2}}~\sqrt{\bar\rho^{2}+\bar\eta^{2}},
\end{equation}
where $\bar\rho$ and $\bar\eta$ are determined by Eq. (\ref{125}).
Mass differences $\Delta m_{d}$ and $\Delta m_{s}$ are given by
(\ref{105}). For the ratio $\frac{\Delta m_{d}}{\Delta m_{s}}$ we
have
\begin{equation}\label{297}
\frac{\Delta m_{d}}{\Delta
m_{s}}=\lambda^{2}[(1-\bar\rho)^{2}+\bar\eta^{2}]~\frac{m_{B_{d}}}{m_{B_{s}}}~\frac{f^{2}_{B_{d}}\hat{B}_{B_{d}}}
{f^{2}_{B_{s}}\hat{B}_{B_{s}}}
\end{equation}
The expression for the parameter $\epsilon$ is given by (\ref{239}).

Let us obtain $\sin 2 \beta$ as a function of $\bar\rho$ and
$\bar\eta$. From Fig.1 we find that
\begin{equation}
\sin\beta= \frac{\bar \eta}{\sqrt{(1-\bar \rho)^{2}+ \bar
\eta^{2}}},~~~ \cos\beta= \frac{  (  1-\bar \rho) }{\sqrt{(1-\bar
\rho)^{2}+ \bar \eta^{2}}}. \label{298}
\end{equation}
From these relations we have
\begin{equation}
\sin2\beta= \frac{2\bar \eta~ (  1-\bar \rho)}{(1-\bar \rho)^{2}+
\bar \eta^{2}}. \label{299}
\end{equation}
From the fit of the experimental data the unique region in the plane
of the parameters $\bar \rho,~\bar \eta$ was found (see Fig 2).

\begin{figure}[htbp]
  \begin{center}
    \leavevmode
\epsfxsize=5in
    \epsfbox{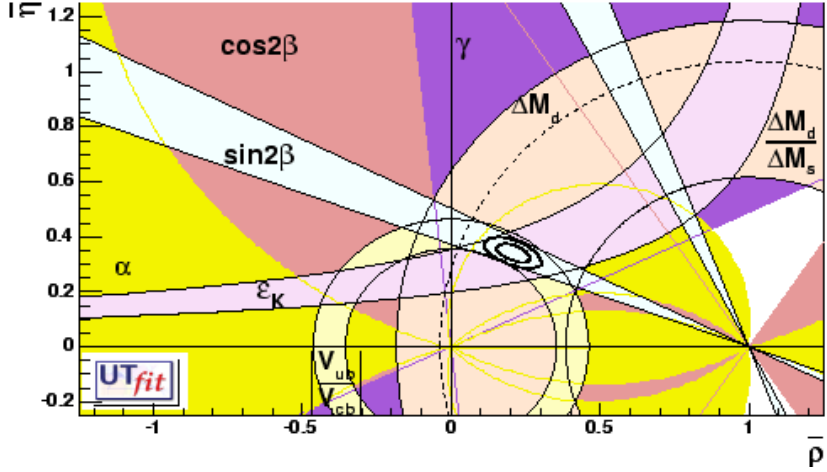}
  \end{center}
  \caption{68\% and 95 \% total probability regions of the
allowed values of the parameters $\bar\rho$ and $\bar\eta$. The
values of the quantities $|V_{ub}|/|V_{cb}|$, $\Delta m_{d}$,
$\Delta m_{s}$, $\epsilon$ and $\sin2\beta$ were used in the fit
\cite{UTfitI}.} \label{fig:2}
\end{figure}

For the parameters $\bar \rho$ and $\bar \eta$ the following values
were obtained
\begin{equation}\label{300}
\bar \rho=0.196 \pm 0.045,~~\bar \eta=0.347 \pm 0.025.
\end{equation}
These values determine the vertex B of the triangle in  Fig 1.

The values of the parameters (\ref{295}) overconstrain the unitarity
triangle. For example, position of the vertex B can be obtained if
only  the parameters $|\frac {V_{ub}} {V_{cb}}|$, $\Delta m_{d}$ and
$\Delta m_{s}$, which determine lengths of the sides of the
triangle, are used in the fit. From the result of such fit the value
of the parameter $\sin 2 \beta$ can be predicted. In \cite{UTfitI}
it was found
\begin{equation}\label{301}
\sin 2 \beta=0.734 \pm 0.043
\end{equation}
We can compare  (\ref{301}) with the measured values of the
parameter $\sin 2 \beta$, given by (\ref{293}) and (\ref{294}). This
comparison illustrates the evidence in favor of the correctness of
the Standard Model.

Recently at the Tevatron in the Fermilab the mass difference $\Delta
m_{s}$ was measured by D0 \cite{AbazovD0} and $CDF$ \cite{CDFdeltas}
collaborations. In the $CDF$ experiment it was found
\begin{equation}\label{302}
\Delta m_{s}=(17.77 \pm 0.10 \pm 0.07)~ps^{-1}.
\end{equation}
The Belle collaboration measured the branching ratio for the
leptonic decay $B\to \tau+\nu_{\tau}$:
\begin{equation}\label{303}
    BR(B_{d}\to \tau+\nu_{\tau})= (1.06^{0.34}_{-0.28}\pm 0.18)\cdot
    10^{-4}.
\end{equation}
From this measurement the value of the constant $f_{B_{d}}$ can be
determined.

The BaBar and Belle Collaborations by the investigation  of the
decays $B_{d}\to \pi\pi$, $B_{d}\to \rho\rho$, $B_{d}\to \pi\pi\pi$
and $B_{d}\to D^{*} K^{*}$ obtained the information about the values
of the angles $\alpha$ and $\gamma$ (see \cite{Ramaangles}).

In the new analysis  of the UTfit collaboration \cite{UTfitII}  all
these data were used. If in the analysis only the values of the
angles $\alpha$, $\beta$ and $\gamma$ are used, for parameters $\bar
\rho$ and $\bar \eta$ the following values were found
\begin{equation}\label{304}
\bar \eta=0.204\pm 0.055,~~~\bar \eta=0.317\pm 0.025.
\end{equation}
If the quantities $|\frac {V_{ub}} {V_{cb}}|,~\Delta m_{d},~\Delta
m_{s},~\epsilon$ and results of the lattice calculations are used,
in this case it was obtained
\begin{equation}\label{305}
\bar \eta=0.197\pm 0.035,~~~\bar \eta=0.380\pm 0.025.
\end{equation}
From the fit of all data it was found
\begin{equation}\label{306}
\bar \eta=0.197\pm 0.031,~~~\bar \eta=0.351\pm 0.020.
\end{equation}
The fit of all data is presented in Fig.3.

\begin{figure}[htbp]
  \begin{center}
    \leavevmode
\epsfxsize=6in
    \epsfbox{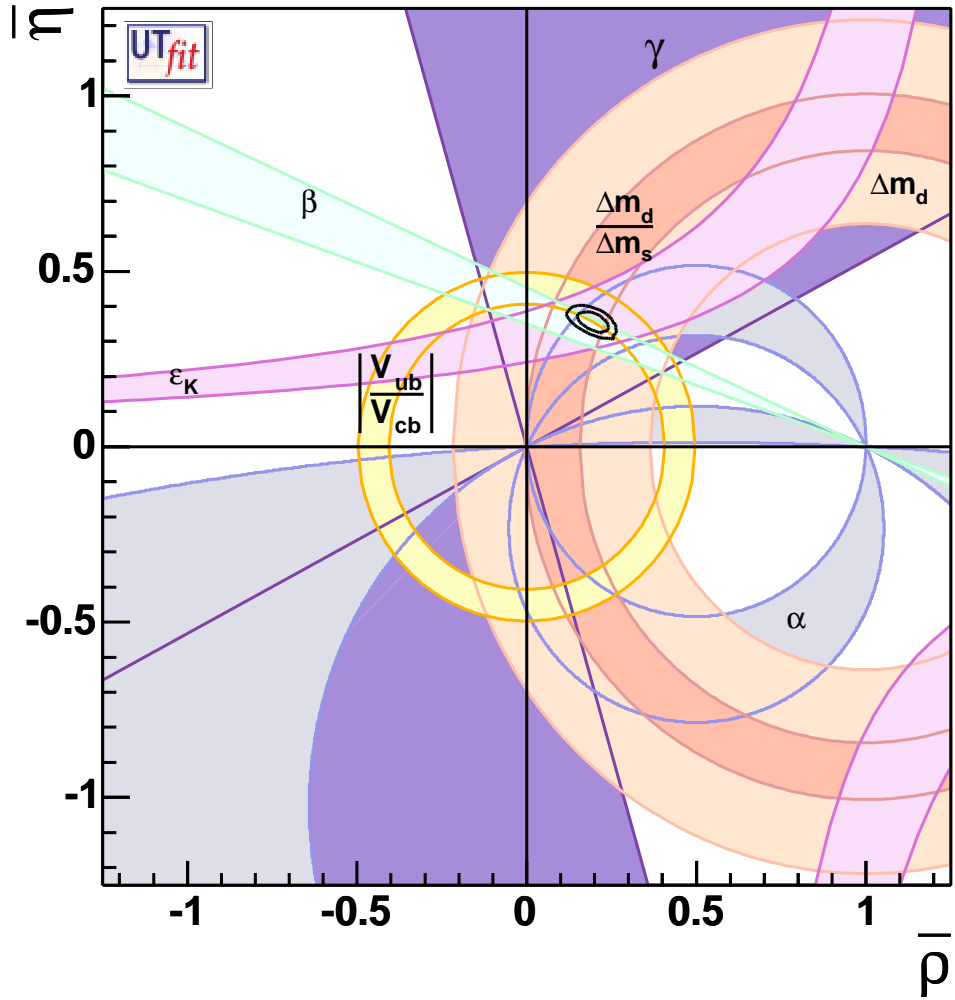}
  \end{center}
  \caption{Allowed values of the parameters $\bar\rho$ and $\bar\eta$
(68\% and 95 \% total probability regions are shown \cite{UTfitII}).
The values of the quantities $|V_{ub}|/|V_{cb}|$, $\Delta m_{d}$,
$\Delta m_{s}$, $\epsilon$, $\beta$, $\gamma$ and $\alpha$ were used
in the fit .} \label{fig:3}
\end{figure}

From this analysis (and analysis performed by other groups) we can
conclude that existing data are in agreement with the Standard
 Model. However, it is necessary to stress that accuracy of the
 experimental data are  limited and  complicated QCD
 calculations are used in the analysis. There is still a room for
 beyond the SM physics (see,for example,\cite{Fleisher}). In order to  reveal it
 more precise data and progress in theoretical calculations are
 mandatory.

\section{Conclusion}
The Glashow-Weinberg-Salam unified theory of weak and
electromagnetic interactions  is outstanding achievement of the XX
century physics. It was created as a result of a long period of the
development of the phenomenological theory during which violation of
$P$, $C$ \cite{Wu57} and later $CP$ \cite{Cronin} were discovered
and universal $V-A$ current $\times$ current theory of the weak
interaction were proposed \cite{FeyGel,MarSud}.

The SM predicted a new class of the weak interactions (neutral
currents), vector $W^{\pm}$ and $Z^{0}$ bosons and their masses.
After $\tau$-lepton was discovered the SM predicted that other
members of the third family of  quarks and leptons (b and t quarks
and the third type of neutrino $\nu_{\tau}$) must exist. All
 predictions of the SM were perfectly confirmed by experiment.

Only one prediction of the SM, existence of the scalar neutral Higgs
boson, is still waiting for its confirmation. The search for the
Higgs boson will be one of the primary goal of the future LHC
collider at CERN.

With  the experiments at the LEP and  SLD in the nineties next step
in the testing of the SM started. The precision of these data
required calculations of radiative corrections. At present numerous
data of LEP, SLD, BaBar, Belle, CDF, D0 and other experiments are in
a good agreement with the prediction of the SM. The fit of all
electroweak data allows to predict the upper bound of the mass of
the Higgs boson (see, for example,\cite{PDG}): $m_{H}\leq 235$ GeV
at 99 \% CL.

 In 1999 with the beginning of BaBar and Belle experiments at
asymmetrical B-factories at SLAC and KEK a new stage in the testing
of the SM started. In the framework of the SM violation of the $CP$
invariance is due to one physical phase in the unitary 3$\times$3
$CKM$ mixing matrix. This phase enter into the unitarity triangle
relation which is a consequence of  the orthogonality of different
columns (or lines) of the mixing matrix. The numerous tests of this
relation became possible with new B-factory data.  Existing data are
in a good agreement with the Standard Model. This agreement confirms
\begin{itemize}
    \item The basic assumption of the SM that two sets of
    quarks fields, fields $q'_{L}$ which possess definite
    transformation property  and fields of quarks with definite masses
    $q_{L}$, are connected by unitary transformation.

\item The assumption that only three quarks families exist in nature.

\end{itemize}
It is necessary, however, to stress that the unitarity triangle test
is based not only on experimental data but also on nonperturbative
$QCD$ calculations of relevant matrix elements. A room for a new
physics is still open. Improvement of the accuracy of data and
improvement of the accuracy of the lattice and other calculations is
a key problem for the future progress.

It is a pleasure for me to acknowledge the  ILIAS program for the
support and theory department of the TRIUMF for the hospitality.

 \renewcommand{\theequation}{A-\arabic{equation}}
 \setcounter{equation}{0}
\appendix

\begin{center}
{\bf Appendix A. Comparison of $M^{0}\leftrightarrows \bar M^{0}$
oscillations with neutrino oscillations }
\end {center}

It is of interest to compare $M^{0}\leftrightarrows \bar M^{0}$
oscillations with neutrino oscillations recently discovered in the
Super Kamiokande \cite{SK}, SNO  \cite{SNO} , KamLAND \cite{Kamland}
and other neutrino oscillation experiments
\cite{Cl,Gallex,Sage,K2K,Minos}.

Particles with definite flavor $M^{0}$ and
 $\bar M^{0}$ ($M^{0}=K^{0},B_{d,s}^{0},...$)
are produced in strong interaction processes. States of these
particles  are eigenstates of the Hamiltonian of the strong and
electromagnetic interactions which conserve the quark flavor. In
processes of production of $M^{0}$ and $\bar M^{0}$ effects of  the
weak interaction, in which quark flavor is violated,  are negligibly
small and can be neglected. After $M^{0}$ and $\bar M^{0}$ are
produced weak interaction  play the major role. Because of the weak
interaction $M^{0}$ and
 $\bar M^{0}$ decay and eigenstates of the total Hamiltonian $|M_{H}^{0}\rangle$ and
$|\bar M_{L}^{0}\rangle$ have
 different masses and widths. Thus, $M^{0}\leftrightarrows \bar M^{0}$
 oscillations are due to the existence of  strong interaction in
 which quark flavor is conserved and  weak interaction  in which quark
 flavor is changed.

Neutrinos have only weak interaction. However, neutrino masses are
very small. In neutrino production and neutrino detection processes
neutrino masses can be safely neglected. This means that in such
processes lepton flavor numbers $L_{e}$, $L_{\mu}$ and $L_{\tau}$
are conserved: together with lepton $l^{+}$ flavor neutrino
$\nu_{l}$ is produced , flavor neutrino $\nu_{l}$ in a charge
current  process  produce $l^{-}$ etc . After flavor neutrino is
produced small neutrino masses (and neutrino mixing) play the key
role. Because of neutrino masses, in  the neutrino propagation
different mass components of the mixed flavor neutrino state acquire
different phases. This is a physical reason for the neutrino
oscillations $\nu_{l}\leftrightarrows \nu_{l'}$ (see, for example,
\cite{BilGiunti,BilMateev}).

In the framework of the Standard Model $CP$ violation is connected
with the physical phase in the mixing matrix. As we have seen, in
the case of two families there are no physical phases in the mixing
matrix. The $CP$ invariance can be violated if (at least) three
families of the quark and leptons exist.

In the case of neutral bosons oscillations take place  only between
two particles $M^{0}$ and
 $\bar M^{0}$. In order to reveal $CP$ violation in  decays of
these particles one must  observe
 effects of all three families of quarks.
Because observable particles are hadrons and not quarks Jarlskog
invariant $J$ does not enter in the quantities which characterize
the $CP$ violation.

Neutrinos are stable particles. In order to observe effects of the
$CP$ violation in neutrino oscillations all three  neutrinos must be
involved in the transition probability. Only elements of the
neutrino mixing matrix and the neutrino mass-squared differences
enter into the transition probabilities. It is natural to expect
that effects of the $CP$ violation in the neutrino oscillations are
determined by the Jarlskog invariant. We will demonstrate this
below.

The   Lagrangian of neutrino interaction has the form
\begin{equation}\label{A1}
{\mathcal L^{CC}}(x) =-\frac{g}{2\sqrt{2}}~j^{CC}_{\alpha}(x)
W^{\alpha}(x) +\rm{h.c.},~~ {\mathcal L^{NC}}(x)
=-\frac{g}{2\cos\theta_{W}}~j^{NC}_{\alpha}(x) Z^{\alpha}(x) ,
\end{equation}
where the charged current $j^{CC}_{\alpha}(x) $ and the neutral
current $j^{NC}_{\alpha}(x) $ are given by the expressions
\begin{equation}\label{A2}
j^{\mathrm{CC}}_{\alpha}(x) =2\, \sum_{l=e,\mu,\tau} \bar
\nu_{lL}(x) \gamma_{\alpha}l_{L}(x),~~~ j ^{\mathrm{NC}}_{\alpha}(x)
=\sum_{l=e,\mu,\tau} \bar \nu_{lL}(x) \gamma_{\alpha}\nu_{lL}(x).
\end{equation}
Here
\begin{equation}\label{A3}
\nu_{l L}(x)=\,\sum^{3}_{k=1}U_{l i}\,\nu_{i L}(x),
\end{equation}
is  the "mixed  field". In (\ref{A3}) $\nu_{i }(x)$ is the field of
neutrino with mass $m_{i}$ and $U$ is the unitary $3\times 3$   PMNS
mixing matrix \cite{BP,MNS}. For neutrinos, particles with electric
charges
 equal to zero, there
are two possibilities (see \cite{BPet, BGG})
\begin{enumerate}
\item
If the total lepton number $L=L_{e}+L_{\mu}+L_{\tau}$ is conserved,
neutrinos $\nu_{i}$ are {\em Dirac particles}.
\item
If there are no conserved lepton numbers, neutrinos $\nu_{i}$ are
{\em Majorana particles}.
\end{enumerate}
The probabilities of the transition $\nu_{l} \to\nu_{l'}$  and
$\nu_{l} \to\nu_{l'}$ ($l,l'=e, \mu, \tau$) during the time $t$ in
the three-neutrino case, we are considering, is given by the
following expressions (see, for example, \cite{BGG})
\begin{equation}\label{A4}
P(\nu_{l} \to\nu_{l'}) =|\sum^{3}_{i=1} U_{l' i} \,e^{-i\,E_{i}\, t
}\,U^{*}_{l i}~|^{2}
\end{equation}
and
\begin{equation}\label{A5}
P(\bar\nu_{l} \to \bar\nu_{l'}) =|\sum^{3}_{i=1} U^{*}_{l' i}
\,e^{-i\,E_{i}\, t }\,U_{l i}~|^{2}.
\end{equation}
Here $E_{i} \simeq p + \frac{m^{2}_{i} }{2\,p}$ is the energy of
neutrino with mass $m_{i}$ and momentum $p$.

If $CP$ is conserved in the case of the Dirac neutrinos arbitrary
phases of the fields of the leptons  and neutrinos can be chosen in
such a way that
\begin{equation}\label{A6}
    U_{l i}^{*}=U_{l i}.
\end{equation}
If $\nu_{i}$ are Majorana particles from the condition of the $CP$
invariance we have (see \cite{BilNedPet})
\begin{equation}\label{A7}
U^{ *}_{l i}= U_{l i} ~ \eta_{i},
\end{equation}
where $\eta_{i}=\pm i $ is the $CP$-parity of the Majorana neutrino
with the mass $m_{i}$.

From (\ref{A4}), (\ref{A5}), (\ref{A6}) and (\ref{A7}) it follows
that in the case of the  $CP$ invariance for the  Dirac as well as
for the Majorana neutrinos we have
\begin{equation}\label{A8}
P(\nu_{l} \to\nu_{l'})=P(\bar\nu_{l} \to \bar\nu_{l'}).
\end{equation}
If we compare expressions (\ref{A4}) and (\ref{A5}) we come to the
conclusion that transition probabilities satisfy the following
relation
\begin{equation}\label{A9}
 P(\nu_{l} \to\nu_{l'})=P(\bar\nu_{l'} \to \bar\nu_{l}).
\end{equation}
This relation is the consequence of the $CPT$ invariance inherent to
any local quantum field theory. It follows from  (\ref{A9}) that the
equality
\begin{equation}\label{A10}
P(\nu_{l} \to\nu_{l})=P(\bar\nu_{l} \to \bar\nu_{l})
\end{equation}
is a consequence of the $CPT$ invariance. Thus, if the inequality
\begin{equation}\label{A11}
P(\nu_{l} \to\nu_{l'})\neq P(\bar\nu_{l'} \to \bar\nu_{l}),~~~
l'\neq l
\end{equation}
takes place it would be a proof of the $CP$ violation in the lepton
sector.

Let us consider now the expression (\ref{A4}) for the transition
probability $P(\nu_{l} \to\nu_{l'})$. We have
\begin{equation}\label{A12}
P(\nu_{l} \to\nu_{l'}) =\sum_{i,k} U_{l' i} \,U^{*}_{l' k}
\,U^{*}_{l i}\, \,U_{l k} e^{-i\,(E_{i}-E_{k})\, t }.
\end{equation}
Here
\begin{equation}\label{A13}
(E_{i}-  E_{k})\, t \simeq \frac{ \Delta m^2_{ k i }}{ 2 E }~ L,
\end{equation}
where $L$ is the distance between the neutrino production and the
neutrino detection points and $\Delta m^2_{ k i }=m^2_{ i }-m^2_{ k
}$. From
 (\ref{A12}) for the transition probability we find the following expression
\begin{equation}\label{A14}
 P(\nu_{l} \to\nu_{l'}) =\sum_{i} |U_{l' i}
|^{2}\,|U_{l i} |^{2}+ 2\,\rm{Re}\sum_{i>k} U_{l' i} \,U^{*}_{l' k}
\,U^{*}_{l i}\, \,U_{l k} e^{-i\,\frac{ \Delta m^2_{ k i }}{ 2 E }\,
L }.
\end{equation}
Further, from the condition of the unitarity of the mixing matrix
$U$
\begin{equation}\label{A15}
\sum_{i}U_{l' i} ~U^{*}_{l i}=\delta_{l' l }.
\end{equation}
we find
\begin{equation}\label{A16}
\sum_{i} |U_{l' i} |^{2}\,|U_{l i} |^{2}= \delta_{l' l } -
2\,\rm{Re}\sum_{i>k} U_{l' i} \,U^{*}_{l' k} \,U^{*}_{l i}\, \,U_{l
k}.
\end{equation}
From (\ref{A14}) and (\ref{A16}) for the transition probability we
obtain the following expression
\begin{equation}\label{A17}
P(\nu_{l} \to\nu_{l'}) = \delta_{l' l } - 2\,\rm{Re}\sum_{i>k} U_{l'
i} \,U^{*}_{l' k} \,U^{*}_{l i}\, \,U_{l k}\,(1- e^{-i\,\frac{
\Delta m^2_{ k i }}{ 2 E }\, L }).
\end{equation}
From (\ref{A17}) we have
\begin{eqnarray}
 P(\nu_{l} \to\nu_{l'}) &=&\delta_{l' l }
-2\,\sum_{i>k}\,\rm{Re}\, (U_{l' i} \,U^{*}_{l' k} \,U^{*}_{l i}\,
U_{l k})\,(1-\cos\frac{ \Delta m^2_{ k i }}{ 2 E }\, L ) \nonumber\\
&+&2 \sum_{i>k}\rm{Im}\, (U_{l' i} \,U^{*}_{l' k} \,U^{*}_{l i}\,
U_{l k})\,\sin\frac{ \Delta m^2_{ k i }}{ 2 E }\, L \label{A18}
\end{eqnarray}
Analogously,  for the probability of the transition $\bar\nu_{l}
\to\bar \nu_{l'}$ we find
\begin{eqnarray}
 P(\bar \nu_{l} \to \bar\nu_{l'}) &=&\delta_{l' l }
-2\,\sum_{i>k}\,\rm{Re}\, (U_{l' i} \,U^{*}_{l' k} \,U^{*}_{l i}\,
U_{l k})\,(1-\cos\frac{ \Delta m^2_{ k i }}{ 2 E }\, L ) \nonumber\\
&-&2 \sum_{i>k}\rm{Im}\, (U_{l' i} \,U^{*}_{l' k} \,U^{*}_{l i}\,
U_{l k})\,\sin\frac{ \Delta m^2_{ k i }}{ 2 E }\, L \label{A19}
\end{eqnarray}
Let us introduce the quantity
\begin{equation}\label{A20}
J^{ik}_{l'l}=\rm{Im}\, U_{l' i}~U_{l k} \,U^{*}_{l' k} \,U^{*}_{l
i}.
\end{equation}
In the case of the $CP$ invariance from (\ref{A6}) for the Dirac
neutrinos and from (\ref{A7}) for the  Majorana neutrinos we find
\begin{equation}\label{A21}
J^{ik}_{l'l}=0.
\end{equation}
From the definition (\ref{A20}) we  have
\begin{equation}\label{A22}
J^{ik}_{l'l}=-J^{ki}_{l'l},~~~J^{ik}_{l'l}=-J^{ik}_{ll'}.
\end{equation}
Further, from the unitarity of the $3\times3$ mixing matrix $U$ we
find
\begin{equation}\label{A23}
\sum_{i}J^{ik}_{l'l}=\delta_{l'l}~\rm{Im}~U^{*}_{l' k}\,U_{l k}=0,~~
\sum_{l'} J^{ik}_{l'l}=\delta_{ik}~\rm{Im}~ U^{*}_{l i}\,U_{l k}=0.
\end{equation}
From (\ref{A22}) and the first equation (\ref{A23}) we have
\begin{equation}\label{A24}
J^{21}_{l'l} =J^{13}_{l'l} = J^{32}_{l'l}.
\end{equation}
Further, from (\ref{A22}) and the second equation (\ref{A23}) we
find
\begin{equation}\label{A25}
J^{ik}_{e \mu} = J^{ik}_{\mu \tau} = J^{ik}_{\tau e}.
\end{equation}
If we introduce the following notation
\begin{equation}\label{A26}
J^{21}_{e \mu} = J
\end{equation}
from (\ref{A24}) and (\ref{A25}) we have
\begin{equation}\label{A27}
J^{ik}_{l'l}=\pm J,~~~l' \neq l, ~~i\neq k.
\end{equation}
Thus, in the neutrino case, as in the quark case (see section 3),
exist only one independent  Jarlskog invariant.

Let us consider now the last term of the expression (\ref{A18}) for
$l'\neq l$. Taking into account (\ref{A22}), we find
\begin{equation}\label{A28}
2 \sum_{i>k}J_{l'l}^{ik}~\sin\frac{ \Delta m^2_{ k i }}{ 2 E }\,
L=2J_{l'l}^{21}~(\sin\frac{ \Delta m^2_{ 12 }}{ 2 E }\, L
+\sin\frac{ \Delta m^2_{ 23 }}{ 2 E }\, L -\sin\frac{ \Delta m^2_{
13}}{ 2 E }\, L ).
\end{equation}
It is obvious that
\begin{equation}\label{A29}
    \Delta m^2_{ 13}= \Delta m^2_{ 12} +\Delta m^2_{ 23}.
\end{equation}
Further, for any $a$ and $b$ we have
\begin{equation}\label{A30}
\sin a +\sin b - \sin (a+b)= 4\, \sin \frac{ a}{2}\, \sin
\frac{b}{2}\, \sin \frac{( a+b)}{2}
\end{equation}
From (\ref{A28}), (\ref{A29}) and (\ref{A30}) we find
\begin{equation}\label{A31}
2 \sum_{i>k}J_{l'l}^{ik}~\sin\frac{ \Delta m^2_{ k i }}{ 2 E }\,
L=8J_{l'l}^{21}~\sin\frac{ \Delta m^2_{ 12 }}{ 4E }\, L \,\sin\frac{
\Delta m^2_{ 23 }}{ 4 E }\, L \, \sin\frac{ (\Delta m^2_{ 12}  +
\Delta m^2_{ 23} )      }{ 4 E }\, L .
\end{equation}
Let us determine $CP$ asymmetry
\begin{equation}\label{A32}
A^{CP}_{l'l}=  P(\nu_{l} \to\nu_{l'})-P(\bar\nu_{l}
\to\bar\nu_{l'})~~~l\neq l'
\end{equation}
From the unitarity of the mixing matrix and the $CPT$ invariance it
easy to obtain the following relations
\begin{equation}\label{A33}
A^{CP}_{e \mu }=  A^{CP}_{\tau  e} = -
 A^{CP}_{\tau  \mu}.
\end{equation}
In fact, from the unitarity of the mixing matrix we find
\begin{equation}\label{A34}
    \sum_{l'=e,\mu,\tau}A^{CP}_{l'l}=0
\end{equation}
Further, from  the relation (\ref{A9}) we  have
\begin{equation}\label{A35}
A^{CP}_{l'l}= - A^{CP}_{l~l'}.
\end{equation}
From (\ref{A34}) and (\ref{A35}) we obtain the following relations
\begin{equation}\label{A36}
A^{CP}_{\mu e}+ A^{CP}_{\tau e}=0,~ A^{CP}_{e \mu }+  A^{CP}_{\tau
\mu}=0,~ A^{CP}_{e \tau}+  A^{CP}_{\mu \tau }=0.
\end{equation}
From (\ref{A36}) we easily find the relations (\ref{A33}). Thus, in
the case of three families exist only one independent asymmetry.

From (\ref{A18}), (\ref{A19}), (\ref{A31}) and (\ref{A32}) we find
\begin{equation}\label{A37}
A^{CP}_{e \mu }=
  16 ~J~
\sin\frac{ \Delta m^2_{ 12 }}{ 4E }\, L \,\sin\frac{ \Delta m^2_{ 23
}}{ 4 E }\, L \, \sin\frac{ (\Delta m^2_{ 12}  + \Delta m^2_{ 23} )
}{ 4 E }\, L .
\end{equation}
Thus, the $CP$ asymmetry is proportional to the invariant $J$.  Let
us comment this connection.

The transition probabilities (\ref{A4}) and (\ref{A5}) are invariant
under the  phase transformation
\begin{equation}\label{A38}
U_{li}\to e^{-i\alpha_{l}}~U_{li}~e^{-i\beta_{i}},
\end{equation}
where $\alpha_{l}$ and $\beta_{i}$ are arbitrary constant phases. It
is obvious that
\begin{enumerate}
  \item
The  $CP$ asymmetry is also invariant under the transformation
(\ref{A38}).
  \item
The $CP$ asymmetry is equal to zero in the case of the $CP$
invariance.
\end{enumerate}
 The Jarlskog invariant $J$ satisfies both these
conditions.

As we have seen  there exists only one Jarlskog invariant in the
case of the three-neutrino mixing. This is connected with the fact
that only one phase characterizes the mixing matrix. Let us comment
now this last statement. In the case of the Dirac neutrinos, like in
the quark case,  one physical $CP$ phase characterizes mixing
matrix. In the case of the Majorana neutrinos 3$\times$3 mixing
matrix is characterized by three  $CP$ phases \cite{BHP}. However,
two additional Majorana phases do not enter into expressions for
neutrino and antineutrino transition probabilities \cite{BHP}.

The $3\times3$ $PMNS$ neutrino mixing matrix  can be parameterized
in the same way as $CKM$  quark mixing matrix (see (\ref{78})). For
the Jarlskog invariant we have in this case
\begin{equation}\label{A39}
 J= -\, s_{12\, }s_{13}\, s_{23}\, \sin\delta ~ c^{2}_{13}\, c_{12}\, c_{23}
\end{equation}
It follows from (\ref{A37}) and (\ref{A39})   that in order the $CP$
asymmetry is different from zero it is necessary that not only the
$CP$ phase  but also three mixing angles $\theta_{12}$,
$\theta_{23}$ and $\theta_{13}$ and two mass-squared differences
$\Delta m^2_{ 23 }$ and $\Delta m^2_{1 2 } $ are different from
zero. Thus, in order to reveal violation of the $CP$ invariance in
the lepton sector all three families  must be involved in
oscillations.

 \renewcommand{\theequation}{B-\arabic{equation}}
 \setcounter{equation}{0}
\appendix

\begin{center}
{\bf Appendix B. Evolution equation for $M^{0}-\bar M^{0}$ system}
\end {center}
The physics of the $M^{0}-\bar M^{0}$ system
($M^{0}=K^{0},B_{d,s}^{0},...$) is based on the evolution equation.
We will show here that wave functions of such systems satisfy the
Schrodinger equation with non hermitian Hamiltonian.

Let us consider, as an example, $K^{0}-\bar K^{0}$ system. $K^{0}$
and $\bar K^{0}$ mesons, are particles with the strangeness +1 and
-1, correspondingly. They  are produced in strong interaction
processes in which strangeness is conserved. After $K^{0}$ ( $\bar
K^{0}$) is produced weak interaction in which strangeness is changed
plays the major role: due to weak interaction particles decay and
transitions $K^{0}\leftrightarrows \bar K^{0}$ take place.

 We will present the
total Hamiltonian in the form
\begin{equation}\label{B1}
H=H_{0}+H_{W},
\end{equation}
 where $H_{0}$ is a sum of the free Hamiltonian and the Hamiltonian of
strong and electromagnetic interactions and $ H_{W}$ is the
Hamiltonian of the weak interaction.

Let $|K^{0}\rangle$ and $|\bar K^{0}\rangle$ be the states of
$K^{0}$ and $\bar K^{0}$ in their rest systems. These states  are
eigenstates of the Hamiltonian $H_{0}$ and of the operator of the
strangeness S. Assuming $CPT$ invariance of the Hamiltonian of the
strong and electromagnetic interactions we have
\begin{equation}\label{B2}
    H_{0}\,|K^{0}\rangle=m\,|K^{0}\rangle,~~H_{0}\,|\bar
K^{0}\rangle=m\,|\bar K^{0}\rangle
\end{equation}
and
\begin{equation}\label{B3}
    S~|K^{0}\rangle=|K^{0}\rangle,~~S~|\bar
K^{0}\rangle=-|\bar K^{0}\rangle,
\end{equation}
where $m$ is the mass. Due to the $CPT$ invariance the masses of
$K^{0}$ and $\bar K^{0}$ are the same.

Because of the conservation of the strangeness by the Hamiltonian
$H_{0}$  the  vectors $|K^{0}\rangle$ and $\bar |K^{0}\rangle$ can
not be distinguished from vectors
\begin{equation}\label{B4}
    |K^{0}\rangle'=e^{iS\alpha}|K^{0}\rangle=e^{i\alpha}|K^{0}\rangle,~~
    |\bar K^{0}\rangle'=e^{iS\alpha}|\bar K^{0}\rangle=e^{-i\alpha}|\bar K^{0}\rangle,
\end{equation}
where $\alpha$ is an arbitrary constant phase.

The operators of the $CP$ conjugation and  the strangeness
anticommute with each other
\begin{equation}\label{B5}
CP~S +S~CP=0
\end{equation}
 From this relation we have
\begin{equation}\label{B6}
S~CP|K^{0}\rangle=- CP|K^{0}\rangle.
\end{equation}
 Thus, we have
\begin{equation}\label{B7}
CP \, |K^{0}\rangle=~ \eta_{CP}~|\bar K^{0}\rangle,
\end{equation}
where $\eta_{CP}$ is a $CP$ phase factor. Taking into account the
freedom in the choice of the phases of the vectors $|K^{0}\rangle$
and $|\bar K^{0}\rangle$ we  can put  $\eta_{CP}=1$. In this case we
have
\begin{equation}\label{B8}
CP \, |K^{0}\rangle=|\bar K^{0}\rangle.
\end{equation}
In this review we  used this  choice. However, we demonstrated that
measurable quantities does not depend on the choice of arbitrary
phase factors.

Let us consider now the  Schrodinger  equation for a vector
 $|\Psi(t)\rangle$. We have
\begin{equation}\label{B9}
i\frac{\partial |\Psi(t)\rangle}{\partial t}= H~|\Psi(t)\rangle.
\end{equation}
The formal general solution of the equation (\ref{B9}) has the form
\begin{equation}\label{B10}
|\Psi(t)\rangle=e^{iHt}~|\Psi(0)\rangle,
\end{equation}
where $|~\Psi(0)\rangle$ is the vector of the state at the initial
time $t=0$.

It will be convenient  to present the solution (\ref{B10}) in
another form. Let us denote $|n\rangle$ the normalized eigenvector
of the total Hamiltonian. We have
\begin{equation}\label{B11}
H~|n\rangle= E_{n}~|n\rangle,~~~~~\langle n'|n\rangle=\delta_{n'n}.
\end{equation}

The vector $|\Psi(0)\rangle$ can be developed over the vectors
$|n\rangle$. We have
\begin{equation}\label{B12}
|\Psi(0)\rangle=\sum_{n}|n\rangle~\langle n|\Psi(0)\rangle
\end{equation}
From (\ref{B10}), (\ref{B11}) and (\ref{B12}) we find
\begin{equation}\label{B13}
|\Psi(t)\rangle=\sum_{n}e^{-iE_{n}t}~|n\rangle~\langle
n|\Psi(0)\rangle
\end{equation}
Further for $t\geq 0$ we have
\begin{equation}\label{B14}
e^{-iE_{n}t}=\frac{-1}{2\pi
i}\int^{\infty}_{-\infty}\frac{e^{-iEt}}{E-E_{n}+i\epsilon}~dE
\end{equation}
From (\ref{B11}), (\ref{B13}) and  (\ref{B14}) we find that the
solution of the equation (\ref{B9}) can be presented in the form

\begin{equation}\label{B15}
|\Psi(t)\rangle=\frac{-1}{2\pi i}~\int^{\infty}_{-\infty}G_{+}(E)
~e^{-iEt}~d E~|\Psi(0)\rangle,
\end{equation}
where
\begin{equation}\label{B16}
G_{+}(E)=\frac{1}{E-H+i\epsilon}
\end{equation}
We assume now that the initial state $|\Psi(0)\rangle$ is a
superposition of the states of $K^{0}$ and $\bar K^{0}$ mesons. We
have
\begin{equation}\label{B17}
|\Psi(0)\rangle=\sum_{\alpha=1,2}a_{\alpha}(0)~|\alpha\rangle,
\end{equation}
where $|K^{0}\rangle\equiv |1\rangle$ and $|\bar K^{0}\rangle\equiv
|2\rangle$.

At $t\geq 0$ we have
\begin{equation}\label{B18}
|\Psi(t)\rangle=\sum_{\alpha=1,2}a_{\alpha}(t)~|\alpha\rangle
+\sum_{i}b_{i}(t)~|i\rangle,
\end{equation}
where $a_{1}(t)(a_{2}(t))$ is the amplitude of the probability to
find $K^{0}$ ($\bar K^{0}$) at time $t$ and $|i\rangle$ are states
of the particles which are produced in decays of neutral kaons
($\pi\pi$, $\pi\pi\pi$, $\pi l \nu_{l}$ etc).

From (\ref{B15}) and (\ref{B17}) for the wave function
$a_{\alpha}(t)$  we find the following expression
\begin{equation}\label{B19}
a_{\alpha'}(t)=\langle \alpha'|\Psi(t)\rangle  = \frac{-1}{2\pi
i}~\int^{\infty}_{-\infty}\sum_{\alpha}\langle
\alpha'|~G_{+}(E)~|\alpha\rangle ~e^{-iEt}~d E~a_{\alpha}(0)
\end{equation}
Up to now all our equations were exact. Now we will develop
perturbation theory over the weak interaction. From (\ref{B16}) we
have
\begin{equation}\label{B20}
(E-H_{0}-H_{W}+i\epsilon)~G_{+}(E)=1
\end{equation}
If we multiply this equation by the operator
$\frac{1}{E-H_{0}+i\epsilon}$ from the left we obtain the
Lippman-Schwinger equation for the operator $G_{+}(E)$:
\begin{equation}\label{B21}
G_{+}(E)=\frac{1}{E-H_{0}+i\epsilon}+\frac{1}{E-H_{0}+i\epsilon}~H_{W}~G_{+}(E).
\end{equation}
We will obtain now the matrix element $\langle
~\alpha'|~G_{+}(E)~|~\alpha\rangle $ in the form of perturbation
series. From (\ref{B2}) we find\footnote{The sum $\sum_{i}$ means
sum and integration over corresponding variables in the state
$|i\rangle$ and  sum over all possible states $|i\rangle$.}
\begin{eqnarray}\label{B22}
\langle \alpha'|~G_{+}(E)~|\alpha\rangle
&=&\frac{\delta_{\alpha'\alpha}}{E-m+i\epsilon}
+\frac{1}{E-m+i\epsilon} \sum_{\alpha''} \langle
\alpha'|~H_{W}~|\alpha''\rangle~\langle
~\alpha''|~G_{+}(E)~|~\alpha\rangle +{}\nonumber\\
&&{}+\frac{1}{E-m+i\epsilon} ~\sum_{i} \langle
\alpha'|~H_{W}~|i\rangle~\langle i|~G_{+}(E)|\alpha\rangle.
\end{eqnarray}
Now, taking into account that $\langle i|\alpha\rangle=0$ from
(\ref{B21}) we find
\begin{eqnarray}\label{B23}
\langle i|~G_{+}(E)~|\alpha\rangle&=&\frac{1}{E-E_{i}+i\epsilon}[
\sum_{\alpha''}\langle i|~H_{W}~|\alpha''\rangle~\langle
\alpha''|~G_{+}(E)~|\alpha\rangle +{}\nonumber\\
&&{}+\sum_{i'} \langle ~i|~H_{W}~|~i'\rangle~\langle
i'|~G_{+}(E)~|\alpha\rangle  ]
\end{eqnarray}
This equation can be easily solved by iterations.  Its  solution can
be presented in form of the perturbation series over the weak
interaction. We will consider only the first term of the series.

From  (\ref{B22}) and (\ref{B23}) we find
\begin{equation}\label{B24}
\langle \alpha'|~G_{+}(E)~|\alpha\rangle
=\frac{\delta_{\alpha'\alpha}}{E-m+i\epsilon}
+\frac{1}{E-m+i\epsilon} \sum_{\alpha_{1}} \langle
\alpha'|~R(E)~|\alpha''\rangle~\langle
\alpha''|~G_{+}(E)~|\alpha\rangle,
\end{equation}
where up to the terms of the second order of the perturbation theory
we have
\begin{equation}\label{B25}
\langle \alpha'|~R(E)~|\alpha''\rangle=\langle
\alpha'|~H_{W}~|\alpha''\rangle+\sum_{i}\langle
\alpha'|~H_{W}~|i\rangle~\frac{1}{E-E_{i}+i\epsilon} \langle
i|~H_{W}~|\alpha''\rangle+...
\end{equation}
In the matrix form the equation  (\ref{B24}) can be written as
follows
\begin{equation}\label{B26}
G_{+}(E) =\frac{1}{E-m+i\epsilon} +\frac{1}{E-m+i\epsilon} ~
R(E)~G_{+}(E),
\end{equation}
where $G_{+}(E)$ and $R(E)$ are $2\times2$ matrices with elements
$\langle\alpha'|~G_{+}(E)~|\alpha\rangle$   and
$\langle\alpha'|~R(E)~|\alpha\rangle$. This matrix equation can be
easily solved. We have
\begin{equation}\label{B27}
G_{+}(E) =\frac{1}{E-m- R(E)+i\epsilon}
\end{equation}
From  (\ref{B19}) and  (\ref{B27}) for the wave function $a(t)$ we
find
\begin{equation}\label{B28}
 a(t)= \frac{-1}{2\pi
i}~\int^{\infty}_{-\infty}\frac{e^{-iEt}}{E-m- R(E)+i\epsilon} ~d
E~a(0)
\end{equation}
Because $|R(E)| \ll m$ the pole in the integral  (\ref{B28})  is at
the point $E\simeq m$. We have $ R(E)= R(m)+ (E-m)
\frac{dR}{dE}|_{E=m}+...$. The second term of this expansion is much
smaller than the first one. We will neglect it. This approximation
is called the Weisskopf-Wigner approximation \cite{WeissVig}. In
this approximation we have \footnote{ We took into account  that
imaginary parts of the eigenvalues of $\mathcal{H}$ are negative}
\begin{equation}\label{B29}
a(t)\simeq  \frac{-1}{2\pi
i}~\int^{\infty}_{-\infty}\frac{e^{-iEt}}{E-\mathcal{H}+i\epsilon}
~d E~a(0)=e^{-i\mathcal{H}t}~a(0),
\end{equation}
where
\begin{equation}\label{B30}
\mathcal{H}= m+R(m).
\end{equation}

From (\ref{B29}) we come to the conclusion that the wave function of
$K^{0}-\bar K^{0}$ system satisfies the Schrodinger  equation
\begin{equation}\label{B31}
i\frac{\partial a(t) }{\partial t}=\mathcal{ H}~a(t)
\end{equation}
Let us consider now the  effective Hamiltonian $\mathcal{H}$. Taking
into account the relation
\begin{equation}\label{B32}
\frac{1}{m- E_{i}+i\epsilon} = P~ \frac{1}{m- E_{i}} -i\pi
\delta(E_{i}-m)
\end{equation}
from  (\ref{B25}) we find
\begin{equation}\label{B33}
\mathcal{H}=M -\frac{i}{2}~\Gamma,
\end{equation}
where
\begin{equation}\label{B34}
M_{\alpha'\alpha}= m~\delta_{\alpha'\alpha} +
\langle\alpha'|~H_{W}~|\alpha\rangle+P\sum_{i}\langle\alpha'|~H_{W}~|i\rangle
~ \frac{1}{m- E_{i}}\langle i|~H_{W}~|\alpha\rangle
\end{equation}
and
\begin{equation}\label{B35}
\Gamma_{\alpha'\alpha}=2\pi~\sum_{i}\langle\alpha'|~H_{W}~|i\rangle
~ \langle i|~H_{W}~|\alpha\rangle~\delta(E_{i}-m)
\end{equation}
It follows from these expressions that $M$ and $\Gamma$ are
hermitian matrices:
\begin{equation}\label{B36}
M^{\dagger}=M, ~~~\Gamma^{\dagger}=\Gamma.
\end{equation}
Thus, summarizing, wave function  of $K^{0}-\bar K^{0}$ system
satisfies the Schrodinger  equation with  effective non hermitian
Hamiltonian $\mathcal{H}$ which is given by (\ref{B33}).

Let us consider $M_{11}$ ( $M_{22}$). The first term in (\ref{B34})
is the bare mass of $K^{0}$ ($\bar K^{0}$). The second and third
terms are the corrections to mass. Thus, $M_{11}$ ( $M_{22}$) is the
mass of $K^{0}$ ($\bar K^{0}$) with corrections due to the weak
interaction. From the $CPT$ invariance it follows that
\begin{equation}\label{B37}
M_{11}=M_{22}.
\end{equation}
From  (\ref{B35}) follows that $\Gamma_{11}$ ($\Gamma_{22}$) is
total decay width of $K^{0}$ ($\bar K^{0}$). Taking into account the
$CPT$ invariance we have
\begin{equation}\label{B38}
\Gamma_{11}=\Gamma_{22}.
\end{equation}

 Thus, if the $CPT$ invariance holds  we have
\begin{equation}\label{B39}
\mathcal{H}_{11}=\mathcal{H}_{22}.
\end{equation}
In the case of the $CP$ invariance we have
\begin{equation}\label{B40}
\mathcal{H}_{11}=\mathcal{H}_{22}.
\end{equation}
and
\begin{equation}\label{B41}
\mathcal{H}_{12}=\mathcal{H}_{21}.
\end{equation}
If the relation (\ref{B39}) is violated this means that $CPT$ and
$CP$ are violated. The violation of the relation (\ref{B41}) is a
signature of the $CP$ violation.

The relation (\ref{B41}) was obtained under the assumption that the
arbitrary  phases of the states are chosen in such a way that $|\bar
K^{0}\rangle=CP~|K^{0}\rangle$. If we change the basic states and
instead of  $|K^{0}\rangle$ and $|\bar K^{0}\rangle$  will use
$|K^{0}\rangle'=e^{i\alpha}~|K^{0}\rangle$ and $|\bar
K^{0}\rangle'=e^{-i\alpha}~|\bar K^{0}\rangle$ we will have
\begin{equation}\label{B42}
\mathcal{H}'_{12}=e^{-4i\alpha}\mathcal{H}'_{21}.
\end{equation}
Thus, there is no any relations between the phases of the non
diagonal elements of the matrix $\mathcal{H}$ in the case of the
$CP$ invariance. Only the violation of the relation
\begin{equation}\label{B43}
|\mathcal{H}_{12}|=|\mathcal{H}_{21}|
\end{equation}
 is a signature of the $CP$ violation.

Let us notice that in the case of $T$ invariance we have
\begin{equation}\label{A.43}
|\mathcal{H}_{12}|=|\mathcal{H}_{21}|.
\end{equation}

 It is obvious that all relations we
derived here are also valid for $B^{0}_{d,s}- \bar B^{0}_{d,s}$,
$D^{0}- \bar D^{0}$ and other systems.


\begin{thebibliography}{99}
\bibitem{Wu57} C.S. Wu {\it et al}, Phys.
Rev. {\bf 105} (1957) 1413.
\bibitem{Landau57}
L.~Landau, Nucl. Phys. \textbf{3}, 127 (1957).

\bibitem{Lee-Yang57}
T.D. Lee and C.N. Yang, Phys. Rev. \textbf{105}, 1671 (1957).

\bibitem{Salam57}A. Salam, Nuovo Cimento {\bf 5} (1957) 299.


\bibitem{Goldhaber58}
M.~Goldhaber, L.~Grodzins and A.W. Sunyar, Phys. Rev. \textbf{109},
1015 (1958).



\bibitem{Cronin} J.H. Christenson, J.W. Cronin, V. L. Fitch and
R. Turlay, Phys. Rev. Lett. \textbf{13} (1964) 138.

\bibitem{Wolfenstein64} L. Wolfenstein, Phys. Rev. Lett. \textbf{33} (1964)
562.

\bibitem{NA48} J.R. Batley {\em et al} (NA48 Collaboration)
Phys. Lett. \textbf{B544} (2002) 97.

\bibitem{KTeV}A. Alavi-Harati
{\em et al} (KTeV Collaboration), Phys. Rev.  \textbf{D67} (2003)
012005.




\bibitem{Glashow61} S.L. Glashow, Nucl. Phys. \textbf{22}, 597
(1961).

\bibitem{Weinberg67}
S.~Weinberg, Phys. Rev. Lett. \textbf{19}, 1264 (1967).

\bibitem{Salam68}
A.~Salam, Proc. of the 8$^{\mathrm{th}}$ Nobel Symposium on
\textit{Elementary Particle
  Theory, Relativistic Groups and Analyticity}, edited by N. Svartholm, 1969.


\bibitem{BurasI} A. J. Buras, Proceedings of the Intern. School
on Subnuclear Physics, ed. A. Zichichi (World Scientific 2000) pp.
200-237; hep-ph /0101336.




\bibitem{Kobayashi-Maskawa73}
M.~Kobayashi and T.~Maskawa, Prog. Theor. Phys. \textbf{49}, 652
(1973).


\bibitem{CP} Proc. of \textit{CP Violation}, edited by C. Jarlskog, Advanced
  Series in High Energy Physics Vol. 3, p. 3, World Scientific, Singapore,
  1989.

\bibitem{BaBar} The BaBar Physics Book, eds. P.Harrison and H. Quinn
(1998) SLAC-R-0504.

\bibitem{Hflavors} Heavy Flavors II, eds. A. Buras and M. Lindner,
World Scientific.

\bibitem{Branco} G.C. Branco, L.Lavoura and J.P. Silva, "CP
Violation" Clarendon Press-Oxford (1999)

\bibitem{Bigi} I.I. Bigi and A.I. Sanda, "CP
Violation" Cambridge Monographs on Particle Physics, Nuclear Physics
and Cosmology, Cambridge University Press, Cambridge (2000)


\bibitem{Bphysics} B-Physics at the Tevatron:run II and beyond,
hep-ph/0201071.

\bibitem{Kleinknecht} K. Kleinknecht, Springer Tracts Mod. Phys.
\textbf{195} (2003) 1.

\bibitem{Battaglia} M. Battaglia {\em et al}, The CKM matrix and the
unitary triangle (CERN, Geneva, 2003), hep-ph/0304132.



\bibitem{Silva} J.P. Silva, Phenomenological aspects of $CP$
violation, hep-ph/0410351.


\bibitem{Nir} Y. Nir, CP violation in meson decays, hep-ph/0510413.


\bibitem{Fleisher} R. Fleisher, Flavor Physics and $CP$ violation,
hep-ph/0608019.



\bibitem{BurasII} A. J. Buras, Flavor Physics and $CP$ violation, hep-ph/0505175.

\bibitem{SuperB}J. A. Hewett {\em et al}, The Discovery Potential of
a Super B Factory, hep-ph/0503261.


\bibitem{Weinberg} S. Weinberg, The Quantum Theory of Fields v.I
Foundations, Cambridge University Press.

\bibitem{Bilenky} S.M. Bilenky, Introduction to Feynman Diagrams and
Electroweak Interaction Physics, Editions Frontiers, 1964.


\bibitem{Cabibbo} N. Cabibbo, Phys. Lett. \textbf{10} (1963) 531.


\bibitem{Jarlskog} C. Jarlskog, Phys. Rev. Lett. \textbf{55} (1985)
1039; C. Jarlskog, Z. Phys. \textbf{C29} (1985) 491. C. Jarlskog and
R.Stora, Phys. Lett. \textbf{B208} (1988) 268.


\bibitem{PDG}W. M. Yao {\em et al}
(Particle Data Group ), J. Phys.G \textbf{33} (2006) 1.







\bibitem{CeccucciPDG} A. Ceccuci, Z. Ligeti and Y. Sakai "The CKM
quark-mixing matrix" in W. M, Yao {\em et al} (Particle Data Group),
J. Phys.G \textbf{33} (2006) 1.

\bibitem{Hardy} J. C. Hardy and I. S. Towner,  Phys. Rev. Lett. \textbf{94} (2005)
092502.

\bibitem{Savard} G. Savard {\em et al},  Phys. Rev. Lett. \textbf{95} (2005)
102501.

\bibitem{Pocanic} D. Pocanic {\em et al},  Phys. Rev. Lett. \textbf{93} (2004)
181803.



\bibitem{Marciano} E. Blucher and W. J. Marciano, "$V_{ud}, V_{us}$.
The Cabibbo angle and CKM unitarity"  in W. M, Yao {\em et al}
(Particle Data Group ), J. Phys.G \textbf{33} (2006) 1.

\bibitem{Alexopoulos} T.Alexopoulos {\em et al} (KTeV
collaboration), Phys. Rev. Lett. \textbf{93} (2004) 181802;
T.Alexopoulos {\em et al} Phys. Rev.  \textbf{D70} (2004) 092006.

\bibitem{Ambrosino}F. Ambrosino {\em et al} (KLOE collaboration)
, Phys. Lett. \textbf{B632} (2006) 43.

\bibitem{Lai} A. Lai {\em et al} (NA48 collaboration)
, Phys. Lett. \textbf{B602} (2004) 41.




\bibitem{Leutwyler} H. Leutwyler and M. Roos, Z. Phys,
\textbf{C25} (1984) 91.


\bibitem{Bernard} C. Bernard {\em et al}, PoS \textbf{LAT2005}
(2005) 025; hep-lat/0509137.


\bibitem{CabibboI} N. Cabibbo, E. S. Swallow and R. Winston,
Phys. Rev. Lett. \textbf{92} (2004) 251803.


\bibitem{Gamiz} E. Gamiz {\em et al}, Phys. Rev. Lett. \textbf{94} (2005) 011803.


\bibitem{Aubin} C. Aubin {\em et al}, Phys. Rev. Lett. \textbf{94} (2005)
011601.

\bibitem{Artuso} M. Artuso, Int.J.Mod.Phys. \textbf{A21} (2006) 1697;
hep-ex/0510052.


\bibitem{Abren} P. Abren {\em et al}, Phys. Lett. \textbf{B439} (1998)
209.

\bibitem{Bigi} I.I.Y.Bigi {\em et al}, Phys. Rev. Lett. \textbf{71} (1993)
496.

\bibitem{Manohar} A.V. Manohar and M.B. Weise, Phys. Rev. \textbf{D49} (1994)
1310.

\bibitem{KowalewskiPDG} B. Kowalewski and T. Mannel, "Determination of
$V_{cb}$ and $V_{ub}$"  in W. M, Yao {\em et al} (Particle Data
Group ). J. Phys.G \textbf{33} (2006) 1.


\bibitem{Isgur} N. Isgur and M.B. Wise,  Phys. Lett. \textbf{B232} (1989)
113; Phys. Lett. \textbf{B237} (1990) 527.


\bibitem{Shifman} M.A. Shifman and M.B. Voloshin, Sov. J. Nucl. Phys.
 \textbf{47} (1988)
511 (Yad. Phys. \textbf{47} (1988) 801.

\bibitem{Okamoto} M. Okamoto  {\em et al}, Nucl.  Phys. (Proc. Supp.) \textbf{B140}
(2005) 461.

\bibitem{Shigemitsu} J. Shigemitsu {\em et al}, Nucl.  Phys. (Proc. Supp.) \textbf{B140}
(2005) 464.


\bibitem{Barbero} E. Barbero {\em et al} (Heavy Flavor Areraging
Group), hep-ex/0603003.

\bibitem{Gray} A. Gray {\em et al}, Phys. Rev. Lett. \textbf{95} (2005)
212001.

\bibitem{Aoki} S. Aoki {\em et al}, Phys. Rev. Lett. \textbf{91} (2003)
212001.

\bibitem{Okamoto} M. Okamoto {\em et al}, PoS \textbf{LAT2005} (2005)
013.

\bibitem{CDFdeltas} A. Abulencia {\em et al} (CDF collaboration),
 Phys.Rev.Lett.\textbf{97} (2006) 062003; ~ hep-ex/0606027.


\bibitem{Acosta} D. Acosta {\em et al} (CDF collaboration),
Phys. Rev. Lett. \textbf{95} (2005) 102002.

\bibitem{AbazovD0}V.M. Abazov {\em et al} (D0 collaboration),
Phys.Lett.\textbf{B639} (2006) 616;~~hep-ex/0603002.


\bibitem{Wolfensteinparam} L. Wolfenstein, Phys. Rev. Lett. \textbf{51} (1983) 1945.



\bibitem{Burasparam} A. Buras, E.M. Lautenbacher and G. Ostermaier,
Phys. Rev. \textbf{D50} (1994) 3433.

\bibitem{WolfensteinPDG} L. Wolfenstein and T.G. Trippe "$CP$ vilation
in $K_{L}$ decays" in W. M, Yao {\em et al} (Particle Data Group),
J. Phys.G \textbf{33} (2006) 1.


\bibitem{Calangelo} G. Calangelo, J. Gasser and H. Leutwyler, Nucl.
 Phys.\textbf{B603} (2001) 125.


\bibitem{Nierste} U. Nierste, hep-ph/0612310.


\bibitem{Sandra}A.B. Carter and A.I. Sanda, Phys. Rev. Lett. \textbf{45}
(1980) 952;~~ Phys. Rev. \textbf{D23} (1981) 1567;~~I.I. Bigi and
A.I. Sandra, Nucl. Phys. \textbf{193} (1981)85.






\bibitem{BaBarI} B. Aubert {\em et al} (BaBar Collaboration),
Phys.Rev. Lett. \textbf{D 66} (2002) 071102.



\bibitem{BelleI} K. Abe {\em et al} (Belle Collaboration),
Phys.Rev.  \textbf{89} (2002) 201802.


\bibitem{BaBarII} B. Aubert {\em et al} (BaBar Collaboration),
hep-ex/0607107.


\bibitem{BelleII} K. Abe {\em et al} (Belle Collaboration),
hep-ex/0608039.


\bibitem{UTfitI} M. Bona {\em et al} (UTfit Collaboration),
hep-ph/0501199.


\bibitem{UTfitII} M. Bona {\em et al} (UTfit Collaboration),
JHEP \textbf{0610} (2006) 081; hep-ph/0606167.


\bibitem{CKMfitter} J. Charles {\em et al} (CKMfitter group),
Eur. Phys. J. \textbf{41} (2005)1; hep-ph/0406184; hep-ph/0606046.

\bibitem{HFAG} E. Barberio {\em et al} (Heavy Flavor Averaging
Group), hep-ex/0603003.


\bibitem{Ramaangles} M. Rama, Proceedings of XXI Rencontres de Physique
de la Vallee d'Aoste, March 4-10, 2007.



\bibitem{Wu57}C.S. Wu {\it et al}, Phys.
Rev. {\bf 105} (1957) 1413.


\bibitem{FeyGel} R.P.Feynman and M.Gell-Mann, Phys.
Rev.{\bf 109} (1958) 193.


\bibitem{MarSud}E.C.G. Sudarshun and R. Marshak, Phys. Rev.
{\bf 109} (1958) 1860.





\bibitem{SK} Y. Ashie {\em et al.,} (Super-Kamiokande Collaboration),
Phys. Rev. Lett. \textbf{93} (2004) 101801; Phys. Rev. \textbf{D71}
(2005) 11205.


\bibitem{SNO}SNO Collaboration, Phys. Rev. Lett.
\textbf{87} (2001) 071301;~\textbf{89} (2002) 011301 ;~\textbf{89}
(2002) 011302;~\textbf{92} (2004) 181301.

\bibitem{Kamland} T.Araki {\em et al.} (KamLAND Collaboration),
T.Araki {\em et al.}, Phys. Rev. Lett. \textbf{94} (2005) 081801;~
hep-ex/0406035.


\bibitem{Cl}B. T. Cleveland {\it et al.}, Astrophys. J. {\bf
496} (1998) 505.

\bibitem{Gallex} W. Hampel
{\it et al.} (GALLEX Collaboration), Phys. Lett. {\bf B 447} (1999)
127 ;\, GNO Collaboration, M. Altmann {\it et al.}, Phys. Lett. {\bf
B 490} (2000) 16 ;\, Nucl. Phys. Proc. Suppl. {\bf 91} (2001) 44;
Phys. Lett. \textbf{B 616} (2005) 174.

\bibitem{Sage}
J. N. Abdurashitov {\it et al.} (SAGE Collaboration),
 Phys. Rev. {\bf C 60} (1999) 055801; \,Nucl. Phys. Proc. Suppl. {\bf 110}
(2002) 315;~\,Nucl. Phys. Proc. Suppl. {\bf 118} (2003) 39.

\bibitem{SKsol} S.~Fukuda {\it et al.} (Super-Kamiokande Collaboration),  Phys. Rev. Lett.
 {\bf 86} (2001) 5651.

\bibitem{K2K} M.H. Ahn {\em et al.} (K2K Collaboration),
Phys. Rev. Lett. \textbf{90} (2003) 041801.

\bibitem{Minos}  D.G. Michael {\it et al.} ( MINOS Collaboration),
 Phys.Rev.Lett. {\bf97} (2006) 191801; arXiv:hep-ex/0607088.

\bibitem{BilGiunti} S. Bilenky and  C. Giunti,  Int.J.Mod.Phys. {\bf A16} (2001) 3931;
  hep-ph/0102320.
\bibitem{BilMateev} S.M. Bilenky and M.D. Mateev,
Phys.Part.Nucl.{\bf 38} (2007) 117; hep-ph/0604044.

\bibitem{BP}
B.~Pontecorvo, J. Exptl. Theoret. Phys. \textbf{33} (1957) 549.
[Sov. Phys. JETP \textbf{6} (1958) 429 ]; J. Exptl. Theoret. Phys.
\textbf {34} (1958) 247 [Sov. Phys. JETP \textbf {7} (1958) 172 ].


\bibitem{MNS} Z. Maki, M. Nakagawa and S. Sakata, Prog. Theor.
Phys. {\bf 28} (1962) 870.


\bibitem{BPet}
S.M. Bilenky and S.T. Petcov, Rev. Mod. Phys. \textbf{59}, 671
(1987).


\bibitem{BGG} S.M.\, Bilenky, C.\, Giunti, and
W.\,Grimus. Prog. Part. Nucl. Phys. {\bf 43} (1999) 1.




\bibitem{BilNedPet} S.M. Bilenky, N.P. Nedelcheva and S.T. Petcov,
Nucl. Phys.\textbf{B247}(1984) 61;\,B. Kayser, Phys.
Rev.\textbf{D30} (1984) 1023.



\bibitem{BHP}S.M. Bilenky, J. Hosek and S.T. Petcov,
Phys. Lett.\textbf{B94} (1980) 495.


\bibitem{WeissVig}V.Weisskopf and E. Wigner, Z. fur Physik, \textbf{63} (1930)
54,~~Z. fur Physik, \textbf{65} (1930) 18.



























































\end{thebibliography}
\end{document}